\documentclass[11pt]{article}
\usepackage{amsfonts,amsmath,amssymb}
\usepackage{enumerate}
\usepackage[hidelinks]{hyperref}
\usepackage{bbm}
\usepackage{nicefrac}
\usepackage[all]{xy}
\usepackage{graphicx}
\usepackage{bm}
\usepackage{makecell}
\usepackage[table]{xcolor}

\usepackage{upgreek}
\usepackage{slashed}

\newcommand{\gr}{\mathring{g}}

\usepackage{float}			% NEW Helps position tables right here [H]
\usepackage{lscape}                 % NEW creates pages in landscape format

\usepackage{booktabs}

\newcommand{\be}{\begin{equation}}
\newcommand{\ee}{\end{equation}}

\newcommand{\bea}{\begin{eqnarray}}
\newcommand{\eea}{\end{eqnarray}}

\newcommand{\bes}{\begin{subequations}}
\newcommand{\ees}{\end{subequations}}

\newcommand{\cN}{{\cal N}}

\newcommand{\cR}{{\cal R}}

\newcommand{\cA}{{\cal A}}

\def\sst#1{{\scriptscriptstyle #1}}

\def\0{{\sst{(0)}}}
\def\1{{\sst{(1)}}}
\def\2{{\sst{(2)}}}
\def\3{{\sst{(3)}}}
\def\4{{\sst{(4)}}}
\def\5{{\sst{(5)}}}
\def\6{{\sst{(6)}}}
\def\7{{\sst{(7)}}}
\def\8{{\sst{(8)}}}

\def\cA{{{\cal A}}}

\def\cV{{{\cal V}}}
\def\cM{{{\cal M}}}
\def\cY{{{\cal Y}}}

\newcommand{\cT}{{\cal T}}

\newcommand{\cL}{{\cal L}}

\newcommand{\vol}{\textrm{vol}}

\allowdisplaybreaks  %Introduces page breaks inside \eqnarray%

\usepackage{multirow}
\usepackage{rotating}

\newcommand{\ba}{\begin{align}}
\newcommand{\ea}{\end{align}}

\newcommand{\bse}{\begin{subequations}}
\newcommand{\ese}{\end{subequations}}

\allowdisplaybreaks

\newcommand{\fT}{\mathcal{T}}
\newcommand{\gM}{\mathcal{M}}
\newcommand{\Aa}{\mathcal{A}}

\newcommand{\Fa}{\mathcal{F}}

\newcommand{\SO}[1]{\mathrm{SO}( #1 )}

\usepackage{lipsum}

\newlength\Colsep
\newcommand{\uA}{{\underline{A}}}
\newcommand{\uB}{{\underline{B}}}
\newcommand{\uC}{{\underline{C}}}
\newcommand{\uD}{{\underline{D}}}
\newcommand{\uE}{{\underline{E}}}
\newcommand{\uF}{{\underline{F}}}
\newcommand{\uG}{{\underline{G}}}

\newcommand{\uM}{\underline{M}}
\newcommand{\uN}{\underline{N}}
\newcommand{\uP}{\underline{P}}
\newcommand{\uQ}{\underline{Q}}
\newcommand{\uR}{\underline{R}}
\newcommand{\uS}{\underline{S}}

\begin{document}

\makeatletter
\renewcommand{\theequation}{\thesection.\arabic{equation}}
\@addtoreset{equation}{section}
\makeatother

\begin{titlepage}

\begin{flushright}
IFT-UAM/CSIC-26-42
%
%\today
\end{flushright}

\vspace{20pt}

   \begin{center}
   \baselineskip=16pt

   \begin{Large}\textbf{
Kaluza-Klein trombone mass matrices \\[8pt]
\; \ and universal class $\mathcal{R}$ operator spectra
}
   \end{Large}

\vspace{25pt}

{\large Martín Pico$^{1,2}$ and  Oscar Varela$^{3,1}$ }

\vspace{30pt}

	\begin{small}

	{\it $^1$  Instituto de F\'\i sica Te\'orica UAM/CSIC, 28049 Madrid, Spain} 

	\vspace{10pt}

	{\it $^2$  Departamento de F\'\i sica Te\'orica, Universidad Aut\'onoma de Madrid, \\
	Cantoblanco, 28049 Madrid, Spain} 

	\vspace{10pt}
          
   {\it $^3$ Department of Physics, Utah State University, Logan, UT 84322, USA} 
		
	\end{small}

\vskip 70pt

\end{center}

\begin{center}
\textbf{Abstract}
\end{center}

\begin{quote}

We determine universal sectors of the light operator spectrum that are present in all 
A$_{N-1}$ superconformal field theories $T_N[\Sigma_3]$ of class $\mathcal{R}$, with $\Sigma_3$ a compact hyperbolic three-manifold. We do this at large $N$ by holographically mapping the problem to a Kaluza-Klein spectral analysis over previously known dual anti-de Sitter backgrounds related to wrapped M5-brane configurations. Associated to these, $D=4$ $\mathcal{N}=8$ supergravities have been recently constructed that allow for the application of spectral techniques based on exceptional generalised geometry/field theory. We derive the Kaluza-Klein mass matrices that arise when the seed lower-dimensional maximal supergravity involves gaugings of the trombone scaling symmetry, as pertains to the case at hand, and diagonalise them to find the universal Kaluza-Klein states. In general, these universal spectra are only defined locally, and we give a prescription to isolate globally-defined sectors therein.

\end{quote}

\vfill

\end{titlepage}

\tableofcontents

%%%%%%%%%%%%%%%%%%%%%%%%%%%%%%%%%%%%%%%%%%%%%%%%%%%%%%%%%%%%%%%%%%%%%%%%%%

%%%%%%%%%%%%%%%
%%%%%%%%%%%%%%%

\section{Introduction}

%%%%%%%%%%%%%%%
%%%%%%%%%%%%%%%

Three-dimensional (3d) $\mathcal{N}=2$ superconformal field theories (SCFTs) $T_N[\Sigma_3]$ of class $\mathcal{R}$ \cite{Dimofte:2011ju,Dimofte:2011py} with gauge group A$_{N-1}$ arise from twisted compactification of the six-dimensional $(2,0)$ theory that resides on a stack of $N$ M5-branes wrapped on three-manifolds $\Sigma_3$. These theories are generically strongly coupled and do not admit conventional Lagrangian descriptions. Nevertheless, protected observables of $T_N[\Sigma_3]$ can be related to topological invariants of $\Sigma_3$ through the so-called 3d--3d correspondence \cite{Dimofte:2011ju,Dimofte:2011py,Dimofte:2014zga}. This has enabled the computation of supersymmetric partition functions and special limits of the superconformal index for broad classes of $T_N[\Sigma_3]$ theories \cite{Gang:2014ema,Gang:2014qla}. The 3d--3d correspondence  has also played a central role in establishing large-$N$ results relevant for holography, like the characteristic $N^3$ scaling of degrees of freedom of M5-brane related SCFTs.

When $\Sigma_3 = H^3/\Gamma$ is realised as a special Lagrangian (SLAG) hyperbolic cycle in a Calabi--Yau threefold, $T_N[\Sigma_3]$ is holographically dual to the anti-de Sitter AdS$_4\times(\Sigma_3\rtimes S^4)$ solutions of $D=11$ supergravity constructed in \cite{Gauntlett:2006ux} following \cite{Maldacena:2000mw}. Here, $\Gamma$ is a discrete group of isometries that renders $\Sigma_3$ compact, and we have used the symbol $\rtimes$ to signify that the four-sphere $S^4$ is fibred over $\Sigma_3$ as a result of the topological twist. A related class of $\cN=1$ solutions of $D=11$ supergravity, of the same schematic form $\mathrm{AdS}_4 \times (\Sigma_3 \rtimes S^4)$ but with a different twist of $S^4$ over $\Sigma_3 = H^3/\Gamma$, was similarly constructed in \cite{Acharya:2000mu}. The latter $\cN=1$ solutions arise near the horizon of a large number $N$ of M5-branes wrapped around a hyperbolic associative three-cycle (A3C) $\Sigma_3 = H^3/\Gamma$ of a G$_2$-holonomy seven-dimensional manifold. These configurations should be dual to 3d $\cN=1$ SCFTs which are, however, much less understood than their class $\cR$ counterparts. 

Holographic studies of class $\mathcal{R}$ theories have largely focused on thermodynamic observables. A notable development in this regard is the microscopic description of the entropy of supersymmetric rotating black holes in AdS$_4$ \cite{Bobev:2019zmz,Benini:2019dyp} building on the 3d-3d correspondence. At large $N$ and in a Cardy-like limit, the class $\cR$ superconformal index remarkably accounts for the leading Bekenstein-Hawking entropy and some subleading corrections. These computations probe the theory at 
the level of its partition functions and index and, thus, capture generic features of the class $\cR$ operator spectrum in the relevant limits. However, they are unable to provide a detailed description of the operator spectrum itself. In fact, the spectrum of local operators of class $\mathcal{R}$ theories and, in particular, the conformal dimensions of light operators that remain of order one in the large-$N$ limit, has remained largely uncharacterised in the literature. This is a significant gap, as the spectrum of local operators constitutes the fundamental data of any given conformal field theory, together with the operator product expansion coefficients. 

In this paper, we partially fill that gap. We characterise, holographically and at large $N$, a universal sector of the light operator spectrum of A$_{N-1}$ class $\cR$ SCFTs $T_N[\Sigma_3]$ associated to hyperbolic $\Sigma_3 =H^3/\Gamma$. By universal we mean that these operators are present for all such $T_N[\Sigma_3]$ theories, regardless of the choice of $\Sigma_3 =H^3/\Gamma$. This sector of the light operator spectrum of $T_N[\Sigma_3]$ is thus insensitive to the details of the specific discrete group $\Gamma$ chosen to compactify the hyperbolic cycle $\Sigma_3$ wrapped by the M5-branes. Similarly, we also determine a universal sector of the light operator spectrum of the holographic 3d $\cN=1$ SCFT cousins of the class $\cR$ SCFTs mentioned above. Only the conformal dimensions of a handful of light universal operators were previously known for these 3d $\cN=2$ \cite{Donos:2010ax} and $\cN=1$ \cite{Acharya:2000mu} SCFTs. Here, we will extend those results by providing infinite families of operators in representations of the relevant OSp$(4|\cN)$ supersymmetry algebras that we will describe in detail. 

Recall that the light operator spectrum of a holographic SCFT is in one-to-one correspondence with the spectrum of Kaluza-Klein (KK) excitations above the dual AdS bulk solutions \cite{Maldacena:1997re,Gubser:1998bc,Witten:1998qj}. Clear-cut as this prescription is, it is also a bottle-neck in practice, as the face-value KK spectral problem is usually quite hard \cite{Kim:1985ez}. Most of the difficulties stem from two sources. Firstly, the gauge symmetries of the higher-dimensional supergravity must be fixed in a compatible way with the lower-dimensional physics. Secondly, and perhaps most importantly, the higher-dimensional supergravity fields must undergo a complicated expansion in different scalar, spinor and vector harmonics of the compactification space, in suitable representations of the residual symmetry.

Powerful KK spectral methods have been recently introduced \cite{Malek:2019eaz,Malek:2020yue} (see also \cite{Varela:2020wty,Cesaro:2020soq}) that, under some assumptions, bypass both problems altogether. These methods are based on exceptional field theory (ExFT) \cite{Hohm:2013pua,Hohm:2013vpa,Hohm:2013uia}. Like its closely related exceptional generalised geometry (ExGG) formalism \cite{Coimbra:2011nw,Coimbra:2011ky,Coimbra:2012af}, ExFT is a reformulation of the higher-dimensional supergravities that makes manifest the exceptional symmetry E$_{d(d)}$ of their maximally supersymmetric lower $D$-dimensional counterparts, see \cite{Berman:2020tqn} for a review. Roughly speaking, ExFT/ExGG unearth E$_{d(d)}$ by introducing suitable gauge-fixings, thus addressing the first issue highlighted above. The second, multiple harmonic, problem is dramatically simplified when there is an underlying consistent truncation to a $D$-dimensional maximal supergravity that contains the AdS solutions of interest as vacua. In this case, the linearised maximal supergravity can be interpreted as level $k=0$ in the KK expansion, and the higher KK levels are obtained by expanding the ExFT fields only in scalar harmonics of the compactification space or, more precisely, in KK graviton mass eigenstates. Under these conditions, the  (differential) KK spectral problem drastically reduces to the (algebraic) diagonalisation of (infinite-dimensional but block-diagonal with finite-dimensional blocks) mass matrices. 

Recent developments \cite{Pico:2025cmc,Pico:2026rji} render in principle the AdS$_4 \times (\Sigma_3 \rtimes S^4)$ M5-brane solutions of \cite{Gauntlett:2006ux,Acharya:2000mu} amenable to the ExFT KK spectral techniques of \cite{Malek:2019eaz,Malek:2020yue,Varela:2020wty,Cesaro:2020soq}. A $D=4$ $\cN=8$ gauged supergravity \cite{deWit:2007mt,LeDiffon:2008sh,LeDiffon:2011wt} with gauge group TCSO$(5,0,0;\mathrm{V})$ \cite{Pico:2025cmc}, arises by consistent truncation \cite{Pico:2026rji} of $D=11$ supergravity on the internal spaces $\Sigma_3 \rtimes S^4$ of the SLAG and A3C AdS$_4 \times (\Sigma_3 \rtimes S^4)$ solutions of \cite{Gauntlett:2006ux,Acharya:2000mu}. From a $D=4$ perspective, these solutions manifest themselves as supersymmetry breaking, $\cN=2$ for SLAG and $\cN=1$ for A3C, vacua of the $\cN=8$ TCSO$(5,0,0;\mathrm{V})$ theory. By the above arguments, the linearisation of the $\cN=8$ supergravity at these vacua provides the lowest level in the KK expansion, and the higher levels are obtained by expanding the ExFT fields about these vacua in graviton eigenstates. As emphasised in \cite{Pico:2026rji}, the $D=11$ truncation on $\Sigma_3 \rtimes S^4$ only holds locally, though. This is like the similar constructions of \cite{Bhattacharya:2024tjw,Varela:2025xeb,BKV2026} but critically different from most cases dealt with the ExFT spectral technology of \cite{Malek:2019eaz,Malek:2020yue,Varela:2020wty,Cesaro:2020soq}, where the truncations are globally well defined.

The local character of the underlying truncation can be traced back to the fact that it proceeds \cite{Pico:2026rji} by replacing $\Sigma_3 = H^3/\Gamma$ with the locally diffeomorphic group manifold $G_3$ of Bianchi type V. Being non-unimodular, the latter is non-compact and, thus, globally different from $\Sigma_3 = H^3/\Gamma$. In turn, this induces a gauging of the trombone scaling symmetry $\mathbb{R}^+$ of the underlying $D=4$ $\cN=8$ TCSO$(5,0,0;\mathrm{V})$ supergravity \cite{Pico:2025cmc}. This local feature will lead us to analyse the applicability of the ExFT KK techniques to the $\Sigma_3 \rtimes S^4$ solutions at hand. We will argue that TCSO$(5,0,0;\mathrm{V})$ is still helpful to implement this programme, as long as one considers the resulting local spectra as a larger pool from which to extract bonafide physical modes by additional criteria. As such, we will refer to these KK spectra as putative. One such additional criterion, though arguably not the only one, is provided by requiring global definiteness on the total bundles $\Sigma_3 \rtimes S^4$.

Thus, the mass matrices of \cite{Malek:2019eaz,Malek:2020yue,Varela:2020wty,Cesaro:2020soq} must first be adapted to the case where the underlying lower-dimensional maximal supergravity includes trombone gaugings. We derive these new trombone-augmented KK mass matrices from ExFT in section \ref{sec:KKMassMat}. In section \ref{sec:Spectra}, we use these  to compute the putative local KK spectra of the SLAG \cite{Gauntlett:2006ux} and A3C \cite{Acharya:2000mu} AdS$_4 \times (\Sigma_3 \rtimes S^4)$ M5-brane solutions. Their global spectra are also obtained in that section by imposing appropriate globally defined generalised structures following \cite{Cassani:2019vcl}, see also \cite{Cassani:2020cod,Blair:2024ofc,Josse:2025uro}. Section \ref{sec:Discussion} concludes with some discussion. Further details and technicalities are contained in the appendices.

%%%%%%%%%%%%%%%
%%%%%%%%%%%%%%%

\section{Kaluza-Klein trombone mass matrices} \label{sec:KKMassMat}

%%%%%%%%%%%%%%%
%%%%%%%%%%%%%%%

We start by presenting the KK mass matrices that arise when the embedding tensor of the underlying lower-dimensional maximal supergravity contains non-vanishing trombone components, and postpone a discussion of their applicability to section~\ref{sec:Spectra}. Here we simply list the mass matrices, leaving the details of their derivation by linearisation of the E$_{d(d)}$ ExFT equations of motion to appendix~\ref{eq:DerMassMat}. While we are mostly interested in $d=7$, corresponding to an underlying $D=4$ $\cN=8$ supergravity, our results are generic for any $d$. For example, these tromboneful mass matrices also apply to the $d=6$, $D=5$ $\cN=8$, cases of \cite{Bhattacharya:2024tjw,Varela:2025xeb,BKV2026}.

%%%%%%%%%%%%%%%
\subsection{Expansion in KK perturbations} \label{sec:FlucExp}
%%%%%%%%%%%%%%%

In order to set up our notation, let us first review some preliminary material including the particularisation of E$_{d(d)}$ ExFT/ExGG \cite{Hohm:2013pua,Hohm:2013vpa,Hohm:2013uia,Coimbra:2011nw,Coimbra:2011ky,Coimbra:2012af} to a background $D$-dimensional maximal gauged supergravity \cite{Berman:2012uy,Lee:2014mla,Hohm:2014qga}, and the expansion in perturbations \cite{Malek:2019eaz,Cesaro:2020soq} over that background. 

The bosonic field content of ExFT includes external and internal metrics, $\overline{g}_{\mu\nu}(x,y)$, $G_{MN}(x,y)$, and gauge field one-forms, $\cA_\mu{}^{M}(x,y)$, with respect to external GL$(D)$ diffeomorphisms. Depending on $d$, the bosonic field content may include other fields, like GL$(D)$ two-forms, that we will subsequently ignore. The fermionic sector includes a gravitino, $\psi_\mu^i (x,y)$, and spin-$1/2$ fields, $\chi^{ijk}(x,y)$ \cite{Godazgar:2014nqa,Musaev:2014lna}. Curved indices $\mu$, $M$ and $i$ respectively label the fundamental representations of GL$(D)$, E$_{d(d)}$ (the $\bm{56}$ for $d=7$ or the $\bm{27}$ for $d=6$) and the maximal compact subgroup of the latter (SU$(8)$ or USp$(8)$ for $d=7$ or $d=6$). Also, we have used $x$ and $y$ to generically denote external and internal coordinates, respectively. We now put ExFT on a background internal space for which there exists a maximally supersymmetric consistent truncation down to $D$-dimensional maximal gauged supergravity. This truncation is governed by the generalised Scherk-Schwarz relations \cite{Berman:2012uy,Lee:2014mla,Hohm:2014qga}
\begin{eqnarray} \label{eq:GenSSBos}
&  \Delta(x,y) =  \Delta(y) \; , \qquad
\overline{g}_{\mu \nu}(x,y)=  \overline{g}_{\mu \nu}(x) \; , \nonumber \\[4pt]
& \cA_\mu{}^M (x,y) = \hat{E}^M{}_{\uN}(y) \, A_\mu{}^{\uN}(x) \; , \quad
(G^{-1})^{MN} (x,y) = \hat{E}^M{}_{\uP}(y) \hat{E}^N{}_{\uQ}(y) \, M^{\uP\uQ}(x) \; ,  \;
\end{eqnarray}
for the bosons, and \cite{Hohm:2014qga} 
\begin{eqnarray} \label{eq:GenSSFer}
\psi_\mu{}^i = \Delta^{\frac12}  \psi_\mu{}^{i}(x)  \; , \qquad
\chi^{ijk} = \Delta^{-\frac12}  \chi^{ijk}(x) \; . 
\end{eqnarray}
for the fermions. In these equations, $\Delta(y)$ is a warp factor, $\hat{E}^M{}_{\uN}(y)$ is a generalised frame on the internal background space, while $\overline{g}_{\mu \nu}(x)$, $A_\mu{}^{\uM} (x)$, $M_{\uM\uN}(x) = (\cV(x) \cV(x)^{\textrm{T}})_{\uM\uN}$ (with inverse  $M^{\uM\uN}(x)$) and $\psi_\mu^i(x)$, $\chi^{ijk}(x)$ are the bosonic (metric, vector and scalar) and fermionic (gravitino and spin-$1/2$) fields of $D$-dimensional $\cN=8$ supergravity, with flat, underlined indices again labelling the fundamental of E$_{d(d)}$. The underlying gauged supergravity is equipped with an embedding tensor \cite{deWit:2007mt,LeDiffon:2008sh,LeDiffon:2011wt,deWit:2004nw,Varela:2025xeb} 
\begin{eqnarray} \label{eq:defX}
X_{\uM \uN}{}^{\uP} \equiv
\Theta_{\uM \uN}{}^{\uP}
 +  \Big(  \alpha \tfrac{D-2}{D-1} \,  \mathbb{\uP}^{\uQ}{}_{\uM}{}^{\uP}{}_{\uN}  - \delta_{\uM}^{\uQ}\delta_{\uN}^{\uP} \Big)
\vartheta_{\uQ}  \; , \quad 
\textrm{with \, $\Theta_{\uM \uN}{}^{\uP} \equiv \Theta_{\uM}{}^\alpha (t_\alpha)_N{}^P$,}
\end{eqnarray}
given by the (constant) intrinsic torsion of the inverse ExFT generalised frame \cite{Lee:2014mla,Cassani:2019vcl},
\begin{equation} \label{eq:CTCon}
L_{\hat E_{\uM} (y)} \hat E_{\uN}(y) = - X_{\uM\uN}{}^{\uP} \, \hat E_{\uP} (y)  \; .
\end{equation}
Here, $L$ is the generalised Lie derivative \cite{Coimbra:2011ky}. In (\ref{eq:defX}), $(\Theta_{\uM \uN}{}^{\uP}, \vartheta_{\uM})$ are the irreducible components of the $\cN=8$ embedding tensor, in the $(\bm{912},\bm{56})$ of E$_{7(7)}$ or the $(\overline{\bm{351}},\bm{27})$ of E$_{6(6)}$. In particular, the components $\vartheta_{\uM}$ are responsible for gauging the $\mathbb{R}^+$ trombone scaling symmetry of the $D$-dimensional supergravity. Finally, $\mathbb{\uP}^{\uQ}{}_{\uM}{}^{\uP}{}_{\uN} \equiv (t^\alpha)_{\uM}{}^{\uQ} (t_\alpha)_{\uN}{}^{\uP} $ in (\ref{eq:defX}) is the projector to the adjoint of E$_{d(d)}$. The generators of the latter are denoted $(t_\alpha)_{\uM}{}^{\uQ}$, and adjoint E$_{d(d)}$ indices $\alpha$ are raised and lowered with the Killing-Cartan form $\kappa_{\alpha\beta} \equiv \textrm{tr} (t_\alpha t_\beta)$. \mbox{The constant $\alpha$ that appears in (\ref{eq:defX}) is defined in (\ref{eq:ExFTConstants}).}

Next, we freeze the $D$-dimensional supergravity fields at a vacuum configuration, 
\begin{equation} \label{eq:VacCond}
\overline{g}_{\mu\nu}(x) = \bar{g}^{(0)}_{\mu\nu}(x) \; , \qquad
M_{\uM\uN} (x) = M^{(0)}_{\uM\uN} \; , 
\end{equation}
with $\bar{g}^{(0)}_{\mu\nu}(x) $ a maximally symmetric metric, scalars in  $M^{(0)}_{\uM\uN}$ constant, and all other fields vanishing, and let the ExFT fields fluctuate about this vacuum. We do this by expanding (\ref{eq:GenSSBos}) in bosonic perturbations $h_{\mu \nu}^\Lambda(x)$, $A_\mu{}^{\uM \Lambda} (x)$, $\phi^{\uM\uN \Lambda}(x)$ as \cite{Malek:2019eaz,Malek:2020yue}
\begin{eqnarray} \label{eq:KKExpansionBos}
&& \bar{g}_{\mu \nu}(x,y)=  \bar{g}^{(0)}_{\mu \nu}(x) + h_{\mu\nu}^\Lambda (x) \, \cY_\Lambda(y) \; , \nonumber \\[5pt]
&& \cA_\mu{}^M (x,y) = \hat{E}^M{}_{\uN}(y) \, A_\mu{}^{\uN \Lambda}(x) \, \cY_\Lambda (y)  \; ,  \\[5pt]
&& (G^{-1})^{MN} (x,y) = \hat{E}^M{}_{\uP}(y) \hat{E}^N{}_{\uQ}(y) \big( M^{(0)\uP\uQ} + \phi^{\uP\uQ \Lambda} (x) \cY_\Lambda(y) \big)  \; , \nonumber
\end{eqnarray}
with $h_{\mu\nu}^\Lambda (x)$ transverse and traceless, and $\phi^{\uM\uN \Lambda} $ subject to the constraint
\begin{equation} \label{eq:ProjCoset}
\phi^{\uM\uN \Lambda} = (P_{\textrm{coset}})_\alpha{}^{\uM\uN} \, \phi^{\alpha  \Lambda} \; , \qquad 
\textrm{with} \quad 
(P_{\textrm{coset}})_\alpha{}^{\uM\uN} \equiv (t_\alpha)^{(\uM}{}_{\uP} M^{\uN)\uP} \; .
\end{equation}
The latter is written in terms of the projector $(P_{\textrm{coset}})_\alpha{}^{\uM\uN}$ \cite{Berman:2019izh} to the coset $\textrm{E}_{7(7)}/\textrm{SU}(8)$ or $\textrm{E}_{6(6)}/\textrm{USp}(8)$, namely, the projector to the $\bm{70}$ of $\textrm{SU}(8)$ or the $\bm{42}$ of $\textrm{USp}(8)$. The quantity $\phi^{\alpha  \Lambda}$ formally sits in the tensor product of the adjoint of E$_{d(d)}$ and the representation space labelled by the infinite-dimensional index $\Lambda$. However, since $\phi^{\alpha  \Lambda}$ will always appear affected by the coset projector, its index $\alpha$ can be safely taken to run only over the 70 or 42 non-compact directions along the relevant coset. The fermions are similarly perturbed by expanding (\ref{eq:GenSSFer}) in fluctuations $\psi_\mu^{i\Lambda}(x)$, $\chi^{ijk\Lambda}(x)$ as \cite{Cesaro:2020soq}
\begin{eqnarray} \label{eq:KKExpansionFer}
\psi_\mu{}^i = \Delta^{\frac12}  \psi_\mu{}^{i \Lambda}(x) \cY_\Lambda(y) \; , \qquad
\chi^{ijk} = \Delta^{-\frac12}  \chi^{ijk \Lambda}(x) \cY_\Lambda(y) \; . 
\end{eqnarray}
In (\ref{eq:KKExpansionBos}), (\ref{eq:KKExpansionFer}), $\cY_\Lambda(y)$ is an infinite tower, labelled by $\Lambda$, of harmonics on the internal compactification space. We will take them to be subject to the differential condition \cite{Malek:2019eaz,Malek:2020yue}
\begin{equation} \label{eq:CurlyTDef}
\hat E^M{}_{\uN} (y) \partial_M \cY_\Lambda (y) = R^{-1} (\cT_{\uN}){}^\Sigma{}_\Lambda \, \cY_\Sigma (y) \; , 
\end{equation}
where the constant matrices $(\cT_{\uN}){}^\Sigma{}_\Lambda$ are the generators of the gauge group of the $D$-dimensional $\cN=8$ supergravity on the infinite dimensional-representation generated by $\cY_\Lambda (y)$, \cite{Malek:2019eaz,Malek:2020yue}
\begin{equation} \label{eq:LieAlgT}
[\cT_{\uM} , \cT_{\uN}]=- R \, X_{[\uM\uN]}{}^{\uP} \cT_{\uP} \; , 
\end{equation}
and the constant $R$ simply sets a scale.

With these ingredients, the KK mass matrices are then obtained by linearising the ExFT field equations  \cite{Hohm:2013pua,Hohm:2013vpa,Hohm:2013uia,Godazgar:2014nqa,Musaev:2014lna} around the configuration (\ref{eq:VacCond}),  (\ref{eq:KKExpansionBos}), (\ref{eq:KKExpansionFer}), while retaining terms linear in the fluctuations. For tromboneless, $\vartheta_{\uM} =0$, lower-dimensional embedding tensor $X_{\uM\uN}{}^{\uP}$, these KK mass matrices were presented in \cite{Malek:2019eaz,Malek:2020yue,Varela:2020wty,Cesaro:2020soq}. Here, we will extend these to the $\vartheta_{\uM} \neq 0$ case. We will simply list the KK mass matrices in sections \ref{sec:BosMatMat} and \ref{sec:FerMatMat}, leaving the details of their derivation to appendix~\ref{eq:DerMassMat}.

%%%%%%%%%%%%%%%
\subsection{Bosonic mass matrices} \label{sec:BosMatMat}
%%%%%%%%%%%%%%%

The bosonic KK mass matrices can be written as quadratic combinations of the $D$-dimensional $\cN=8$ embedding tensor (\ref{eq:defX}) and the generators $(\cT_{\uN}){}^\Sigma{}_\Lambda$ defined by (\ref{eq:CurlyTDef}), contracted with the scalar matrix, $M_{\uM\uN}$, or its inverse, $M^{\uM\uN}$. The latter should be evaluated at the vacuum (\ref{eq:VacCond}) but, to alleviate the notation, we drop the zero superscript on it. All the symbols that enter the mass matrices have already been defined in section~\ref{sec:FlucExp}, except for the constants $\alpha$, $\kappa$, which are inherited from the E$_{d(d)}$ ExFT equations of motion. These have the $d$-dependent values
\begin{equation} \label{eq:ExFTConstants}
d=7 \; (D=4) \; : \quad \alpha =12 \; , \kappa = 7 \; , \qquad \qquad 
d=6 \; (D=5) \; : \quad \alpha =6 \; , \kappa = 5 \; . 
\end{equation}

With these conventions, for the KK graviton mass matrix we find 
\begin{equation} \label{eq:KKgraviton}
(\cM^2_{\textrm{graviton}})^\Sigma{}_\Lambda = - R^{-2} M^{ \uM \uN } \big( \fT_{\uM}{} \fT_{\uN} \big)^\Sigma{}_\Lambda -R^{-1} (D-2) M^{\uM\uN}  \vartheta_{\uM} (\fT_{\uN})^\Sigma{}_\Lambda \; .
\end{equation}
The KK gauge field mass matrix is, in turn, 
{\setlength\arraycolsep{0pt}
\begin{eqnarray} \label{eq:KKvector}
(\cM^2_{\textrm{vector}})^{\uM\Sigma}{}_{\uN\Lambda} & = & M^{\uM\uA'} \bigg[  \tfrac{2}{\alpha} \left( X_{\uA'\uC}{}^{\uD} - \alpha (D-2) \delta_{\uA'}^{\uD} \vartheta_{\uC} \right) M^{\uC \uE} \left( X_{\uN(\uD}{}^{\uF} M_{\uE)\uF} + \vartheta_{\uN} M_{\uD \uE} \right) \delta^\Lambda_\Sigma \nonumber  \\[5pt]
&& \qquad \quad \; +2 R^{-1} \left( X_{\uA' \uC}{}^{\uD} - \alpha (D-2) \delta_{\uA'}^{\uD} \theta_{\uC} \right) M^{\uC \uE}  M_{\uF( \uD} \mathbb{P}^{\uF}{}_{\uE)}{}^{\uG}{}_{\uN} \,  ( \fT_{\uG})^\Sigma{}_\Lambda \nonumber  \\[5pt] 
&& \qquad  \quad \;   - 2 R^{-1} M^{\uC \uD} \left(  X_{\uN(\uA'}{}^{\uE} M_{\uD) \uE} + \vartheta_{\uN} M_{\uA' \uD} \right) ( \fT_{\uC})^\Sigma{}_\Lambda \nonumber  \\[5pt] 
&&  \qquad   \quad \;  - \alpha R^{-2}  M^{\uC \uD}  M_{\uE(\uA'} \mathbb{P}^{\uE}{}_{\uD)}{}^{\uF}{}_{\uN}  ( \fT_{\uF} \fT_{\uC})^\Sigma{}_\Lambda  \bigg] \; . 
\end{eqnarray}
}Finally, the KK scalar mass matrix reads
{\setlength\arraycolsep{1pt}
\begin{eqnarray} \label{eq:KKscalar}
%\begin{split}
 (\cM^2_{\textrm{scalar}})^{\alpha\Sigma}{}_{\beta\Lambda} &=&
\left( P_{\textrm{coset}} \right)^{\alpha \uA\uB} \Big[ \Theta_{\uA\uE}{}^{\uF} \Theta_{\uC\uF}{}^{\uE} M_{\uB\uD} \, \delta_\Lambda^\Sigma \nonumber  \\[4pt]
&& +\tfrac{1}{\kappa} M_{\uB\uD} \Big( M_{\uF\uF'} M^{\uE\uE'}\Theta_{\uA\uE}{}^{\uF} \Theta_{\uC\uE'}{}^{\uF'}  + M_{\uF\uF'} M^{\uE\uE'}\Theta_{\uE\uA}{}^{\uF} \Theta_{\uE'\uC}{}^{\uF'} \nonumber   \\
&& \qquad  \qquad \quad + M_{\uA\uA'} M_{\uC\uC'} M^{\uF\uF'} M^{\uE\uE'}\Theta_{\uE\uF}{}^{\uA'} \Theta_{\uE\uF'}{}^{\uC'} \Big) \delta_\Lambda^\Sigma \nonumber  \\[4pt]
&&   +\tfrac{2}{\kappa} \Big(  M_{\uE\uF}\Theta_{\uA\uC}{}^{\uE} \Theta_{\uB\uD}{}^{\uF} - M_{\uC\uC'} M_{\uD\uD'} M^{\uE\uF}\Theta_{\uA\uE}{}^{\uC'} \Theta_{\uB\uF}{}^{\uD'} \nonumber  \\
&& \qquad  \qquad \quad - M_{\uC\uC'} M_{\uD\uD'} M^{\uE\uF}\Theta_{\uE\uA}{}^{\uC'} \Theta_{\uF\uB}{}^{\uD'} \Big) \delta_\Lambda^\Sigma \nonumber \\[4pt]
&& + 2 (D-2) \Big( \vartheta_{\uC}  \Theta_{\uD\uA}{}^{\uB'} M_{\uB'\uB} - \vartheta_{\uE} \Theta_{\uF\uA}{}^{\uC'}  M_{\uB\uD}  M^{\uE\uF}  M_{\uC'\uC} \Big) \delta_\Lambda^\Sigma 
\nonumber  \\[4pt]
&& -2R^{-1} M_{\uE\uD} \Theta_{\uA\uC}{}^{\uE} (\mathcal{T}_{\uB})^\Sigma{}_\Lambda  +2R^{-1} M_{\uE\uB}  \Theta_{\uC\uA}{}^{\uE}  ( \mathcal{T}_{\uD})^\Sigma{}_\Lambda  \nonumber  \\[4pt]
&&  -4R^{-1} M^{\uE\uF} M_{\uB\uD} M_{\uC\uC'} \Theta_{\uF\uA}{}^{\uC'}    (\mathcal{T}_{\uE})^\Sigma{}_\Lambda  \nonumber  \\[4pt]
&&  +2\alpha (D-2) R^{-1}  M_{\uB\uD} \vartheta_{\uC}  (\mathcal{T}_{\uA})^\Sigma{}_\Lambda   +2\alpha R^{-2} M_{\uB\uD}   (\mathcal{T}_{\uC} \mathcal{T}_{\uA})^\Sigma{}_\Lambda  \Big] \left( P_{\textrm{coset}}  \right)_\beta{}^{\uC\uD}  \nonumber  \\[4pt]
&&  -\delta^\alpha_\beta \,  \big[ R^{-1} (D-2)  M^{\uA\uB} \vartheta_{\uA} (\fT_{\uB})^\Sigma{}_\Lambda + R^{-2} M^{\uA\uB} ( \fT_{\uA} \fT_{\uB})^\Sigma{}_\Lambda \big] \; . 
%
%\end{split}
\end{eqnarray}
}For $(\cT_{\uM})^\Sigma{}_\Lambda =0$, and ignoring the KK indices $\Lambda$, $\Sigma$, the mass matrices (\ref{eq:KKgraviton})--(\ref{eq:KKscalar}) reduce to those for $D$-dimensional maximal gauged supergravity as given in \cite{Pico:2025cmc,Varela:2025xeb}. This is consistent with the expectation that KK level $k=0$ should reproduce the (linearised) $D$-dimensional theory. For $(\cT_{\uM})^\Sigma{}_\Lambda \neq 0$ and $\vartheta_{\uM} = 0$, they reduce to the tromboneless KK bosonic mass matrices of \cite{Malek:2019eaz,Malek:2020yue,Varela:2020wty}. When $(\cT_{\uM})^\Sigma{}_\Lambda \neq 0$ and $\vartheta_{\uM} \neq 0$ these matrices are new, and extend the results of the latter references to the case where the $D$-dimensional gauged supergravity involves trombone gaugings.

The eigenvalues of (\ref{eq:KKgraviton})--(\ref{eq:KKscalar}) correspond to square masses. All the eigenvalues of (\ref{eq:KKgraviton}) are physical and correspond to graviton eigenstates. In contrast, both (\ref{eq:KKvector}) and (\ref{eq:KKscalar}) contain spurious modes. For $D=4$ and at fixed KK level, the former always contains vanishing eigenvalues corresponding to the magnetic gauge fields. In addition, for all $D$, (\ref{eq:KKvector}) contains Goldstone modes that are eaten by the massive gravitons. Goldstone modes are also present in the scalar mass matrix (\ref{eq:KKscalar}), corresponding to the Higgsing of graviton and vector states.

%%%%%%%%%%%%%%%
\subsection{Fermionic mass matrices} \label{sec:FerMatMat}
%%%%%%%%%%%%%%%

Similarly, we can obtain the KK fermion mass matrices by linearisation of the ExFT fermion equations of motion \cite{Godazgar:2014nqa,Musaev:2014lna}. The resulting mass matrices are linear in the generators $(\cT_{\uM})^\Sigma{}_\Lambda$ and in the $D$-dimensional embedding tensor $X_{\uM\uN}{}^{\uP}$. The dependence on the latter is introduced through the $\cN=8$ fermion shifts $A_1^{ij}$, $A_{2 h}{}^{ijk}$ and $B^{ij}$. These are contractions of the $\textrm{E}_{7(7)}/\textrm{SU}(8)$ or $\textrm{E}_{6(6)}/\textrm{USp}(8)$ (inverse) coset representative $\cV$ with the embedding tensor: $A_1^{ij}$, $A_{2 h}{}^{ijk}$ with the $\bm{912}$ or $\overline{\bm{351}}$ components $\Theta_{\uM\uN}{}^{\uP}$, and $B^{ij}$ with the trombone components $\vartheta_{\uM}$: see \cite{deWit:2007mt,LeDiffon:2011wt,deWit:2004nw,Varela:2025xeb} for the details. 

With these definitions, and focusing in $D=4$ for definiteness and for later reference, we find the KK gravitino mass matrix to be
\begin{equation} \label{eq:KKgravitino}
(\cM_{\textrm{gravitino}})_i{}^\Sigma{}_{j\Lambda} = 2\left( A_{1 ij} -2 B_{ij} \right) \delta^\Sigma_\Lambda - 4 R^{-1} \, (\cV^{-1})_{ij}{}^{\uM}  ( \fT_{\uM})^\Sigma{}_\Lambda  \; , 
\end{equation}
while for the KK spin-$1/2$ mass matrix we obtain 
\begin{equation} \label{eq:KKspin12} 
(\cM_{\textrm{spin-$\frac12$}})_{ijk}{}^\Sigma{}_{lmn\Lambda} \hspace{-5pt}  =  \hspace{-2pt} \left(   \epsilon_{ijkpqr[lm} A_{2 n]}{}^{pqr} \hspace{-3pt} +  \hspace{-3pt} 2  \,   \epsilon_{ijklmnpq} B^{pq} \right) \delta^\Sigma_\Lambda - 2 R^{-1} \, \epsilon_{ijklmnpq} (\cV^{-1})^{pq\uM} (\fT_{\uM})^\Sigma{}_\Lambda \,. 
\end{equation}
Again, for $(\cT_{\uM})^\Sigma{}_\Lambda =0$ and ignoring the KK indices $\Lambda$, $\Sigma$, these reduce to the $D=4$ $\cN=8$ gauged supergravity fermion mass matrices \cite{LeDiffon:2011wt}. When $(\cT_{\uM})^\Sigma{}_\Lambda \neq 0$ and $\vartheta_{\uM} = 0$, (\ref{eq:KKgravitino}), (\ref{eq:KKspin12}) reduce to the tromboneless KK fermion mass matrices of \cite{Cesaro:2020soq} and, for $\vartheta_{\uM} \neq 0$ extend the latter to the tromboneful case. The eigenvalues of these mass matrices correspond to linear masses. All states arising from diagonalisation of (\ref{eq:KKgravitino}) are physical, but (\ref{eq:KKspin12}) contains goldstino modes that must be removed from the physical spectrum.

%%%%%%%%%%%%%%%
%%%%%%%%%%%%%%%

\section{Light operator spectra of M5-brane 3d SCFTs} \label{sec:Spectra}

%%%%%%%%%%%%%%%
%%%%%%%%%%%%%%%

We now turn our attention to the $\textrm{AdS}_4 \times (\Sigma_3 \rtimes S^4)$ M5-brane solutions of \cite{Gauntlett:2006ux,Acharya:2000mu}. We will start by posing the KK problem in full generality, and then argue how the mass matrices that we derived in section \ref{sec:KKMassMat} are helpful to extract universal sectors of the spectra.

%%%%%%%%%%%%%%%
\subsection{Setting up the problem: the spin-2 spectra} \label{sec:SpecGen}
%%%%%%%%%%%%%%%

The complete KK spectrum of $D=11$ supergravity on the seven-dimensional geometries $\Sigma_3 \rtimes S^4$, SLAG \cite{Gauntlett:2006ux} and A3C \cite{Acharya:2000mu}, can in principle be determined by applying the programme of \cite{Kim:1985ez} on a case by case basis for each distinct $\Sigma_3 = H^3 / \Gamma$. We will be interested in extracting universal spectral sectors valid for all such $\Sigma_3$. As we will argue in section \ref{sec:PutUniv}, the traditional approach of \cite{Kim:1985ez} can be by-passed in this case by building on \cite{Pico:2025cmc,Pico:2026rji} and deploying the technology developed in section~\ref{sec:KKMassMat} following \cite{Malek:2019eaz,Malek:2020yue,Varela:2020wty,Cesaro:2020soq}. It is still informative to examine specific sectors of the spectrum using \cite{Kim:1985ez}, as this will help motivate our strategy going forward.

More concretely, it is helpful to start by inspecting the massive graviton spectrum. This is obtained by diagonalising the modified Laplacian $\cL_7$ defined on the seven-dimensional internal spaces of interest as \cite{Bachas:2011xa} 
\begin{equation} \label{eq:MassGravOp}
\cL_7 = - e^{-2 \tilde \Delta}  g^{-\frac12} \partial_m \big(g^{\frac12} g^{mn} e^{4 \tilde \Delta} \partial_n \big) \; .
\end{equation}
Here, $g^{mn}$ is the inverse metric on either seven-dimensional geometry $\Sigma_3 \rtimes S^4$ \cite{Gauntlett:2006ux,Acharya:2000mu}, with $\Sigma_3$ hyperbolic and compact, $\Sigma_3 = H^3/\Gamma$. The function $e^{-2 \tilde \Delta}$ in (\ref{eq:MassGravOp}) is a warp factor and $g \equiv \textrm{det} \, g_{mn}$. 

Let $y=(y_3,y_4)$ collectively denote local coordinates for $\Sigma_3 \rtimes S^4$, $\Sigma_3$, and $S^4$, and consider separated eigenfunctions $h(y) = f(y_3) {\cal Y}(y_4)$ for $\cL_7$. We further choose ${\cal Y}$ to be the $S^4$ spherical harmonics,
\begin{equation} \label{eq:harmS4}
\cY^\Lambda = \{ 1 \, , \, y^i \, , \, y^{\{i_1} y^{i_2\}} \, , \, \ldots , \, y^{\{i_1} \cdots y^{i_k\}} \, , \, \ldots \} \; ,
\end{equation}
with $y^i (y_4)$ coordinates on $\mathbb{R}^5$ constrained as $\delta_{ij} y^i y^j =1$ and curly brackets denoting traceless symmetrisation so that $y^{\{i_1} \cdots y^{i_k\}}$ lies in the SO$(5)$ representation with Dynkin labels $[k0]$, $k= 0 ,1, 2, \ldots$  As we show in appendix~\ref{sec:GravOp}, acting on 
\begin{equation} \label{eq:FullEF}
h_{k} (y) = f_{i_1 \cdots i_k} (y_3) \, y^{\{i_1} \cdots y^{i_k\}} \; , \qquad k=0,1,2,\ldots \; ,
\end{equation}
the massive graviton operator (\ref{eq:MassGravOp}) splits for both $\Sigma_3 \rtimes S^4$ geometries, SLAG and A3C, as 
\begin{equation} \label{eq:MassGravOpSplit}
\cL_7 = \oplus_j \cL_{3(j)} + \cL_4  \; .
\end{equation}
Here, $\cL_4$ is a modified Laplacian on $S^4$ that depends on the details of either solution \cite{Gauntlett:2006ux,Acharya:2000mu}, see appendix~\ref{sec:GravOp}. The $S^4$ harmonics $y^{\{i_1} \cdots y^{i_k\}}$ (\ref{eq:harmS4}) are still eigenfunctions of $\cL_4$. These are not all $\textrm{dim} \, [k0]$-fold mass degenerate, though, as the metrics on $S^4$ are not round \cite{Gauntlett:2006ux,Acharya:2000mu}. Instead, they come further branched out in representations of the following SO(5) subgroups in either case:
\begin{equation} \label{eq:SO5subgroups}
\textrm{SLAG:}  \;\;
\textrm{SO}(5)  \supset
\textrm{SO}(2) \times \textrm{SO}(3)^\prime \; , \qquad
\textrm{A3C:}  \;\;
\textrm{SO}(5) \supset 
\textrm{SO}(4)  \sim 
\textrm{SO}(3)_- \times \textrm{SO}(3)_+ \;   ,
\end{equation}
so that 
\begin{equation} \label{eq:SO5subgroupsBranch}
\textrm{SLAG:}  \;\;
\bm{5} \rightarrow \bm{1}_2 +\bm{1}_{-2} +\bm{3}_0 \; , \qquad
\textrm{A3C:}  \;\;
\bm{5} \rightarrow (\bm{1},\bm{1}) + (\bm{2},\bm{2}) \;   ,
\end{equation}
respectively. 

The operator $\cL_{3(j)}$ in (\ref{eq:MassGravOpSplit}) is the Bochner SO(3)-connection Laplacian on $\Sigma_3$, 
\begin{equation} \label{eq:Bochner}
\cL_{3(j)} = - \square_{3} - 2 z \left( S_{1(j)} \, \partial_x + S_{2(j)} \, \partial_y \right)  - (S_{1(j)}^2 + S_{2(j)}^2) \; .
\end{equation}
Here, we have used Poincar\'e-like coordinates $y_3 \equiv (x,y,z)$ in which the usual Einstein metric on $\Sigma_3 = H^3/\Gamma$ takes on the form (\ref{eq:H3MetricUHP}). The operator $\square_{3} =  z^2 ( \partial_x^2 + \partial_y^2+ \partial_z^2) - z \partial_z $ is the usual Laplacian on $\Sigma_3 = H^3/\Gamma$, and $S_{1(j)}$, $S_{2(j)}$ are constant $(2j+1)$-dimensional matrices corresponding to two of the SO$(3)$ generators $S_{a(j)}$, $a=1,2,3$, in the spin-$j$ representation, normalised as $[S_{a(j)}, S_{b(j)} ] = \epsilon_{ab}{}^c \, S_{c(j)}$. This SO$(3)$ is the structure group on the $\Sigma_3 \rtimes S^4$ bundle for either solution. We will denote it by SO$(3)_S$ for SLAG and SO$(3)_S^\prime$ for A3C. These are defined by taking suitable diagonals of the $\textrm{SO}(3)_{\Sigma_3}$ structure on $\Sigma_3$ defined by its spin connection, and different $\textrm{SO}(3) \subset \textrm{SO}(5)$ subgroups acting on $S^4$. More concretely,
\begin{equation} \label{eq:GroupTh}
\textrm{SO}(5) \times \textrm{SO}(3)_{\Sigma_3} \,  \supset  \, 
\hspace{-8pt}
\begin{array}{c}
\textrm{SO}(2)  \times \textrm{SO}(3)^\prime \times \textrm{SO}(3)_{\Sigma_3} \,  \supset  \,   \textrm{SO}(2) \times \textrm{SO}(3)_S \quad $\textrm(SLAG)$, \\[6pt]
\textrm{SO}(3)_- \times \textrm{SO}(3)_+ \times \textrm{SO}(3)_{\Sigma_3} \,  \supset  \,  \textrm{SO}(3)_+ \times \textrm{SO}(3)_S^\prime   \quad $\textrm(A3C)$,
 \end{array}
\end{equation}
where the first branching follows (\ref{eq:SO5subgroups}), and then SO$(3)_S$ is the diagonal of $\textrm{SO}(3)^\prime \times \textrm{SO}(3)_{\Sigma_3}$, and SO$(3)_S^\prime$ the diagonal of $\textrm{SO}(3)_- \times \textrm{SO}(3)_{\Sigma_3}$. Finally, for each fixed $k$, the sum $\oplus_j$ in (\ref{eq:MassGravOpSplit}) is extended to all spins $j$ that appear in the branching of the $[k0] \times \bm{1}$ representation of $\textrm{SO}(5) \times \textrm{SO}(3)_{\Sigma_3}$ in (\ref{eq:GroupTh}) under SO$(3)_S$ or SO$(3)_S^\prime$. For example, for $k=1$, it follows from (\ref{eq:SO5subgroupsBranch}) that $j=0$ (twice) and $j=1$ (once) for SLAG, while $j=0$ (once) and $j=\tfrac12$ (twice) for A3C. 

To summarise, the eigenfunctions of the graviton mass operator (\ref{eq:MassGravOp}) on the SLAG and A3C internal spaces $\Sigma_3 \rtimes S^4$ factorise as in (\ref{eq:FullEF}) in terms of $S^4$ spherical harmonics (\ref{eq:harmS4}), and $(2j+1)$-dimensional spinor eigenstates $f_{i_1 \cdots i_k}(y_3)$ of the Bochner, SO$(3)_S$- or SO$(3)_S^\prime$-connection Laplacian (\ref{eq:Bochner}) on $\Sigma_3$ at spin $j$. Here, $j$ ranges along all possibilities allowed at fixed $k=0,1,2, \ldots$ by the branching of the  $[k0] \times \bm{1}$ representation of $\textrm{SO}(5) \times \textrm{SO}(3)_{\Sigma_3}$ under the SO$(3)_S$ or SO$(3)_S^\prime$ groups defined in (\ref{eq:GroupTh}), for the SLAG and A3C cases. The Bochner Laplacian (\ref{eq:Bochner}) is an elliptic operator defined on the compact space $\Sigma_3 = H^3/\Gamma$. Thus, its eigenvalues are discrete with finite-dimensional eigenspaces. The specific spectrum of the operator $\cL_7$ on $\Sigma_3 \rtimes S^4$ needs to be determined for each choice of compactifying $\Gamma$ in $\Sigma_3 = H^3/\Gamma$. The Bochner Laplacian (\ref{eq:Bochner}) is the natural counterpart in $\Sigma_3 = H^3/\Gamma$ of the Maass Laplacian on a genus $g$ Riemann surface $\Sigma_2$ encountered in \cite{Bhattacharya:2024tjw,BKV2026}.

For each $k=0, 1, 2, \ldots $ and value of $j$ allowed by $k$, the constant ({\it i.e.}~independent of the $\Sigma_3 = H^3/\Gamma$ coordinates $y_3 = (x,y,z)$) function   $f_{i_1 \cdots i_k}$, trivially is an eigenfunction of the Bochner Laplacian (\ref{eq:Bochner}). The resulting eigenfunction (\ref{eq:FullEF}) of the full graviton mass operator $\cL_7$ in (\ref{eq:MassGravOp}) is constant over $\Sigma_3$ and has only a profile on $S^4$. Thus, these eigenstates are universally valid for all $\Sigma_3$. However, their global or local character critically depends on whether they arise at $j=0$ or $j \neq 0$, thus coupling to SO$(3)_S$ (or SO$(3)_S^\prime$) singlets or non-singlet $S^4$ spherical harmonics (\ref{eq:harmS4}) at given $k$. The SO$(3)_S$-singlet eigenfunctions are globally well-defined in the full fibred space $\Sigma_3 \rtimes S^4$, as they are in fact insensitive to the twist. In contrast, the non-singlets are only defined locally on $\Sigma_3 \rtimes S^4$. This is because, in general, the coefficients $f_{i_1 \cdots i_k}$ can only be chosen to be constant in a local trivialisation of the bundle, and are not usually well defined globally. Indeed, as one moves around different patches of the $\Sigma_3$ base, the fibre will accordingly rotate with SO$(3)_S$, generating a $y_3 = (x,y,z)$ coordinate dependence, $f_{i_1 \cdots i_k}(y_3)$, for the coefficients. The global definiteness of the non-singlets needs to be assessed individually for each $\Sigma_3 = H^3/\Gamma$. 

We have computed the eigenvalues of the graviton mass operator (\ref{eq:MassGravOp}) with (\ref{eq:MassGravOpSplit}), (\ref{eq:Bochner}),  corresponding to the (typically) local, universal eigenstates (\ref{eq:FullEF}) with (\ref{eq:harmS4}) and constant $f_{i_1 \cdots i_k}$, for both SLAG and A3C solutions and for various low-lying values of $k$. A pattern emerges that allows one to infer the result for all $k$. This graviton spectra  can be read off from equations (\ref{eq:SLAGN8Form}), (\ref{eq:LGRAVFormula}) and (\ref{eq:A3CGravs}), (\ref{eq:A3CN8Form}) of appendix \ref{sec:PutSpec}, by choosing the relevant spin-2 states from the supermultiplets listed therein. Similarly, the globally defined SO$(3)_S$ or SO$(3)_S^\prime$ singlet modes  can be read off from equations (\ref{eq:GRAVSLAGUniv}), (\ref{eq:SLAGForm}) and table \ref{tab:GlobalA3CMultiplets} below. In principle, one may push the formalism of \cite{Kim:1985ez} to obtain the KK spectrum of supergravity fields with all other spins $s=0, \frac12, 1, \frac32$ above both backgrounds $\textrm{AdS}_4 \times (\Sigma_3 \rtimes S^4)$ \cite{Gauntlett:2006ux,Acharya:2000mu} for a given choice of $\Sigma_3 = H^3 /\Gamma$. Alternatively, the preceding discussion opens up an ExFT/ExGG-based strategy to determine the universal KK spectrum for all spin states, valid for all $\Sigma_3 = H^3 /\Gamma$, to which we now turn.

%%%%%%%%%%%%%%%
\subsection{Putative and global universal spectra} \label{sec:PutUniv}
%%%%%%%%%%%%%%%

In ExGG, the SLAG \cite{Gauntlett:2006ux} and A3C \cite{Acharya:2000mu} solutions are equipped with various generalised structures \cite{Pico:2026rji} in the sense of \cite{Cassani:2019vcl}, see also \cite{Cassani:2020cod,Josse:2025uro}. More concretely, these solutions respectively enjoy generalised SO$(3)_S$ and SO$(3)_S^\prime$ structures that arise as reductions of the generic SU(8) structure group of E$_{7(7)}$ ExGG down to the subgroups defined through the embedding
\begin{equation} \label{eq:SO3SBreaking}
\textrm{SU}(8) \; \supset \; 
\textrm{SO}(8) \; \supset \; 
\textrm{SO}(5) \times \textrm{SO}(3)  \; ,
\end{equation}
followed by (\ref{eq:GroupTh}). These can be further reduced to generalised identity structures (GISs) through the further inclusion $\textrm{SO}(3)_S \supset \bm{1}$, and similarly for $\textrm{SO}(3)_S^\prime$. The generalised SO$(3)_S$ and SO$(3)_S^\prime$ structures are defined in terms of globally defined geometric objects on $\Sigma_3 \rtimes S^4$ and are thus globally well defined. In contrast, the GISs are only local: they involve the replacement of $\Sigma_3 =H^3/\Gamma$ with the locally equivalent group manifold $G_3$ of Bianchi type V. This replacement does not hold globally, though: being non-unimodular, $G_3$ cannot be compactified, unlike $\Sigma_3 =H^3/\Gamma$.

As we recently pointed out \cite{Pico:2026rji}, all these generalised structures turn out to have constant generalised intrinsic torsion, again in the sense of \cite{Cassani:2019vcl}. For that reason, consistent truncations of $D=11$ supergravity on $\Sigma_3 \rtimes S^4$ down to $D=4$ gauged supergravities can be expected \cite{Cassani:2019vcl}, with either maximal \cite{Lee:2014mla} or submaximal \cite{Cassani:2019vcl} supersymmetry depending on whether $\Sigma_3 \rtimes S^4$ is equipped with its GIS or with its generalised $\textrm{SO}(3)_S$ (or $\textrm{SO}(3)_S^\prime$) structure. In fact, all these consistent truncations were either previously known (as they had already been constructed with different, though similar, methods) or have been recently obtained. A globally-defined SO$(3)_S$-invariant truncation on the $\Sigma_3 \rtimes S^4$ SLAG geometry \cite{Gauntlett:2006ux} down to $D=4$ $\cN=2$ gauged supergravity was presented in \cite{Donos:2010ax}. Similarly, a global $D=4$ $\cN=1$ truncation on the A3C geometry $\Sigma_3 \rtimes S^4$ was partially built in \cite{Gauntlett:2002rv}. More recently, we have given the local $D=4$ $\cN=8$ truncation on both SLAG and A3C geometries in \cite{Pico:2026rji}. The resulting $D=4$ $\cN=8$ supergravity is the same in both cases: it is the TCSO$(5,0,0;\mathrm{V})$-gauged theory of \cite{Pico:2025cmc}. Since $\textrm{SO}(3)_S \supset \bm{1}$, the submaximal truncations of \cite{Donos:2010ax,Gauntlett:2002rv} arise as (globally defined) subsectors of the (local) maximal truncation of \cite{Pico:2026rji}. Appendix~\ref{sec:MaxTrunc} summarises some useful expressions from  \cite{Pico:2025cmc,Pico:2026rji}.

The results of \cite{Pico:2025cmc,Pico:2026rji} place the M5-brane $\textrm{AdS}_4 \times (\Sigma_3 \rtimes S^4)$ solutions of \cite{Gauntlett:2006ux,Acharya:2000mu} right at the sweet spot where the ExFT KK spectral techniques of \cite{Malek:2019eaz,Malek:2020yue,Varela:2020wty,Cesaro:2020soq} can be applied. As briefly reviewed in the introduction, these ExFT spectral methods rely on the existence of a maximally supersymmetric consistent truncation, whose linearisation can be regarded as level $k=0$ in the KK expansion. With this maximal truncation in place, all higher KK modes of all spins $s=0, \frac12, 1, \frac32, 2$ turn out to arise by expanding the ExFT fields in eigenstates (\ref{eq:FullEF}) of the graviton mass operator (\ref{eq:MassGravOp}). In particular, the subsector of graviton eigenstates that are universally valid for all $\Sigma_3$ are constant (over $\Sigma_3$) linear combinations of the $S^4$ spherical harmonics (\ref{eq:harmS4}). As also discussed in section \ref{sec:SpecGen}, these universal graviton eigenstates will be globally or only locally defined, depending on their charges under SO$(3)_S$ or SO$(3)_S^\prime$.

Thus, the (local, in general) universal KK fluctuations of all spins $s=0, \frac12, 1, \frac32, 2$ above the $\textrm{AdS}_4 \times (\Sigma_3 \rtimes S^4)$ solutions of \cite{Gauntlett:2006ux,Acharya:2000mu} can be obtained by linearisation of the E$_{7(7)}$ ExFT equations of motion \cite{Hohm:2013uia,Godazgar:2014nqa}, reviewed in appendix~\ref{eq:DerMassMat}, over the backgrounds at hand, and expanding the linearised perturbations (\ref{eq:KKExpansionBos}), (\ref{eq:KKExpansionFer}) in the graviton eigenstates (\ref{eq:FullEF}) with (\ref{eq:harmS4}) and constant coefficients $f_{i_1 \cdots i_k}$. These expansion functions can be verified to satisfy (\ref{eq:CurlyTDef}) with GIS $\hat{E}^M{}_{\uN}$ on $\Sigma_3 \rtimes S^4$ given by equations (\ref{eq:GenFrameSplit})--(\ref{eq:GenFrameG3S4}) of appendix~\ref{sec:MaxTrunc}, and matrices $(\cT_{\uM})^\Sigma{}_\Lambda$ with non-vanishing components given by
\begin{equation} \label{eq:TMS4}
(\cT_{\uM})^\Sigma{}_\Lambda = \{ 0 \, , \, (\cT_{ij})^h{}_\ell \, , \, 2 (\cT_{ij})^{ \{ h_1}{}_{ \{\ell_1}  \delta^{h_2 \}}_{\ell_2 \} } \, , \, \ldots , \, k \, (\cT_{ij})^{ \{ h_1}{}_{ \{\ell_1} \cdots \delta^{h_k \}}_{\ell_k \} }  \, , \, \ldots \} \; .
\end{equation}
Here, only the components $(\bm{1} , \bm{10})$ are active in the decomposition (\ref{eq:56Branching}) of the flat index $\uM$ under $\textrm{E}_{7(7)} \supset \textrm{GL}(3) \times \textrm{SL}(5) \supset \textrm{SO}(5)$, and $(\cT_{ij})^h{}_\ell \equiv 2 \delta^h_{[i} \delta_{j] \ell} $ are the SO$(5)$ generators in the fundamental representation. Altogether, $(\cT_{\uM})^\Sigma{}_\Lambda$ in (\ref{eq:TMS4}) are the SO(5) generators in the $\oplus_{k=0}^{\infty} [k0]$ representation. These can be also checked to satisfy the Lie algebra (\ref{eq:LieAlgT}) with structure constants $X_{\uM\uN}{}^{\uP}$ given by the embedding tensor, (\ref{eq:ETtromboneSplit}) with (\ref{eq:ETtrombone}), (\ref{eq:SCBianchiV}), of $D=4$ $\cN=8$ TCSO$(5,0,0;\mathrm{V})$-gauged supergravity. Under these provisions, the universal KK spectral problem reduces to the diagonalisation of the infinite-dimensional mass matrices (\ref{eq:KKgraviton})--(\ref{eq:KKspin12}) evaluated on either vacuum, as shown in appendix~\ref{eq:DerMassMat} and section \ref{sec:KKMassMat}. Thanks to (\ref{eq:TMS4}), these mass matrices are block diagonal with blocks at KK level $k =0 ,1 ,  2 , \ldots$ of size $(\textrm{dim} \, \bm{r} \,  \textrm{dim}[k0] ) \times (\textrm{dim} \, \bm{r} \,  \textrm{dim}[k0] )$, acting on the $\bm{r} \times [k0]$ representation of $\textrm{SU}(8) \times \textrm{SO}(5)$. Here, $\bm{r}$ is the SU$(8)$ representation of the $k=0$ mode at the bottom of the tower regarded as a linearised $D=4$ $\cN=8$ supergravity field, namely, $\bm{1}$, $\overline{\bm{8}}$, $\bm{28} + \bm{28}^\prime$, $\overline{\bm{56}}$ and $\bm{70}$ for gravitons, gravitini, vectors, spin-$1/2$ fermions and scalars, respectively. Thus, the mass matrices (\ref{eq:KKgraviton})--(\ref{eq:KKspin12}) can be diagonalised KK level by KK level. 

As we have repeatedly emphasised, the universal KK spectra thus obtained are, in general, only defined locally on the bundles $\Sigma_3 \rtimes S^4$. This local character is due to various interrelated reasons. Firstly, the graviton eigenstates (\ref{eq:FullEF}) with (\ref{eq:harmS4}) and constant coefficients $f_{i_1 \cdots i_k}$ that we used in the expansions (\ref{eq:KKExpansionBos}), (\ref{eq:KKExpansionFer}) are in general only defined locally on $\Sigma_3 \rtimes S^4$. Secondly, the GIS (\ref{eq:GenFrameSplit})--(\ref{eq:GenFrameG3S4}) that underlies our construction is  only locally defined on $\Sigma_3 \rtimes S^4$, as it locally trades $\Sigma_3$ with the Bianchi type V group manifold $G_3$. This, in turn, induces a gauging $\vartheta_{\uM}$ of the trombone scaling symmetry in the $D=4$ $\cN=8$ TCSO$(5,0,0;\mathrm{V})$-gauged supergravity, a landmark of dimensional reductions on non-compact internal spaces. Further, the KK mass matrices (\ref{eq:KKgraviton})--(\ref{eq:KKspin12}) with $\vartheta_{\uM} \neq 0$ are not symmetric. Yet, except for $k=0$ \cite{Pico:2025cmc}, we find them to be not only perfectly diagonalisable, but also diagonalisable over the reals. Moreover, the resulting spectra are discrete rather than continuous thanks to the compactness of $S^4$ and the constancy of the eigenfunctions over $\Sigma_3$ (or $G_3$) though, again, only locally defined on the bundle $\Sigma_3 \rtimes S^4$. Reassuringly enough,  these local KK spectra pass a couple of consistency tests. For both SLAG and A3C solutions, the universal graviton spectrum computed with the mass matrix (\ref{eq:KKgraviton}) and with the standard approach of section~\ref{sec:SpecGen}, agree. Also, the individual eigenstates of the mass matrices (\ref{eq:KKgraviton})--(\ref{eq:KKspin12}) close into supermultiplets of the supersymmetry algebra, OSp$(4|\cN)$, with $\cN=2$ for SLAG and $\cN=1$ for A3C. The local character of these spectra nevertheless advises caution about their physical validity. Even if the $k \geq 1$ mass matrices are free from the pathologies observed at $k=0$ like Jordan chains or complex eigenvalues \cite{Pico:2025cmc}, that does not necessarily imply that all these states are physical.

All things considered, we will adopt a pragmatic attitude. We will regard the universal, typically local KK spectrum on either $\textrm{AdS}_4 \times (\Sigma_3 \rtimes S^4)$ solution obtained with the ExFT-derived, trombone-enhanced, TCSO$(5,0,0;\mathrm{V})$-gauged supergravity-based,  KK mass matrices of section~\ref{sec:KKMassMat} as a useful pool from which to extract physical KK modes. As such, we will refer to these KK spectra as putative. A complementary criterion is needed to extract physical modes above suspicion out of these putative spectra. One such criterion is provided by global definiteness: the SO$(3)_S$ or SO$(3)_S^\prime$-invariant modes are globally well defined on either bundle $\Sigma_3 \rtimes S^4$ and are therefore perfectly physical. The global, universal KK spectrum above either solution is obtained by retaining modes that preserve the globally defined generalised SO$(3)_S$ or SO$(3)_S^\prime$ structures defined on $\Sigma_3 \rtimes S^4$ by (\ref{eq:SO3SBreaking}), (\ref{eq:GroupTh}). In practice, this is achieved by selecting the SO$(3)_S$ or SO$(3)_S^\prime$-singlets in the branching of the $\textrm{SU}(8)\times \textrm{SO}(5)$ representation $\bm{r} \times [k0]$ where the blocks of (\ref{eq:KKgraviton})--(\ref{eq:KKspin12}) lie, under
\begin{equation} \label{eq:SO3SBreaking2}
\textrm{SU}(8) \times \textrm{SO}(5) \; \supset \; 
\textrm{SO}(8) \times \textrm{SO}(5) \; \supset \; 
\textrm{SO}(5) \times \textrm{SO}(3)  \times \textrm{SO}(5) \; \supset \; 
\textrm{SO}(5) \times \textrm{SO}(3) \; ,
\end{equation}
with $\textrm{SO}(5)$ in the final step the diagonal of $\textrm{SO}(5) \times \textrm{SO}(5) $, followed by (\ref{eq:GroupTh}). 

The universal putative KK spectrum of both SLAG and A3C solutions can be found in appendix \ref{sec:PutSpec}. Sections \ref{sec:ClassRSpec} and \ref{sec:N=1Spec} below present the universal global spectra.

\newpage 

%%%%%%%%%%%%%%%
\subsection{Global class $\cR$ spectrum} \label{sec:ClassRSpec}
%%%%%%%%%%%%%%%

Let us particularise the discussion of section~\ref{sec:PutUniv} to obtain the universal global KK spectrum of the $\cN=2$ $\mathrm{AdS}_4 \times (\Sigma_3 \rtimes S^4)$ solution \cite{Gauntlett:2006ux} corresponding to M5-branes wrapped on a compact, negatively curved SLAG cycle $\Sigma_3 =H^3/\Gamma$ of a Calabi-Yau three-fold, for any choice of $\Sigma_3$. This is holographically dual to a sector of the light operator spectrum that is present in any 3d $\mathcal{N}=2$ SCFT of class $\mathcal{R}$, $T_N[H^3/\Gamma]$, with gauge group A$_{N-1}$.

The $D=4$ $\textrm{E}_{7(7)}/\textrm{SU}(8)$ coset representative on the SLAG solution is \cite{Pico:2025cmc}
\begin{equation} \label{eq:SugraSectorSLAG}
 \textrm{SLAG: } \quad \cV = e^{-\frac12 \varphi_0 H_0}  \; , \quad \textrm{with $e^{2\varphi_0}= 2$}, \quad  H_0= -2 (t_4{}^4 + t_5{}^5 ) \; ,
\end{equation}
with $t_4{}^4$, $t_5{}^5$ specific generators of $\textrm{E}_{7(7)}$ in the conventions of \cite{Guarino:2015qaa,Josse:2025uro}. This AdS vacuum occurs in $D=4$ $\cN=8$ TCSO$(5,0,0;\mathrm{V})$ supegravity with radius $L^2=  \sqrt 2 \,  g^{-2}$, where $g \equiv g_1 = \sqrt{2} \, g_2$, and $g_1$, $g_2$ are the coupling constants of the $\cN=8$ embedding tensor. The scalar dependent quantities $M_{\uM\uN}$, $A_1^{ij}$, $A_{2 h}{}^{ijk}$ and $B^{ij}$ that enter the KK mass matrices are constructed using (\ref{eq:SugraSectorSLAG}). As written in that equation, the SLAG vacuum is attained in a specific duality frame of the  $\cN=8$ supergravity that the mass matrices must respect. More concretely, the embedding tensor $\tilde{X}_{\uM\uN}{}^{\uP}$ that must be used is (\ref{eq:TransXSymbol}) with $U_{\uM}{}^{\uN}$ there given by (\ref{eq:E7GroupElementFlat}) with the SLAG value of (\ref{eq:TransMat}), and $X_{\uM\uN}{}^{\uP}$ given by (\ref{eq:ETtromboneSplit}) with (\ref{eq:ETtrombone}), (\ref{eq:SCBianchiV}). Similarly, SO(5) generators of the form
\begin{equation} \label{eq:TransCurlyT}
(\tilde{\cT}_{\uM})^\Sigma{}_\Lambda = U_{\uM}{}^{\uN} \, (\cT_{\uN})^\Sigma{}_\Lambda 
\end{equation}
must be used, with $(\cT_{\uN})^\Sigma{}_\Lambda$ given by (\ref{eq:TMS4}). The resulting KK mass matrices (\ref{eq:KKgraviton})--(\ref{eq:KKspin12}) can be diagonalised KK level by KK level to obtain the putative spectrum of the SLAG solution. The putative $k=0$ spectrum was already obtained in \cite{Pico:2025cmc}, and we reproduce that result. Now we have gone up a few more KK levels, from which a pattern can be found for all  $k \geq 0$. Please see appendix \ref{sec:PutSpec}.

{
\renewcommand{\arraystretch}{1.4} % Adjust row spacing globally

\begin{table}%[H]
\centering

\begin{tabular}{l | l } 
\hline
\hline
\textbf{Multiplet} & $k=3$   \\
\hline
\hline

SGRAV & $[8,\pm6] $ \\
\hline

LGRAV & $[\frac{1}{2} \left( 1 + \sqrt{161} \right), \pm 2] $ \\
\hline

LVEC 
& \hspace{-9pt} $\begin{array}{l}
[\tfrac12  (1 + \sqrt{233}), \pm 6] \oplus  [\tfrac12 (1 + \sqrt{193}), \pm 4]  \oplus \left( 2 \times [7,\pm2] \right)
\oplus [\tfrac12  (1 + \sqrt{161}),0] 
\end{array}$ \\
\hline

HYP 
& $ [10,\pm10] $ \\
\hline
\hline
\textbf{Multiplet}  & $k=2$    \\
\hline

SGRAV 
& $[6,\pm4] $ 
        \\
\hline

LGRAV   
& $[\tfrac12 (1 + \sqrt{89}),0]$ 
        \\
\hline

LVEC  
&  \hspace{-9pt} $\begin{array}{l}
[\tfrac12 (1 + \sqrt{129}),\pm4] \oplus [\tfrac12(1 +  \sqrt{105}),\pm2] \oplus
 \left( 2 \times [\tfrac12 (1 + \sqrt{97}),0] \right)
\end{array}$
\\
\hline

HYP 
& $ [8,\pm8] $ 
 \\
\hline
\hline
\textbf{Multiplet} & $k=1$     \\
\hline

SGRAV 
& $[4,\pm2] $ 

        \\
\hline

LVEC 
&   \hspace{-9pt} $\begin{array}{c}
[\tfrac12 (1 + \sqrt{57}),\pm2]  \oplus [4,0]
\end{array}$ 
\\
\hline

HYP  
& $ [6,\pm6] $ 

 \\
\hline
\hline
\textbf{Multiplet} & $k=0$    \\
\hline

MGRAV &  $ [2,0] $    \\
\hline

LVEC 
& $ [\tfrac12 (1 + \sqrt{17}),0]$ 
\\
\hline

HYP 
& $ [4,\pm4] $  
 \\
\hline
\hline
\end{tabular}
\caption{\footnotesize{Universal global spectrum of $\textrm{OSp}(4|2)$supermultiplets for the SLAG solution, for the first few KK levels $k=0,1,2,3$, as follows from (\ref{eq:GRAVSLAGUniv})--(\ref{eq:SLAGForm}). The OSp$(4|2)$ supermultiplets are denoted with the acronyms of \cite{Klebanov:2008vq}. The entries collect the superconformal primary dimension and R-charge for each supermultiplet as $[E_0,y_0]$. An entry with $n\times$ in front means that the corresponding multiplet is $n$ times repeated. An entry with multiple $\pm$ signs contains all possible sign combinations. 
}\normalsize}
\label{tab:UnivSLAGMultiplets}
\end{table}
}

Here, we are interested in extracting the universal, global, SO$(3)_S$-neutral KK spectrum above the SLAG solution. Proceeding as described in section \ref{sec:PutUniv}, we have retained the SO$(3)_S$-singlets contained in the first few KK levels $k$ of the putative spectrum, from where a pattern valid for all $k$ is again found. The individual SO$(3)_S$-invariant KK states arrange themselves in infinite towers of OSp$(4|2)$ supermultiplets. Specifically, we find a (unique, at $k=0$) massless graviton multiplet along with short and long graviton multiplets,
\begin{eqnarray} \label{eq:GRAVSLAGUniv}
\textrm{MGRAV}[2,0] \oplus \bigoplus_{k=1}^{\infty} \textrm{SGRAV}[2k+2, \pm 2k] \oplus  \bigoplus_{k=2}^{\infty}  \bigoplus_{q=0}^{k-2} \textrm{LGRAV}[E_0,4q-2k+4] \; ,
\end{eqnarray}
respectively starting at $k=1$ and $k=2$. We also find three towers of vector multiplets, all of them long, each of them starting at $k=0$, $k=1$ and $k=2$,
\begin{equation} \label{eq:VECSLAGUniv}
\bigoplus_{k=0}^{\infty}  \bigoplus_{q=0}^{k} \textrm{LVEC}[E_0,4q-2k]
 \oplus \bigoplus_{k=1}^{\infty}  \bigoplus_{q=0}^{k-1} \textrm{LVEC}[E_0,4q-2k +2 ]
 \oplus \bigoplus_{k=2}^{\infty}  \bigoplus_{q=0}^{k-2} \textrm{LVEC}[E_0,4q-2k+4] \,.
\end{equation}
Finally, there is a tower of hypermultiplets starting at $k=0$,
\begin{equation} \label{eq:HYPSLAGUniv}
\bigoplus_{k=0}^{\infty} \textrm{HYP}[2k+4, \pm(2k+4)] \; .
\end{equation}
In (\ref{eq:GRAVSLAGUniv})--(\ref{eq:HYPSLAGUniv}) we have used the    supermultiplet notation of \cite{Klebanov:2008vq}, LABEL$[E_0 , y_0]$, where $E_0$ and $y_0$ are the dimension and R-charge of the superconformal primary and the LABELs indicate the top spin component. An entry with a $\pm$ sign for the R-charge means that multiplets with both signs are present, for example, $\textrm{SGRAV}[2k+2, \pm 2k] \equiv \textrm{SGRAV}[2k+2, 2k] \oplus \textrm{SGRAV}[2k+2, -2k]$. Finally, $E_0$ is given in all cases, long and short, by
\begin{equation} \label{eq:SLAGForm}
E_0=\tfrac12 + \sqrt{\tfrac{17}{4}-s_0(s_0+1)+2k(k+3)+\tfrac12 y_0^2 }\,.
\end{equation}
Here, $s_0$ is the spin of the superconformal primary ($s_0=1, 0, 0$ for graviton, vector and hypermultiplets, respectively), $k$ is the KK level at which the multiplet arises, and $y_0$ is the indicated R-charge. For example, $E_0=\tfrac12 + \sqrt{\tfrac{9}{4}+2k(k+3)+\tfrac12 (4q-2k+4 )^2  }$, $q=0, 1, \ldots , k-2$, for the long graviton multiplets at level $k$ in (\ref{eq:GRAVSLAGUniv}). The formula (\ref{eq:SLAGForm}) also reproduces the appropriate short multiplet dimensions whenever the relevant superconformal bound is saturated. 

For quick reference, table~\ref{tab:UnivSLAGMultiplets} explicitly lists the superconformal multiplets (\ref{eq:GRAVSLAGUniv})--(\ref{eq:SLAGForm}) for the first few KK levels. The universal, global SLAG spectrum at KK level $k=0$ was first given in \cite{Gauntlett:2006ux} as 
\begin{equation} \label{eq:SLAGKK0}
\textrm{MGRAV} [2,0] \oplus \textrm{LVEC}  [\tfrac12 (1 + \sqrt{17}),0] \oplus \textrm{HYP} [4,\pm4] \; .
\end{equation}
Equations (\ref{eq:GRAVSLAGUniv})--(\ref{eq:SLAGForm}) reproduce that result and extend it to all KK levels $k \geq 0$. 

%%%%%%%%%%%%%%%
\subsection{Vanishing of the superconformal index contribution} \label{sec:SCI}
%%%%%%%%%%%%%%%

No charge saturation is produced for any of the long multiplets present in (\ref{eq:GRAVSLAGUniv})--(\ref{eq:SLAGForm}), which therefore stay long as indicated for all values of their charges. This makes it straightforward to extract the universal, globally defined short spectrum as 
\begin{eqnarray} \label{eq:ShortSLAGUniv}
\textrm{MGRAV}[2,0] \oplus \bigoplus_{k=1}^{\infty} \textrm{SGRAV}[2k+2, \pm 2k] \oplus  \bigoplus_{k=0}^{\infty} \textrm{HYP}[2k+4, \pm(2k+4)] \; .
\end{eqnarray}
This allows us to compute holographically the contribution of this sector of the short operator spectrum to the superconformal index \cite{Kinney:2005ej}, 
\begin{equation} \label{eq:SCI}
{\cal I} = \textrm{Tr} \,  (-1)^F x^{\Delta +s} \, e^{-\beta \{ Q, S \} } \; ,
\end{equation}
where the trace is taken over the Hilbert space of the large-$N$ $T[H^3/\Gamma]$ theory on $S^1 \times S^2$, $x$ is a superconformal fugacity, and $F$, $\Delta$ and $s$ are the fermion number, the conformal dimension and the spin of the contributing states. Only states in short multiplets that saturate the bound
\begin{equation} \label{eq:SCIBound}
\{ Q, S \} = \Delta -R -s = 0 \; , 
\end{equation}
where $R$ is the R-charge, contribute to the index, so that it does not depend on $\beta$. See \cite{Kim:2021zlh} for the calculation of the superconformal index of a different 3d $\cN=2$ SCFT in a similar holographic setup.

Enforcing (\ref{eq:SCIBound}), the contributions to the single particle index of the three different multiplets present in (\ref{eq:ShortSLAGUniv}) are found to be
\begin{equation}
\textrm{MGRAV}[2,0] \, : \, -\frac{x^4}{1-x^2} \; , \quad
\textrm{SGRAV}[|y_0|+2 , y_0] \, : \, -\frac{x^{|y_0|+4}}{1-x^2} \; , \quad
\textrm{HYP}[|y_0| , y_0] \, : \, +\frac{x^{|y_0|}}{1-x^2} \; ,
\end{equation}
where the $1/(1-x^2)$ terms account for derivatives. Bringing in the explicit superconformal primary R-charges, the contribution of (\ref{eq:ShortSLAGUniv}) to the single-particle superconformal index can be seen to vanish,
\begin{equation} \label{eq:ZeroIndex}
{\cal I}_{\textrm{sp}} = \frac{1}{1-x^2} \Big( -x^4  -\sum_{k=1}^\infty x^{2k+4} + \sum_{k=0}^\infty x^{2k+4}  \Big) =0 \; .
\end{equation}
These cancellations occur between states that do not recombine into long multiplets. Vanishing still holds if the calculation is extended to include contributions coming from the locally defined SGINO and SVEC multiplets present in the short putative spectrum (\ref{eq:SLAGN8ShortKKZero}), (\ref{eq:SLAGN8ShortKK}). In the latter case, though, the vanishing can be interpreted from the fact that all those short multiplets arise from long multiplets at threshold, as discussed in appendix \ref{sec:PutSpec}. 

This vanishing result is slightly surprising in the light of \cite{Benini:2019dyp,Bobev:2020zov}, where order $N^0$ corrections are found to the entropy of M5-brane-related black holes in the context of class $\cR$. This is not contradictory, though, because in \cite{Benini:2019dyp,Bobev:2020zov} the physics is dominated by the black hole saddle, not the AdS vacuum saddle, in a Cardy-like limit. In our case, the superconformal index simply turns out to be too coarse to provide a non-vanishing result. Of course, the short operators (\ref{eq:ShortSLAGUniv}) are still perfectly physical and may contribute to other observables, like conformal blocks or higher point correlators.

%%%%%%%%%%%%%%%
\subsection{Global $\cN=1$ spectrum} \label{sec:N=1Spec}
%%%%%%%%%%%%%%%

We now turn to present the universal KK spectrum of the $\cN=1$ $\mathrm{AdS}_4 \times (\Sigma_3 \rtimes S^4)$ solution of  \cite{Acharya:2000mu} associated to M5-branes wrapped on a negatively curved A3C cycle $\Sigma_3 = H^3/\Gamma$ of a seven-dimensional G$_2$-holonomy manifold. The process is similar to that described in section \ref{sec:ClassRSpec}, so we will now be brief. In $D=4$ $\cN=8$ TCSO$(5,0,0;\mathrm{V})$-gauged supergravity, the A3C vacuum is attained with $\textrm{E}_{7(7)}/\textrm{SU}(8)$ coset representative \cite{Pico:2025cmc}
\begin{eqnarray} \label{eq:SugraSectorA3C}
 \textrm{A3C} &:& \cV = e^{-\frac12 (\varphi_0 H_0 +\varphi_1 H_1 )}  \; , \quad \textrm{with $e^{2\varphi_0} =e^{-2\varphi_1} = \tfrac85 $}, \nonumber \\
 && H_0 = t_1{}^1+t_2{}^2+t_3{}^3+t_4{}^4 \; , \quad 
H_1 = -t_1{}^1-t_2{}^2-t_3{}^3+3t_4{}^4 \; ,
\end{eqnarray}
and radius $L^2 = \frac{25}{32}\sqrt{\frac{5}{2}} \,  g^{-2}$. The KK mass matrices  (\ref{eq:KKgraviton})--(\ref{eq:KKspin12}) must be evaluated on (\ref{eq:SugraSectorA3C}), and with generators $\tilde{X}_{\uM\uN}{}^{\uP}$ and $(\tilde{\cT}_{\uM})^\Sigma{}_\Lambda$, respectively given by (\ref{eq:TransXSymbol}) and (\ref{eq:TransCurlyT}) with $U_{\uM}{}^{\uN}$ there now given by (\ref{eq:E7GroupElementFlat}) with the A3C value of (\ref{eq:TransMat}). 

{
\renewcommand{\arraystretch}{1.4} % Adjust row spacing globally

\begin{table}%[H]
\centering
\begin{tabular}{l | l} 
\hline
\hline

\textbf{Multiplet} &$k \geq  4 $  \\
\hline

GRAV  
& $ \left[ 1 + \sqrt{\tfrac52 k(k+3) + \tfrac94  } \right] \otimes [0]$
\\
\hline

GINO    
& $\left( 2 \times  \left[ 1 + \sqrt{\tfrac52 k(k+3) - 1 } \right] \otimes [1]  \right)$
\\
\hline

VEC 
& $\left( 3 \times  \left[  1 + \sqrt{\tfrac52 k(k+3) + \tfrac14  } \right] \otimes [1]  \right) \oplus \left[  1 + \sqrt{\tfrac52 k(k+3) - \tfrac{39}{4}  }  \right] \otimes [2]$
\\
\hline

CHIRAL 
&  $\left( 4 \times  \left[ 1 + \sqrt{\tfrac52 k(k+3) + 6  } \right] \otimes [0] \right) \oplus \left( 2 \times \left[ 1 + \sqrt{\tfrac52 k(k+3) -9  }  \right] \otimes [2] \right)$
\\
\hline
\hline

\textbf{Multiplet} & $k=3$    \\
\hline

GRAV
&  $ \left[ 1 + \frac{3 \sqrt{21}}{2} \right] \otimes [0]$
        \\
\hline

GINO 
& $\left( 2 \times  \left[ 1 + 2 \sqrt{11} \right] \otimes [1]  \right)$
        \\
\hline

VEC 
& $\left( 3 \times  \left[  1 + \frac{\sqrt{181}}{2} \right] \otimes [1]  \right) \oplus \left[  1 + \frac{\sqrt{141}}{2}  \right] \otimes [2]$
\\
\hline

CHIRAL 
&  $\left( 4 \times  \left[ 1 + \sqrt{51} \right] \otimes [0] \right) \oplus \left[ 7 \right] \otimes [2]$
 \\
\hline
\hline

\textbf{Multiplet} & $k=2$    \\
\hline

GRAV 
& $ \left[  1 + \frac{\sqrt{109}}{2} \right] \otimes [0]$
        \\
\hline

GINO 
& $\left( 2 \times  \left[ 1 + 2 \sqrt{6} \right] \otimes [1]  \right)$ 
        \\
\hline

VEC 
& $\left( 2 \times  \left[ 1 + \frac{\sqrt{101}}{2} \right] \otimes [1]  \right) $
\\
\hline

CHIRAL 
&  $\left( 4 \times  \left[ 1 + \sqrt{31} \right] \otimes [0] \right) \oplus \left[ 5 \right] \otimes [2]$
\\
\hline
\hline

\textbf{Multiplet} & $k=1$    \\
\hline

GRAV 
& $ \left[ \tfrac92 \right] \otimes [0]$  
        \\
\hline

GINO
&  $  \left[4  \right] \otimes [1]  $ 
\\
\hline

VEC 
& $\left[ 1 + \frac{\sqrt{41}}{2}  \right] \otimes [1]$
\\
\hline

CHIRAL 
&  $\left( 3 \times  \left[ 5 \right] \otimes [0] \right)$
\\
\hline
\hline

\textbf{Multiplet} & $k=0$    \\
\hline

MGRAV 
&  $ \left[ \tfrac52 \right] \otimes [0]$  
\\
\hline

MVEC 
& $\left[ \tfrac32 \right] \otimes [1]$ 
\\
\hline

CHIRAL 
&  $\left( 2 \times  \left[1+\sqrt{6} \right] \otimes [0] \right)$
\\
\hline
\hline
\end{tabular}
\caption{\footnotesize{Universal global spectrum of $\textrm{OSp}(4|1) \times \textrm{SO}(3)_+$ supermultiplets for the A3C solution at all KK level $k$. The OSp$(4|2)$ supermultiplets are denoted with the acronyms of \cite{Cesaro:2020soq}. The entries collect the superconformal primary dimension $E_0$, (\ref{eq:A3CUnivForm}), for each OSp$(4|1)$ supermultiplet and the spin $\ell$ of $\textrm{SO}(3)_+$ as $[ E_0 ] \otimes [\ell]$. An entry with $n\times$ in front means that the corresponding multiplet is $n$ times repeated.
}\normalsize}
\label{tab:GlobalA3CMultiplets}
\end{table}
}

The resulting matrices (\ref{eq:KKgraviton})--(\ref{eq:KKspin12}) can again be diagonalised KK level by KK level. The resulting putative spectrum can be found in appendix~\ref{sec:PutSpec}. The globally defined KK spectrum can be obtained by extracting the singlets under the SO$(3)_S^\prime$ defined in (\ref{eq:SO3SBreaking2}), (\ref{eq:GroupTh}). The $\cN=1$ A3C solution preserves the SO$(3)_+$ that appears in (\ref{eq:GroupTh}). Accordingly, the global KK modes arrange themselves in representations of OSp$(4|1) \times \textrm{SO}(3)_+$. The dimension $E_0$ of the superconformal primary with spin $s_0$ of an OSp$(4|1)$ multiplet arising at KK level $k$ with SO$(3)_+$ spin $\ell$ turns out to be given by
\begin{equation} \label{eq:A3CUnivForm}
E_0= 1 + \sqrt{6-s_0(s_0+1)+\tfrac52 k(k+3) -\tfrac52 \ell (\ell+1) }\,.
\end{equation}
The global multiplets present in the spectrum have been listed in table~\ref{tab:GlobalA3CMultiplets}. The table uses the multiplet notation of \cite{Cesaro:2020soq}, LABEL$[E_0]$, with $E_0$ the superconformal primary dimension and LABEL one of the acronyms (M)GRAV, GINO, (M)VEC and CHIRAL. These respectively have $s_{0} =\frac32 , 1, \frac12 , 0$ in (\ref{eq:A3CUnivForm}). The number of multiplets increases with $k$ and, as shown in table~\ref{tab:GlobalA3CMultiplets},  appears to stabilise at $k\geq4$, as we have checked up to $k=6$. The $k=0$ spectrum reproduces the partial result of \cite{Gauntlett:2002rv}.

%%%%%%%%%%%%%%%
%%%%%%%%%%%%%%%

\section{Discussion} \label{sec:Discussion}

%%%%%%%%%%%%%%%
%%%%%%%%%%%%%%%

Trombone gaugings of lower dimensional supergravities typically arise as dimensional reductions on non-compact internal spaces. The trombone-gauged maximal four-dimensional TCSO$(5,0,0;\mathrm{V})$ supergravity that we constructed in \cite{Pico:2025cmc} is no exception. It arises by dimensional reduction of $D=11$ supergravity on seven-dimensional spaces of the form $G_3 \rtimes S^4$ \cite{Pico:2026rji}. Here, $G_3$ is the (non-compact) Bianchi type V group manifold, equipped with its usual Einstein metric (\ref{eq:BachasEstesSimp}). The $S^4$ factor is the four-sphere, endowed with metrics deformed from the usual round one, which can be read off from (\ref{eq:SLAGMetric}), (\ref{eq:A3CMet}). The $S^4$ is fibred over $G_3$ by the spin connection of the latter following either of two distinct possibilities. Altogether, $G_3 \rtimes S^4$ are locally equivalent \cite{Pico:2026rji} to the honestly compact spaces $\Sigma_3 \rtimes S^4$, with $\Sigma_3 = H^3/\Gamma$, that arise near the AdS$_4$ throats of large stacks of M5-branes wrapped on SLAG \cite{Gauntlett:2006ux} or A3C \cite{Acharya:2000mu} three-cycles $\Sigma_3$ of special holonomy manifolds.

The linearised $D=4$ $\cN=8$ TCSO$(5,0,0;\mathrm{V})$-gauged supergravity contains modes that sit at the bottom of the continuous spectrum of $D=11$ supergravity on the non-compact spaces $G_3 \rtimes S^4$. In this paper we have shown that one can still use both $G_3 \rtimes S^4$ and TCSO$(5,0,0;\mathrm{V})$ supergravity to extract significant information about the universal, discrete KK spectrum of $D=11$ supergravity on the actual M5-brane-related compact geometries $\Sigma_3 \rtimes S^4$ of \cite{Gauntlett:2006ux,Acharya:2000mu}, valid for any choice of $\Sigma_3 = H^3/\Gamma$. We have done this by adapting the ExFT-based KK spectral methods of \cite{Malek:2019eaz,Malek:2020yue,Varela:2020wty,Cesaro:2020soq} to the case where the underlying lower-dimensional maximal supergravity contains gaugings of the trombone scaling symmetry. We have referred to the resulting discrete spectra as putative, with the understanding that additional criteria must be imposed in order to extract bonafide physical modes. 

These features make the M5-brane spectral cases in this paper and in 
\cite{Bhattacharya:2024tjw,Varela:2025xeb,BKV2026} significantly differ from those previously treated with these ExFT techniques in, for example, \cite{Malek:2019eaz,Malek:2020yue,Varela:2020wty,Cesaro:2020soq,Malek:2020mlk,Guarino:2020flh,Bobev:2020lsk,Giambrone:2021zvp,Cesaro:2021haf,Cesaro:2021tna,Eloy:2021fhc,Cesaro:2022mbu}. 
In the latter cases, the underlying maximal truncation occurs on spheres or products thereof. Thus, the resulting KK spectra are both complete and automatically globally well defined. In the present $\Sigma_3 \rtimes S^4$ cases, our ExFT methods can only assess  reliably a universal sector of the spectrum, valid for all $\Sigma_3$. This is certainly very helpful, but the complete $\Sigma_3 \rtimes S^4$ spectrum for each choice of $\Sigma_3$ necessitates of further assessment, including the additional diagonalisation of the Bochner Laplacian (\ref{eq:Bochner}). More importantly, this universal sector is, in general,  only locally defined on the bundles $\Sigma_3 \rtimes S^4$. It would be interesting to address the global validity of these modes for specific choices of $\Sigma_3 = H^3/\Gamma$. More generally, it would be interesting to compute the complete spectrum for concrete choices of $\Sigma_3 = H^3/\Gamma$, perhaps by adapting the bootstrap inspired methods of \cite{Bonifacio:2020xoc,Bonifacio:2023ban,Gesteau:2023brw} for the standard hyperbolic Laplacian to the Bochner Laplacian.

A concrete criterion we have used to extract physically meaningful modes out of our putative spectra has been global definiteness. By imposing certain generalised $G$-structures in the sense of \cite{Cassani:2019vcl}, we have been able to extract modes that are globally defined on the bundles $\Sigma_3 \rtimes S^4$ and, thus, physical. In \cite{Blair:2024ofc}, it was shown how to obtain consistent truncations with infinitely many lower-dimensional fields at the full non-linear level. Our global KK spectra may be regarded as a linearisation of the construction of \cite{Blair:2024ofc} for the cases at hand. It would be helpful to understand how to implement this KK prescription directly from \cite{Cassani:2019vcl,Blair:2024ofc}, as a direct way to obtain honestly physical modes without the need to compute the putative spectra first. Having said this, even if these putative spectra are only locally defined in general, they can still be useful and have physical significance. In other words, the global prescription that we employed in sections \ref{sec:ClassRSpec} and \ref{sec:N=1Spec} might not be the only sensible criterion for physicality. For example, one may envisage the existence of a domain wall connecting the SLAG and A3C vacua, dual to a renormalisation group flow of the dual 3d field theory. If it exists, such flow would of course be perfectly physical, and yet must necessarily occur outside the globally defined sector around either solution. The reason is that, geometrically, the SLAG fibres must transition into the A3C fibres along the flow, and that may only happen locally.

Our analysis of the light, globally defined short multiplets illuminates the structural role played by protected operators in the 3d--3d correspondence. Although their total contribution to the superconformal index vanishes, these modes nevertheless furnish a concrete and universal subsector of the theories $T_N[\Sigma_3]$, present for all hyperbolic three-manifolds $\Sigma_3 = H^3/\Gamma$. In particular, the results of section 3.3 provide explicit operator-level data, including scaling dimensions, shortening conditions, and OSp$(4|2)$ representation content, in a class of strongly coupled SCFTs for which no intrinsic field-theoretic construction is available. Our results supply a 
rare foothold into the operator algebra of the $T_N[\Sigma_3]$ class $\cR$ theories at large $N$ beyond the level of partition functions.

%%%%%%%%%%%%%%%
%%%%%%%%%%%%%%%

\section*{Acknowledgements}

%%%%%%%%%%%%%%%
%%%%%%%%%%%%%%%

We would like to thank Jim T.~Liu for discussions, and Ritabrata Bhattacharya, Matt Doniere and Abhay Katyal for collaboration in related projects. MP is supported by predoctoral award FPU22/02084 from the Spanish Government, and partially by Spanish Government grants CEX2020-001007-S, PID2021-123017NB-I00 and PID2024-156043NB-I00, funded by MCIN/AEI/10.13039/ 501100011033, and ERDF, EU. OV is supported by NSF grant PHY-2310223.

%%%%%%%%%%%%%%%
%%%%%%%%%%%%%%%

\appendix

\addtocontents{toc}{\setcounter{tocdepth}{1}}

%%%%%%%%%%%%%%%
%%%%%%%%%%%%%%%

%%%%%%%%%%%%%%%
%%%%%%%%%%%%%%%

\section{Derivation of the mass matrices} 
\label{eq:DerMassMat} 

%%%%%%%%%%%%%%%
%%%%%%%%%%%%%%%

In this appendix, we give some details of the derivation of the KK mass matrices given in section \ref{sec:KKMassMat} by linearisation of the ExFT equations of motion. We will focus for simplicity on the bosonic mass matrices. In order to lighten the (already heavy) index notation, we suppress the underlining of the ExFT flat indices (corresponding to the $D$-dimensional $\cN=8$ supergravity) and use the convention that beginning of alphabet indices, $A$, $B$, $\ldots$, are flat, and middle alphabet indices, $M$, $N$, $\ldots$, are curved. 

Let us also introduce some preliminary definitions. The metric $\overline{g}_{\mu \nu} (x,y)$ used in the main text is related to the metric $g_{\mu \nu} (x,y)$ used in this appendix following \cite{Blair:2024ofc} by
\begin{equation}
g_{\mu \nu} (x,y) = \Delta^{2} (x,y) \overline{g}_{\mu \nu} (x,y) \; ,
\end{equation}
where $\Delta$ here and in (\ref{eq:GenSSBos}), (\ref{eq:GenSSFer}) is related to $\tilde{\Delta}$ in (\ref{eq:MassGravOp}) as
\begin{equation}
\Delta^{D-2}= e^{(D-2) \tilde{\Delta}} {g}^{\frac{1}{D-2}} \; ,
\end{equation}
with $g \equiv \textrm{det} \, g_{mn}$, and $g_{mn}$ the internal metric. We also introduce the rescaled quantities
\begin{equation}
U^M{}_A = \Delta^{-1} \, \hat{E}^M{}_A    \; , \quad 
\cM_{MN}=\Delta^{-2} G_{MN} \; ,
\end{equation}
w.r.t.~$\hat{E}^M{}_A$ and $G_{MN}$ used in the main text. The following relations hold:
\begin{equation}
(\det G_{MN})^{- \tfrac{D-2}{2 \dim E_{d(d)}}}= (\det \hat{E}^M{}_{A})^{ \tfrac{D-2}{ \dim E_{d(d)}}}=\sqrt{g} \, e^{(D-2) \tilde \Delta}\, .
\end{equation}

\subsection{ExFT equations of motion} \label{sec:ExFTeoms}

%%%%%%%%%%%%%%%

We start by collecting the ExFT equations of motion \cite{Hohm:2013pua,Hohm:2013vpa,Hohm:2013uia}, following the conventions of \cite{Blair:2024ofc}. The Einstein equation for the external metric $g_{\mu\nu}$ is 
\be
\begin{split}
0 & = 
\hat R_{\mu \nu} - \tfrac{1}{2} g_{\mu\nu} \left( \hat R + \tfrac{1}{4 \alpha}  g^{\rho\sigma} \mathcal{D}_\rho \mathcal{M}_{MN} \mathcal{D}_\sigma \mathcal{M}^{MN} - \tfrac{c_A}{4} \gM_{MN} \Fa_{\rho \sigma}{}^M \Fa^{\rho \sigma N} \right) 
\\ & \qquad
+ V_{\mu\nu} 
+ \tfrac{1}{4 \alpha} \mathcal{D}_\mu \gM_{MN} \mathcal{D}_\nu \gM^{MN} 
- \tfrac{c_A}{2 } \gM_{MN} \Fa_{\mu \rho}{}^M \Fa_{\nu \sigma}{}^N g^{\rho \sigma}
 \,,
\end{split} 
\label{ExcEin} 
\ee 
with the Ricci tensor and scalar given in terms of the Levi-Civita connection as
\be
\hat R_{\mu\nu} = \mathcal{D}_\rho \Gamma_{\mu\nu}{}^\rho - \mathcal{D}_\mu \Gamma_{\nu \rho}{}^\rho 
+ \Gamma_{\rho\lambda}{}^\rho \Gamma_{\mu\nu}{}^\lambda 
- \Gamma_{\nu\lambda}{}^\rho\Gamma_{\rho \mu}{}^\lambda 
\,,\qquad
\hat R = g^{\mu\nu} \hat R_{\mu\nu} \, ,
\label{ExceptionalRicci}
\ee
and we have defined
{\setlength\arraycolsep{2pt}
\begin{eqnarray} \label{eq:Veff_Ein}
V_{\mu\nu}  &\equiv  & \tfrac12 g_{\mu\nu} V( \mathcal{M},g )
+ \tfrac{1}{2}|g|^{-1/2} g_{\mu\nu} \partial_M \left( |g|^{1/2} ( \partial_N \gM^{MN} + \gM^{MN} \partial_N \ln |g| )\right) 
\\ &&
- \tfrac{1}{2} |g|^{-1/2}\partial_M ( |g|^{1/2} \gM^{MN} ) \partial_N g_{\mu\nu}  
- \tfrac{1}{2} \gM^{MN} g_{\mu \rho} \partial_M g^{\rho \sigma} \partial_N g_{\sigma \nu} 
- \tfrac{1}{2} \gM^{MN} \partial_M \partial_N g_{\mu\nu}\, , \nonumber 
\end{eqnarray}
}where
{\setlength\arraycolsep{2pt}
\begin{eqnarray} 
V( \mathcal{M},g ) & \equiv&  - \tfrac{1}{4 \alpha} \gM^{MN} \partial_M \gM^{KL} \partial_N \gM_{KL} + \tfrac{1}{2} \gM^{MN} \partial_M \gM^{KL} \partial_L \gM_{NK} \\
&& - \tfrac{1}{2} \partial_M \ln |g| \partial_N \gM^{MN} 
- \tfrac{1}{4} \gM^{MN} \partial_M \ln |g| \partial_N \ln |g| - \tfrac{1}{4} \gM^{MN} \partial_M g^{\mu \nu} \partial_N g_{\mu\nu}\; . \nonumber 
\end{eqnarray}
}Also, $\Fa_{\mu\nu}{}^M$ is the field strength of the gauge field $\cA_{\mu}{}^M$, subject, in E$_{7(7)}$ ExFT, to the self-duality relation
\begin{equation}
\mathcal{F}_{\mu\nu}{}^M = - \tfrac12 |g|^{1/2} \epsilon_{\mu\nu\rho\sigma} \Omega^{MN} \gM_{NK} \mathcal{F}^{\rho\sigma K} \; .
\end{equation}
The constant $\alpha$ has been defined in (\ref{eq:ExFTConstants}), while $c_A = \frac12$ for E$_{7(7)}$ and $c_A = 1$ for E$_{6(6)}$. Finally, the covariant derivative is defined as $\mathcal{D}_\mu = \partial_\mu - L_{\cA_\mu}$, in terms of the generalised Lie derivative $L$ \cite{Coimbra:2011ky}.

The gauge field equation of motion is
{\setlength\arraycolsep{0pt}
\begin{eqnarray} \label{eq:Aeom} 
&& c_A
g^{-1/2} \mathcal{D}_\nu \left[ g^{1/2} \gM_{MN} \Fa^{\nu\mu N} \right] + |g|^{-1/2} \frac{\delta \mathcal{L}_{\text{CS}}}{\delta \Aa_\mu{}^M} 
 \\ 
 && 
 + \tfrac{1}{\alpha} \left(
 -\frac{1}{2}  \partial_M \gM_{KL} \mathcal{D}^\mu \gM^{KL} 
+ g^{-1/2} \partial_P \left( \left[ \delta_K^P \gM_{LM} - Y^{QP}{}_{MK} \gM_{LQ} \right] g^{1/2} \mathcal{D}^\mu \gM^{KL} \right)
\right) \nonumber 
\\ 
&& 
 +g^{\mu\nu} \partial_M \mathcal{D}_\nu \ln g + \partial_M \mathcal{D}_\nu g^{\mu\nu} 
 - \frac{1}{2} g^{\mu\lambda} \partial_M g^{\nu\rho} \mathcal{D}_\lambda g_{\nu\rho} 
+g^{\mu\lambda} \partial_Mg^{\nu\rho} \mathcal{D}_\nu g_{\rho\lambda} 
+\frac{1}{2} \partial_M g^{\mu\nu} \mathcal{D}_\nu\ln g =0 \; . \nonumber 
\end{eqnarray}
Here, 
\be
|g|^{-1/2}
\frac{\delta \mathcal{L}_{\text{CS}}}{\delta \Aa_\mu{}^M} = 
 - \tfrac14 |g|^{-1/2} \epsilon^{\mu\nu\rho\sigma} \mathcal{D}_\nu \mathcal{F}_{\rho\sigma}{}^N \Omega_{NM}\, ,
 \label{ChernSimonsContribs}
\ee
is a Chern-Simons contribution that is present for E$_{7(7)}$ only, and 
\begin{equation} \label{YtensorDef}
Y^{AB}{}_{CD} \equiv - \alpha \mathbb{P}^{A}{}_D{}^B{}_C + \delta^A_C \delta^B_D + \tfrac{1}{D-2} \delta ^A_D \delta^B_C \; ,
\end{equation}
with $\mathbb{P}$ the adjoint projector defined below equation (\ref{eq:CTCon}). 

Finally, the scalar equation of motion reads
\begin{equation} \label{eq:ExFTScalar}
 \gM_{P(M} \mathbb{P}^P{}_{N)}{}^K{}_Q \gM^{LQ} \mathcal{K}_{KL} =0 \; ,
\end{equation}
where we have defined 
{\setlength\arraycolsep{2pt}
\begin{eqnarray}
\mathcal{K}_{MN} & \equiv & 
- \tfrac{1}{4\alpha}|g|^{-1/2}  \mathcal{D}_\mu( |g|^{1/2} \mathcal{D}^\mu \gM_{MN} ) 
+ \tfrac{1}{4\alpha}|g|^{-1/2} \gM_{MK} \gM_{NL}  \mathcal{D}_\mu( |g|^{1/2} \mathcal{D}^\mu \gM^{KL} ) \nonumber\\[4pt]
&& + \tfrac{c_A}{2} \gM_{MK} \gM_{NL} \mathcal{F}_{\mu\nu}{}^K \mathcal{F}^{\mu\nu}{}_L  + V_{(MN)}\,,
\end{eqnarray}
}with
{\setlength\arraycolsep{2pt}
\begin{eqnarray} \label{eq:VMN}
V_{MN} & \equiv & 
\tfrac{1}{4\alpha} \partial_M \gM^{KL} \partial_N \gM_{KL} - \tfrac12 \partial_M \gM^{KL} \partial_L \gM_{NK} 
%\nonumber \\ && 
- \tfrac{1}{4\alpha} \partial_K ( \gM^{KL}[ \partial_L \gM_{MN} - 2 \alpha \partial_M \gM_{LN} ) \nonumber \\[5pt] 
&& + \tfrac{1}{4\alpha} \gM_{KM} \gM_{LN} \partial_P ( \gM^{PQ} \partial_Q \gM^{KL} - 2 \alpha \gM^{KQ} \partial_Q \gM^{LP} )  \\[5pt] 
&& 
- \tfrac{1}{4\alpha} \partial_P \ln |g| \gM^{PQ} ( \partial_Q \gM_{MN} - 2 \alpha \partial_M \gM_{QN} ) 
+ \tfrac14 \partial_M g^{\mu\nu} \partial_N g_{\mu\nu} - \tfrac12 \partial_M \partial_N \ln |g| \,. \nonumber
\end{eqnarray}
}

%%%%%%%%%%%%%%%
\subsection{KK trombone mass matrices} \label{sec:KKTrombMMDer}
%%%%%%%%%%%%%%%

We now linearise the ExFT field equations reviewed in appendix \ref{sec:ExFTeoms} to extract the mass matrices for the KK fluctuations (\ref{eq:KKExpansionBos}) above the vacuum (\ref{eq:VacCond}). We focus on keeping track of the new trombone contributions to the tromboneless mass matrices derived in \cite{Malek:2019eaz,Malek:2020yue,Varela:2020wty}.

%%%%%%%%%%%%%%%
\subsubsection{Graviton mass matrix} \label{sec:GravMassMat}
%%%%%%%%%%%%%%%

The linearisation  \cite{Malek:2020yue} of the ExFT Einstein equation (\ref{ExcEin}) produces the eigenvalue problem for the operator  (\ref{eq:MassGravOp}), 
\begin{equation} \label{eq:BachasEstes}
e^{(2-D)\tilde{\Delta}}  {g}^{-\tfrac12} \partial_m ({g}^{\tfrac12} {g}^{mn} e^{D \tilde{\Delta}} \partial_n h(y))=-M^2 h(y) \; ,
\end{equation}
corresponding to the massive graviton fluctuations, of squared mass $M^2$, around a maximally symmetric background spacetime \cite{Bachas:2011xa}. Here, indices $m$, $n$, $\ldots$ are curved indices on the local compactification space. Under the ExFT section condition
\begin{equation} 
{g}^{mn} \partial_m \otimes \partial_n= e^{-2 \tilde{\Delta}} G^{MN}  \partial_M \otimes \partial_N = {g}^{\frac{1}{D-2}} \gM^{MN}  \partial_M \otimes \partial_N \; ,
\end{equation}
equation (\ref{eq:BachasEstes}) reduces to
\begin{equation} \label{eq:BEStep2}
\Delta^{2-D} \partial_M \left( \Delta^D \gM^{MN} \partial_N h(y) \right) = -M^2 h(y) \; .
\end{equation}
Then, on a consistently truncated background, (\ref{eq:GenSSBos}) holds and $\hat \partial_A = \Delta U^M{}_A \partial_M$ have the right weight so that the vector component is a proper vector, not a vector density. Under these conditions, (\ref{eq:BEStep2}) gives
\begin{equation}
\Delta \left( (D-1) U^M{}_A \partial_M \ln \Delta +\partial_M U^M{}_A \right) + M^{AB} \hat \partial_A \hat \partial_B h(y) = -M^2 h(y)\, ,
\end{equation}
with
\begin{equation}
U^M{}_A \partial_M \ln \Delta = \tfrac{1}{D-1} \left( \tfrac{D-2}{\Delta} \, \vartheta_A - \partial_M U^M{}_A \right)\, .
\end{equation}
Thus, the eigenvalue equation for the graviton mass operator takes on the final form
\begin{equation} \label{eq:GravEigenv}
 (D-2) \vartheta_A M^{AB} \hat \partial_B h(y)  + M^{AB} \hat \partial_A \hat \partial_B h(y) = -M^2 h(y)\,.
\end{equation}
Expanding the perturbation $h(y)$, as in (\ref{eq:KKExpansionBos}), in harmonics $\cY^\Lambda(y)$ subject to (\ref{eq:CurlyTDef}), the graviton mass matrix (\ref{eq:KKgraviton}) straightforwardly follows from (\ref{eq:GravEigenv}).

%%%%%%%%%%%%%%%
\subsubsection{Vector mass matrix} \label{sec:VectorMassMat}
%%%%%%%%%%%%%%%

The vector mass matrix is derived by linearisation of (\ref{eq:Aeom}). Under the fluctuation ansatz \eqref{eq:KKExpansionBos}, the covariant derivatives of the external metric become
\begin{equation}
\mathcal{D}_\mu g_{\nu \rho} = \Delta^2 D_\mu \bar g_{\nu \rho}=\Delta^2 \Big( \partial_\mu \bar g_{\nu \rho} - \big(2 A_\mu{}^A \vartheta_A + \tfrac{2}{D-2} \hat \partial_A  A_\mu{}^A \big) \bar g_{\nu \rho} \Big) \; ,
\end{equation}
with $A_\mu{}^A = A_\mu{}^{A \Lambda}(x) \cY(y)_\Lambda$. Consequently,
\begin{equation}
\mathcal{D}_\mu \ln g = \partial_\mu \bar g - 2D \big( A_\mu{}^A \vartheta_A  +\tfrac{1}{D-2} \,  \hat \partial_A  A_\mu{}^A \big) \; ,
\end{equation}
and the last five terms of \eqref{eq:Aeom}, containing covariant derivatives of the external metric, combine into
\begin{equation}
\Delta^{-3} U^A{}_M \bar g^{\mu \nu} (-2)(D-1)\left( \vartheta_B \hat \partial_A A_\nu{}^B + \tfrac{1}{D-2} \hat \partial_A \hat \partial_B A_\nu{}^B \right) \; . 
\end{equation}
These terms are responsible for the Higgsing of the massive gravitons and do not contribute to the vector mass matrix \cite{Malek:2020yue}.

Under (\ref{eq:GenSSBos}), the vector contributions coming from the generalised internal metric read
{\setlength\arraycolsep{0.5pt}
\begin{eqnarray}
\mathcal{D}_\mu \cM_{MN} &=& U^A{}_M U^B{}_N \big( -2 A_\mu{}^C X_{C(A}{}^D M_{B)D} -2  A_\mu{}^C \vartheta_C M_{AB} -2 \alpha M_{C(A} \mathbb{P}^C{}_{B)}{}^E{}_D \hat \partial_E  A_\mu{}^D \big) \; , \nonumber \\[4pt]
\mathcal{D}_\mu \cM^{MN} & = & U^M{}_A U^N{}_B \left( 2 A_\mu{}^C X_{CD}{}^{(A} M^{B)D} +2  A_\mu{}^C \vartheta_C M^{AB} + 2 \alpha M^{C(A} \mathbb{P}^{B)}{}_C{}{}^E{}_D \hat \partial_E  A_\mu{}^D \right) .\nonumber \\
&& 
\end{eqnarray}
}With these prescriptions, the following terms of \eqref{eq:Aeom}},
\begin{equation}
 -\tfrac{1}{2}  \partial_M \gM_{KL} \mathcal{D}^\mu \gM^{KL} 
+ g^{-1/2} \partial_P \left( \left[ \delta_K^P \gM_{LM} - Y^{QP}{}_{MK} \gM_{LQ} \right] g^{1/2} \mathcal{D}^\mu \gM^{KL} \right) \; ,
\end{equation} 
become
\begin{equation} \label{eq:intStepVec}
\begin{split}
\Delta^{-2} M_{CD} D^\mu M^{AB} \partial_P \left( - \tfrac12 \partial_M (U^C{}_K U^D{}_L) U^K{}_A U^L{}_B \right)\\[4pt]
+ \Delta^{-D} M_{CD} D^\mu \partial_P \left( \left[ \delta^P_K U^C{}_L U^D{}_M - Y^{PQ}{}_{MK} U^C{}_L U^D{}_Q \right]\Delta^{D-2} U^K{}_A U^L{}_B \right)\\[4pt]
+\left( \delta^P_K M_{LM} - Y^{PQ}{}_{MK} \right) \partial_P \left( D^\mu M^{AB}\right) U^K{}_A U^L{}_B \Delta^{-2} \; .
\end{split}
\end{equation}
Under generalised Scherk-Schwarz (\ref{eq:GenSSBos}), the first two lines in (\ref{eq:intStepVec}) produce the following contribution to the $D$-dimensional $\cN=8$ gauge field mass matrix
\begin{equation} \label{eq:1st2ndlines}
\Delta^{-3} U^A{}_M M_{CD} D^\mu M^{BD} \left( -X_{AB}{}^C + \alpha (D-2) \delta^C_A \vartheta_B \right) \; .
\end{equation}
Using the E$_{7(7)}$ invariance of $Y$, the third line in (\ref{eq:intStepVec}), can be further manipulated into 
\begin{equation} \label{eq:InterStepVec}
 U^A{}_M \left( M_{AB} \delta^E_C - Y^{DE}{}_{AC} M_{BD} \right) \Delta^{-3} \hat \partial_E \left( D^\mu M^{CB} \right) \; .
\end{equation}
Now, using \eqref{YtensorDef} and
\begin{equation}
\mathbb{P}^{B'}{}_{A'}{}^{D'}{}_{C'} M^{A A'} M_{B B'}M^{C C'}M^{D D'}= \mathbb{P}^{A}{}_B{}^C{}_D \; ,
\end{equation}
together with fact that $M_{AC}D^\mu M^{BC}$ is adjoint projected so that $M_{AC}D^\mu M^{AC}=0$, the contribution (\ref{eq:InterStepVec}) reduces to
\begin{equation} \label{eq:3rdline}
\Delta^{-3} U^A{}_M \alpha M_{AB} \hat \partial_E D^\mu M^{EB} \; .
\end{equation}
Combining (\ref{eq:1st2ndlines}), (\ref{eq:3rdline}), and keeping track of appropriate normalisations, the contributions (\ref{eq:intStepVec}) reduce altogether to
{\setlength\arraycolsep{0pt}
\begin{eqnarray} \label{eq:Anotherintvec}
\Delta^{-3} U^A{}_M \bar{g}^{\mu \nu} \bigg[ && - \tfrac2\alpha \left( X_{AB}{}^C - \alpha (D-2) \delta^C_A \vartheta_B \right)M^{BD} \nonumber \\
&& \times \left(  A_\nu{}^E X_{E(C}{}^F M_{D)F} +  A_\nu{}^E \vartheta_E M_{CD} + \alpha M_{E(C} \mathbb{P}^E{}_{D)}{}^F{}_G \hat \partial_F  A_\nu{}^G\right) \\
&& + 2 M^{BC} \hat \partial_C \left( A_\nu{}^D X_{D(A}{}^E M_{B)E} +  A_\nu{}^D \vartheta_D M_{AB} + \alpha M_{D(A} \mathbb{P}^D{}_{B)}{}^E{}_F \hat \partial_E  A_\nu{}^F \right) \; . \nonumber
\end{eqnarray}
}Finally, expanding the ExFT gauge field $\cA_\mu$ in harmonics $\cY^\Lambda(y)$ subject to (\ref{eq:CurlyTDef}), as in (\ref{eq:KKExpansionBos}), the vector mass matrix (\ref{eq:KKvector}) follows from (\ref{eq:Anotherintvec}).

%%%%%%%%%%%%%%%
\subsubsection{Scalar mass matrix} \label{sec:ScalarMassMat}
%%%%%%%%%%%%%%%

For the KK scalar mass matrix, we simply keep track of the trombone contributions coming from the ExFT scalar equation of motion (\ref{eq:ExFTScalar}). These are of the schematic form $\vartheta \Theta$ and $\vartheta \cT$. In principle, there could also be terms $\vartheta \vartheta$, but these would necessarily contribute to the scalar mass matrix of $D$-dimensional $\cN=8$ supergravity, and the latter is known to lack such terms \cite{LeDiffon:2011wt,Varela:2025xeb}. The terms in $\vartheta \Theta$ also contribute to the $D$-dimesional mass matrix, and are of the form
\begin{equation} \label{eq:contscal}
-\tfrac{(D-2)}{\alpha} \, \vartheta_C M^{CD} M_{E(A|} \Theta_{D|B)}{}^E\, .
\end{equation}
Upon linearisation, projection to adjoint indices using the projector defined in (\ref{eq:ProjCoset}) and normalisation by the kinetic terms (which amounts to multiplication by $-2 \alpha$), (\ref{eq:contscal}) contributes as
\begin{equation}
\begin{split}
& 2 (D-2) \left(P_{\textrm{coset}}\right)^{\alpha AB}  \Big( \vartheta_C  \Theta_{DA}{}^{B'} M_{B'B} - \vartheta_E  \Theta_{FA}{}^{C'} M_{BD} M^{EF} M_{C'C} \Big) \left(P_{\textrm{coset}} \right)_\beta{}^{CD} \, .
\end{split}
\end{equation}

The $\vartheta \cT$ terms come from $\partial_M \ln \Delta$ and $\partial_M \Delta^{-1}= - \Delta^{-1} \partial_M \ln \Delta $. Various pieces of the scalar equation of motion (\ref{eq:ExFTScalar}) provide such terms, including:
\begin{equation} \label{eq:contScalKK1}
\begin{split}
-\tfrac{1}{4 \alpha} \partial_K\left( \gM^{KL} \partial_L \gM_{MN} \right) &= -\tfrac{1}{4 \alpha} \partial_K\left(U^K{}_C M^{CD} \Delta^{-1} U^E{}_M \hat \partial_D \phi_{EF} U^F{}_N + \ldots \right)\\[4pt]
&= \tfrac{1}{4 \alpha} U^E{}_M U^F{}_N   M^{CD} U^K{}_C \partial_K\Delta^{-1}  M_{EA} M_{FB} \hat \partial_D \phi^{AB}  + \ldots \\[4pt]
&=- \tfrac{1}{4 \alpha} \tfrac{D-2}{D-1} \Delta^{-2}U^E{}_M U^F{}_N  \vartheta_C M^{CD}   M_{EA} M_{FB} \hat \partial_D \phi^{AB}  + \ldots \,,\\[4pt]
&=- \tfrac{1}{4 \alpha} \tfrac{D-2}{D-1} \Delta^{-2}U^A{}_M U^B{}_N  \vartheta_C M^{CD}   M_{AE} M_{BF} \hat \partial_D \phi^{EF}  + \ldots 
\end{split}
\end{equation}
Here and below, the ellipses stand for terms that do not involve $\vartheta_A$. Similarly,
{\setlength\arraycolsep{0pt}
\begin{eqnarray} \label{eq:contScalKK2}
&& \tfrac12 \partial_K \left( \cM^{KL} \partial_M \cM_{LN} \right) =  \tfrac12 \tfrac{D-2}{D-1} \Delta^{-2}U^A{}_M U^B{}_N  \vartheta_E M_{BF} \hat \partial_A \phi^{EF}+\ldots \nonumber \\[5pt]
&& \tfrac{1}{4 \alpha}  \cM_{KM} \cM_{LN} \partial_P \left(\cM^{PQ} \partial_Q \cM^{KL} \right) \hspace{-4pt} = \hspace{-3pt}  - \tfrac{1}{4 \alpha} \tfrac{D-2}{D-1} \Delta^{-2}U^A{}_M U^B{}_N  \vartheta_C M_{AE} M_{BF} M^{CD} \hat \partial_D \phi^{EF} \hspace{-3pt}+\ldots \nonumber \\[5pt]
&& -\tfrac{1}{2}  \cM_{KM} \cM_{LN} \partial_P \left(\cM^{KQ} \partial_Q \cM^{LP} \right) =   \tfrac{1}{2}  \tfrac{D-2}{D-1}  \Delta^{-2}U^A{}_M U^B{}_N  \vartheta_F M_{AC} M_{BE} M^{CD} \hat \partial_D \phi^{EF}+\ldots \nonumber \\[5pt]
&& -\tfrac{1}{4 \alpha}  \partial_P \ln g \cM^{PQ} \partial_Q \cM_{MN} = \tfrac{1}{4 \alpha} 2D \tfrac{D-2}{D-1} \vartheta_C M^{CD} M_{AE} M_{BF} \hat \partial_D \phi^{EF}+\ldots \nonumber \\[5pt]
&& \tfrac{1}{2}  \partial_P \ln g \cM^{PQ} \partial_M \cM_{QN} = -D \tfrac{D-2}{D-1} \vartheta_E  M_{BF} \hat \partial_A \phi^{EF}+\ldots % \nonumber \\[5pt]
\end{eqnarray}
}Adding (\ref{eq:contScalKK1}), (\ref{eq:contScalKK2}), factorising out  $\Delta^{-2}U^E{}_M U^F{}_N$, expanding as in (\ref{eq:KKExpansionBos}) and normalising by the kinetic term prefactor $-2\alpha$, we obtain a contribution
\begin{equation}
2\alpha (D-2) R^{-1} \vartheta_E M_{BF} (\cT_A)^\Sigma{}_\Lambda \phi^{EF, \Lambda} \cY_\Sigma - (D-2) R^{-1} M_{AE} M_{BF} \vartheta_C M^{CD} (\cT_D)^\Sigma{}_\Lambda \phi^{EF, \Lambda} \cY_\Sigma\,.
\end{equation}
Finally, using (\ref{eq:CurlyTDef}) and appropriately projecting to the coset, we get the following contribution to the KK scalar mass matrix:
\begin{equation} \label{eq:thetaTCont}
2\alpha(D-2) \left(P_{\textrm{coset}}\right)^{\alpha AB}M_{BD}    R^{-1} (\mathcal{T}_A)^\Sigma{}_\Lambda \vartheta_C \left(P_{\textrm{coset}}\right)_\beta{}^{CD}  - (D-2) \delta^\alpha_\beta   R^{-1}  \vartheta_A M^{AB} (\fT_B)^\Sigma{}_\Lambda \; .
\end{equation}
This, together with the $\Theta \Theta$ and $\Theta \fT$ contributions determined in \cite{Malek:2020yue},  gives rise to the KK scalar mass matrix (\ref{eq:KKscalar}).

%%%%%%%%%%%%%%%
%%%%%%%%%%%%%%%

\section{Massive graviton operator} \label{sec:GravOp}

%%%%%%%%%%%%%%%
%%%%%%%%%%%%%%%

It is helpful to have a careful look at the massive graviton spectrum following \cite{Bachas:2011xa}. Particularising (\ref{eq:BachasEstes}) to $D=4$ we obtain $\cL_7 h = M^2 h$, with  $\cL_7$ defined in (\ref{eq:MassGravOp}). We find it convenient to rewrite this eigenvalue equation using the ``round'' seven-dimensional metric
\begin{equation} \label{eq:BachasEstesRoundMetric}
\gr_7 = R^2 \left( 2 \gr_{H^3} + \gr_{S^4} \right)\, ,
\end{equation}
where $\gr_{H^3}$ is the Einstein metric on $H^3$,
\begin{equation} \label{eq:H3MetricUHP}
\gr_{H^3}= \tfrac{dx^2 + dy^2+dz^2}{z^2} =  \delta_{ab}  e^a e^b \; , \quad
\textrm{with} \; 
e^1 = \tfrac{dx}{z} \,,\quad 
e^2 = \tfrac{dy}{z} \,, \quad 
e^3 = -\tfrac{dz}{z}  \, ,
\end{equation}
and $ \gr_{S^4}$ the round metric on $S^4$. We choose angles adapted to either solution, SLAG or A3C, so that
{\setlength\arraycolsep{0pt}
\begin{eqnarray} \label{eq:SLAGRoundMetS4}
&& \textrm{SLAG} :  \gr_{S^4} =   d\theta^2 + \sin^2 \theta \; d\psi^2  + \cos^2 \theta \; g_{S^2} \, , \quad  \textrm{with} \; 
g_{S^2}= d\chi^2 + \sin^2 \chi \,  d\phi^2 \; , \\[5pt]
&& \label{eq:A3CRoundMetS4}  \textrm{A3C}  :  \gr_{S^4} =   \gr_{S^4} =  d\theta^2 + \sin^2 \theta \, g_{S^3} \, , \;  \textrm{with} \; 
g_{S^3} =  \tfrac{1}{4}  \big(  d\psi^2 +d\chi^2+ d\phi^2 + 2  \cos \psi \, d\psi d\phi \big) . \qquad 
\end{eqnarray}
}With these definitions, and using the fact that $e^{2 \tilde \Delta}=\sqrt{\frac{\gr_7}{g}}$, where $\gr_7$ is defined in (\ref{eq:BachasEstesRoundMetric}) and $g$ is the metric that appears in (\ref{eq:MassGravOp}), the latter equation can be rewritten as
\begin{equation}\label{eq:BachasEstesSimp}
-\frac{e^{-2 \tilde \Delta}}{\sqrt{\gr_7}} \partial_m ( e^{2 \tilde \Delta} \sqrt{\gr_7} g^{mn}_7  \partial_n h(y))=M^2 h(y)\,.
\end{equation}

%%%%%%%%%%%%
\subsection{SLAG}
%%%%%%%%%%%%

The inverse metric on the $\Sigma_3 \rtimes S^4$ SLAG solution of \cite{Gauntlett:2006ux} is, in our conventions,
\begin{equation} \label{eq:SLAGMetric}
g_7^{-1}= R^{-2} \bigg( \frac{e^{-2 \tilde \Delta}}{\sqrt{2}} \delta^{ab} \tilde e_a \otimes \tilde e_b + \sqrt{2} e^{-2 \tilde \Delta} \partial_\theta \otimes \partial_\theta +  \frac{e^{4 \tilde \Delta}}{\sin^2 \theta} \partial_\psi \otimes \partial_\psi + \frac{e^{4 \tilde \Delta}}{2 \cos^2 \theta} g_{S^2}^{-1} \bigg)\,,
\end{equation}
with warp factor
\begin{equation}
e^{2 \tilde \Delta}= 2^{\frac16} (1+\sin ^2(\theta ))^{\frac13}\,,
\end{equation}
and twisted vectors 
\begin{equation} \label{eq:TwistVecH3}
\tilde e_a =  \hat e_a + v_a \; .
\end{equation}
Here, $\hat e_a$ are the vectors dual to the vielbein one-forms in (\ref{eq:H3MetricUHP}), and 
\begin{equation} \label{eq:SLAGS2KVs}
v_1 =   \sin \phi \; \partial_\chi + \cos\phi \cot \chi \; \partial_\phi \,, \quad v_2 =  \partial_\phi \,, \quad v_3= 0 \, .
\end{equation}
are two of the three Killing vectors of the $S^2$ inside $S^4$ that features in (\ref{eq:SLAGRoundMetS4}). With these definitions, some calculation shows that, on eigenfunctions of the form (\ref{eq:FullEF}), equation \eqref{eq:BachasEstesSimp} can be brought to the form
\begin{equation}\label{eq:BachasEstesNotFinalYetSLAG}
\cL_3 \, h(y) +  \cL_4 \, h(y) = M^2 L^2 h(y)\,,
\end{equation}
with $L^2= \sqrt{2} \, R^2 $,
\begin{equation}\label{eq:L3SLAGdef}
\cL_3 \, h(y) \equiv - \frac{1}{\sqrt{\gr_7}} \partial_m \left( \sqrt{\gr_7} \tilde e^{ma} \tilde e^n{}_a  \partial_n h(y)\right)  \,,
\end{equation}
and $\cL_4$ a modified Laplacian on $S^4$,
\begin{equation}
- \tfrac12 \, \cL_4 \, h(y) =  \frac{\partial_\theta \left(  \cos^2 \theta \sin \theta \; \partial_\theta h(y)\right)}{\cos^2 \theta \sin \theta}  + (2 + \cot^2 \theta)  \partial_\psi^2 h(y) +  (\tfrac12 + \tan^2 \theta)\square_{S^2} h(y)\, .
\end{equation}

The operator in (\ref{eq:L3SLAGdef}) can be simplified as follows. Using the fact that $v_a$ are killing vectors of $S^2$, we have
\begin{equation}
0 = \cL_{v_a} \vol_{S^2} = (\text{div } v_a) \vol_{S^2}\,,
\end{equation}
so that
\begin{equation}
(\text{div } v_a) = \frac{1}{\sqrt{\gr_{S^2}}} \partial_m \left( \sqrt{\gr_{S^2}} v^{m}{}_a \right)= \frac{1}{\sqrt{\gr_7}} \partial_m \left( \sqrt{\gr_7} v^{m}{}_a \right)= 0\,,
\end{equation}
and 
\begin{equation}
(\text{div } \hat e_a ) \vol_3 = \cL_{\hat e_a} \vol_3 = -f_{ab}{}^b \vol_3 \, ,
\end{equation}
with $f_{ab}{}^c$ the (traceful) Bianchi type V structure constants (\ref{eq:SCBianchiV}). This implies
\begin{equation}
\text{div } \hat e_a  = \frac{1}{\sqrt{\gr_{H^3}}} \partial_m \left( \sqrt{\gr_{H^3}} \hat e^{m}{}_a \right) = \frac{1}{\sqrt{\gr_7}} \partial_m \left( \sqrt{\gr_7} \hat e^{m}{}_a \right) = (0,0,2)\,.
\end{equation}
A straightforward computation now brings (\ref{eq:L3SLAGdef}) into the form
\begin{equation} \label{eq:BachasEstesH3Operator}
\cL_3 \, h(y) = - \square_{H^3} h(y)
- 2    \hat e^n{}_a \partial_n \left( v^{ma}\partial_m  h(y) \right) -  v^{ma} \partial_m \left(   v^n{}_a  \partial_n h(y)\right)\,. \\
\end{equation}
Finally, the differential action of the $S^2$ Killing vectors (\ref{eq:SLAGS2KVs}) on the eigenfunction can be traded by the algebraic action of the SO(3) generators $S_{a(j)}$ in the spin $j$ representation
\begin{equation} \label{eq:KillingsActionOnHarmonics}
 v^m{}_a  \partial_m h(y) = S_{a(j)} h(y) \,,
\end{equation}
by virtue of the splitting (\ref{eq:FullEF}) and the identity $ v^m{}_a  \partial_m y^{i_1 \ldots i_k} = S_{a(j)} y^{i_1 \ldots i_k}$, with $j$ taking all possible values that arise in the branching of the $[k0] \times \bm{1}$ representation of $\textrm{SO}(5) \times \textrm{SO}(3)_{\Sigma_3}$ in (\ref{eq:GroupTh}) under SO$(3)_S$. Consequently, (\ref{eq:BachasEstesH3Operator}) takes on the form (\ref{eq:Bochner}) for the Bochner Laplacian on $H^3$, summed over all possible values of $j$, as in (\ref{eq:MassGravOpSplit}).

%%%%%%%%%%%%%%
\subsection{A3C}
%%%%%%%%%%%%%%

The inverse A3C metric \cite{Acharya:2000mu} on $\Sigma_3 \rtimes S^4$ is, in our conventions, 
\begin{equation} \label{eq:A3CMet}
g_7^{-1}= L^{-2} \bigg( \frac{5}{4} e^{-2 \tilde \Delta} \delta^{ab} \tilde e_a \otimes \tilde e_b + \frac52  e^{-2 \tilde \Delta} \partial_\theta \otimes \partial_\theta + \frac{625 \sqrt{\frac52}}{2048} \frac{ e^{4 \tilde \Delta}}{ \sin^2 \theta} g_{S^3}^{-1} \bigg)\,,
\end{equation}
where now the warp factor is
\begin{equation}
e^{2 \tilde \Delta}=2^{\frac{5}{2}} \cdot 5^{-\frac{7}{6}} \left(5 \sin ^2(\theta )+8 \cos ^2(\theta )\right)^{\frac13}\,,
\end{equation}
and the twisted vectors $\tilde e_a$ are still given by (\ref{eq:TwistVecH3}) with, now, 
\begin{equation} \label{eq:BachasEstesVectorsFibratingH3}
v_1= \cos \phi \; \partial_\psi + \frac{\sin \phi}{\sin \psi} \partial_\chi - \sin\phi \cot \psi \; \partial_\phi \,, \quad v_2= - \partial_\phi  \,, \quad v_3 = 0 \,.
\end{equation}
These are two of the Killing vectors of the (left) action of $\SO{3}_-$ on the $S^3$ defined inside $S^4$ in (\ref{eq:A3CRoundMetS4}). 

With these definitions, (\ref{eq:BachasEstesSimp}) can be massaged into the form
\begin{equation}\label{eq:BachasEstesNotFinalYetA3C}
\cL_3 \, h(y) + \cL_4 \, h(y) = \tfrac{4}{5} M^2 L^2 h(y)\,,
\end{equation}
with $\cL_3$ formally as in (\ref{eq:L3SLAGdef}), where now $\tilde e_a$ are given by (\ref{eq:TwistVecH3}) with (\ref{eq:BachasEstesVectorsFibratingH3}). In (\ref{eq:BachasEstesNotFinalYetA3C}),  $\cL_4$ is now the following modified Laplacian on $S^4$,
\begin{equation}
- \tfrac12 \, \cL_4 \,  h(y) = \frac{1}{ \sin^3 \theta} \partial_\theta \left(  \sin^3 \theta \; \partial_\theta h(y)\right) + \left(  \frac{1}{\sin^2 \theta} - \frac38\right) \square_{S^3} h(y)\,,
\end{equation}
with $\square_{S^3}$ the standard Laplacian on the round $S^3$. Finally, trading again the differential action of the Killing vectors (\ref{eq:BachasEstesVectorsFibratingH3}) with the algebraic action of the SO$(3)_-$ generators via \eqref{eq:KillingsActionOnHarmonics}, $\cL_3$ in (\ref{eq:BachasEstesNotFinalYetA3C}) again takes on the Bochner form (\ref{eq:Bochner}) summed over all allowed $j$ as in (\ref{eq:MassGravOpSplit}).

%%%%%%%%%%%%%%%
%%%%%%%%%%%%%%%

\section{$D=4$ $\cN=8$ truncation on wrapped M5-brane geometries} \label{sec:MaxTrunc}

%%%%%%%%%%%%%%%
%%%%%%%%%%%%%%%

Locally, $D=11$ supergravity admits a maximally supersymmetric consistent truncation \cite{Pico:2026rji} on the internal geometries $\Sigma_3 \rtimes S^4$ corresponding to the two distinct, SLAG and A3C, wrapped M5-brane configurations of \cite{Gauntlett:2006ux,Acharya:2000mu}. The resulting $D=4$ $\cN=8$ supergravity is, in both cases, the TCSO$(5,0,0;\mathrm{V})$ gauged theory introduced in \cite{Pico:2025cmc}. Here, we will review some aspects of this truncation and the $D=4$ theory that have been used in the main text, referring to \cite{Pico:2025cmc,Pico:2026rji} for full details. The construction regards $\Sigma_3 = H^3/\Gamma$ as locally interchangeable with the group manifold $G_3$ of Bianchi type V.

For both SLAG and A3C twisted geometries $\Sigma_3 \rtimes S^4$, a convenient generalised frame is given by
\begin{equation} \label{eq:TwistedFrame}
\hat{\tilde{E}}^M{}_{\uP} (y)= U_N{}^M (y) \, \hat{E}^N{}_{\uP} (y) \; .
\end{equation}
Here, $\hat{E}^N{}_{\uP} (y)$ is a frame on the direct product manifold $\Sigma_3 \times S^4$. In order to define it in terms of geometric objects on the latter, it is useful to split curved, $M$, and flat, $\uM$, fundamental E$_{7(7)}$ indices under GL$(7)$ and $\textrm{GL}(3) \times \textrm{SL}(5)$, respectively,
\begin{equation} \label{eq:56Branching}
\bm{56} \rightarrow \bm{7}^\prime + \bm{21} + \bm{21}^\prime + \bm{7} \; , \qquad 
\bm{56} \rightarrow (\bm{3} , \bm{1}) + (\bm{3} , \bm{5}) + (\bm{1} , \bm{10}) + (\bm{3}^\prime , \bm{1}) + (\bm{3}^\prime , \bm{5}^\prime)  + (\bm{1} , \bm{10}^\prime) \; ,
\end{equation}
so that
\begin{equation} \label{eq:GenFrameSplit}
\hat{E}^M{}_{\uN}  = \big( \hat{E}^m{}_{\uN}  , \hat{E}_{mn \, \uN}  ,  \hat{E}_{mnpqr \, \uN}  ,  \hat{E}_{m; n_1 \dots n_7 \, \uN}  \big) \; ,
\end{equation}
with $m=1, \ldots , 7$, and
\begin{equation} \label{eq:GenFrameSplit2}
\hat{E}^M{}_{\uN}  = \big(\epsilon_{abc}\hat E^{M \, c} \; , \;  - \tfrac12 \epsilon_{abc} \hat E^{M \, bc}{}_i \; , \;  \hat E^M{}_{ij} \; , \;  \epsilon^{abc}\hat E'^M{}_c \; , \;  - \hat E'^{M \,ia} \; , \;  \hat E'^{M \,ij}   \big) \; ,
\end{equation}
with $a=1,2,3$ and $i = 4, \ldots 8$. With these definitions, the various blocks in (\ref{eq:GenFrameSplit2}) are taken to be
{\setlength\arraycolsep{2pt}
\begin{eqnarray} \label{eq:GenFrameG3S4}
\hat E^a &=& \big( 0 \; , \; 0 \; , \; 0 \; , \; R_3\,  e^a \otimes (\text{vol}_{4} \wedge \text{vol}_{3}) \big) \; , \nonumber \\[5pt]
\hat E^{ab}{}_i \;  & =&   \big( 0 \; , \;  R_3^2 \, y_i \, e^a \wedge e^b \; , \; R_3^2 \, e^a \wedge e^b \wedge \left( R *_4 \, d y_i +y_i \, A \right) \; , \; 0 \big) \; , \nonumber \\[5pt]
\hat E_{ij} \;  &=& \big( v_{ij} \; , \;  R^2 *_4 ( d y_i \wedge d y_j) + \imath_{v_{ij}} A \; , \;  0 \; , \;  0 \big) \; , \nonumber \\[5pt]
 \hat E'_a  \; & =& \big( R_3^{-1}\hat e_a \; , \;  0 \; , \;  0 \; , \;  0 \big) \; ,  \\[5pt]
\hat E'^{ia} \;  &=& \big( 0 \; , \; R \, R_3 \, e^a \wedge d y^i  \; , \; R_3\, e^a \wedge \left( R \, d y^i\wedge A - y^i \, \text{vol}_{4} \right) \; , \; 0  \big) \; , \nonumber \\[5pt]
\hat E'^{ij}&=&  \big( 0 \; , \; 0 \; , \; R^2 (d y^i \wedge d y^j)\wedge \text{vol}_{3} \; , \;  R^2 \, jA \wedge  ( d y^i \wedge d y^j)\wedge \text{vol}_{3} -  v^{ij\,\flat} \otimes (\text{vol}_{4} \wedge \text{vol}_{3}) \big) \; , \nonumber
\end{eqnarray}
}with the curved index $M$ suppressed on the l.h.s.'s and split on the r.h.s.'s as in (\ref{eq:GenFrameSplit}). Here, $e^a$ denotes a left-, say, $G_3$-invariant vielbein on the $G_3$ group manifold, with inverse $\hat{e}_a$, and $\vol_3$ is the volume form on $G_3$. The coordinates $y^i$ are constrained coordinates on $\mathbb{R}^5$ that define $S^4$ as the locus $\delta_{ij} y^i y^j =1$. The quantities $v^{ij}$ are the $\text{SO}(5)$ Killing vectors on $S^4$, and $A$ is a local three-form potential for the volume form on $S^4$, satisfying $dA = 3 R^{-1} \, \vol_4$. The Hodge dual $*_4$ is taken w.r.t.~the round metric on $S^4$, $\imath_{v}$ denotes the interior product w.r.t.~a conventional vector $v$, and $v^{\flat}$ denotes its dual one-form. The constants $R_3$ and $R$ set scales on $G_3$ and $S^4$. In particular, $R$ is the radius of $S^4$. See \cite{Pacheco:2008ps, Coimbra:2011ky} for the definition of the operator $j$.

The matrix $U_N{}^M (y)$ in (\ref{eq:TwistedFrame}) implements the two distinct fibrations of $S^4$ over $\Sigma_3$ corresponding to the two different solutions, SLAG and A3C. This can be written as 
\begin{equation} \label{eq:E7GroupElement}
U_{M}{}^{N} (y) =e^{\Upsilon_{M}{}^{N}(y)} \; ,
\end{equation}
with the $\mathbb{R}^+ \times \textrm{E}_{7(7)}$ Lie algebra element $\Upsilon(y)$ given in either case by 
\begin{equation} \label{eq:ExGGTwist}
\textrm{SLAG:} \; 
\Upsilon_M{}^N= \tfrac{R}{R_3} \, \delta_a^{\bar a} \delta_b^{\bar b}\, \big( \omega^{ab} \times_\text{ad}  \hat{E}_{\bar a \bar b} \big)_M{}^N  , \quad 
\textrm{A3C:} \;
\Upsilon_M{}^N=-\tfrac14 \tfrac{R}{R_3} \, \epsilon_{abc} (J_-^c)^{ij} \big( \omega^{ab} \times_\text{ad}  \hat{E}_{ij} \big)_M{}^N . 
\end{equation}
Here, $\omega^{ab}$ is the spin connection on $\Sigma_3$. In the A3C case, $(J^c_-)^{ij}$ are the generators of the $\textrm{SO}(3)_-$ group defined in (\ref{eq:SO5subgroups}) in the $(\bm{2},\bm{2}) + (\bm{1},\bm{1})$ representation, and $\hat E_{ij}$ is one of the blocks of $\hat{E}^N{}_{\uP} (y)$ defined in (\ref{eq:GenFrameG3S4}). In the SLAG case, $\hat{E}_{\bar a \bar b}$ are the further components of $\hat E_{ij}$ defined by splitting indices as $i=(\alpha , \bar{a})$, with $\alpha =4,5$, $\bar{a} = 6,7,8$. Finally, the symbol $\times_{\textrm{ad}}$ has been defined in (3.8) of \cite{Pico:2026rji} following \cite{Coimbra:2011ky}. Altogether, (\ref{eq:ExGGTwist}) implements the twist of $S^4$ over $\Sigma_3$ by identifying the natural SO$(3)_{\Sigma_3}$-structure on $\Sigma_3$ carried by $\omega^{ab}$ with either subgroup, SO$(3)^\prime$ or SO$(3)_-$, of the SO(5) that rotates $S^4$ defined in either case by (\ref{eq:SO5subgroups}), as in(\ref{eq:GroupTh}).

The frames $\hat{\tilde{E}}^M{}_{\uP} (y)$ in (\ref{eq:TwistedFrame}) with (\ref{eq:GenFrameSplit})--(\ref{eq:ExGGTwist}) have constant intrinsic torsion: they satisfy (\ref{eq:CTCon}) with $\tilde{X}_{\uM\uN}{}^{\uP}$ given by the $D=4$ $\cN=8$ TCSO$(5,0,0;\mathrm{V})$ embedding tensor (hence the consistency of the truncation) in suitable duality frames. More concretely, $\hat{\tilde{E}}^M{}_{\uP} (y)$ obey (\ref{eq:CTCon}) with $\tilde{X}_{\uM\uN}{}^{\uP}$ defined as
\begin{equation} \label{eq:TransXSymbol}
\tilde{X}_{\uM \uN}{}^{\uP} = U_{\uM}{}^{\uQ} \, U_{\uN}{}^{\uR} \, X_{\uQ \uR}{}^{\uS}  \, (U^{-1})_{\uS}{}^{\uP} \; .
\end{equation}
Here, $U_{\uM}{}^{\uN}$ are flattened versions,
\begin{equation} \label{eq:FlatteningCon}
U_{\uM}{}^{\uN} = U_{M}{}^{N}(y) \, \hat{E}^M{}_{\uM} (y) \, E_N{}^{\uN} (y) \; , 
\end{equation}
of the $U_{M}{}^{N}$ transformations defined in (\ref{eq:E7GroupElement}), (\ref{eq:ExGGTwist}). Explicitly, 
\begin{equation} \label{eq:E7GroupElementFlat}
U_{\uM}{}^{\uN} =e^{\Upsilon_{\uM}{}^{\uN}} \; , 
\end{equation}
with
\begin{equation} \label{eq:TransMat}
\textrm{SLAG:}  \;\;
\Upsilon \equiv g_2 g_1^{-1} \big( t_{2368} - t_{1378} \big) \; , \quad
\textrm{A3C:}  \;\;
\Upsilon = \tfrac12   \, g_2 g_1^{-1} \big( t_{1738} + t_{6238}  + t_{1356} + t_{2357}   \big)\;  ,
\end{equation}
where $g_1\equiv R^{-1}$, $g_2\equiv R_3^{-1}$ and we follow the E$_{7(7)}$ generator conventions of \cite{Guarino:2015qaa,Josse:2025uro}. Finally, $X_{\uQ \uR}{}^{\uS}$ in (\ref{eq:TransXSymbol}) is given, suppressing representation indices, by
\begin{equation} \label{eq:ETtromboneSplit}
X_{\uM} = (X_{AB} , X^{AB})= (X_{ab} , X_{ai} , X_{ij} , X^{ab} , X^{ai} , X^{ij}) \; ,
\end{equation}
where $A=1 , \ldots , 8$ in the intermediate step labels the fundamental representation of SL$(8,\mathbb{R})$, and split in the second step as $A= (a,i)$, with $a$, $i$, ranging as below (\ref{eq:GenFrameSplit2}). The various blocks in (\ref{eq:ETtromboneSplit}) are defined as
{\setlength\arraycolsep{0pt}
\begin{eqnarray} \label{eq:ETtrombone}
&& X_{ij} = -2 g_1 \, \delta_{k [i} \, t_{j]}{}^k \; , \qquad 
X_{ai} = g_1 \, t_a{}^j \delta_{ij} -\tfrac{1}{18} g_2 \, f_{ab}{}^{b} \epsilon^{cde} t_{icde}  \; , \qquad 
X_{ab} = X^{ij} = 0  , \quad \\[4pt]
&& 
X^{ai} =  -\frac{1}{2} g_2 \,  \epsilon^{bcd} f_{bc}{}^a \, t_d{}^i    \; , \quad
X^{ab} =  \tfrac12 g_2 \epsilon^{abc} f_{cd}{}^{d} (t_0 - \tfrac43 t_e{}^e)  + g_2 \epsilon^{abc} f_{cd}{}^e ( t_e{}^d - \tfrac13 \delta^d_e t_f{}^f) \; . \nonumber 
\end{eqnarray}
}with $t_0$, $t_\alpha = (t_A{}^B ,  t_{ABCD} ) =  (t_a{}^b , t_a{}^j , t_i{}^b , t_i{}^j , t_{abc\ell}, t_{abk\ell}, t_{ajk\ell} , t_{ijk\ell} )$ the $\mathbb{R}^+ \times \textrm{E}_{7(7)}$ generators and $f_{ab}{}^c$ the Bianchi type V structure constants. The latter have non-vanishing entries
\begin{equation} \label{eq:SCBianchiV}
f_{13}{}^1 = -f_{31}{}^1 = f_{23}{}^{2}= -f_{32}{}^{2} = 1  \; .
\end{equation}
%

%%%%%%%%%%%%%%%
%%%%%%%%%%%%%%%

\section{Putative spectra} \label{sec:PutSpec}

%%%%%%%%%%%%%%%
%%%%%%%%%%%%%%%

This appendix describes the putative KK spectra for the $\textrm{AdS}_4 \times (\Sigma_3 \rtimes S^4)$ solutions of \cite{Gauntlett:2006ux,Acharya:2000mu}, as defined in section~\ref{sec:PutUniv}. 

%%%%%%%%%%%%%%%

\subsection{Putative SLAG spectrum} \label{sec:N=8SpecSLAG}

%%%%%%%%%%%%%%%

We have computed the putative KK masses for the $\cN=2$ $\textrm{AdS}_4 \times (\Sigma_3 \rtimes S^4)$ SLAG solution of \cite{Gauntlett:2006ux}, for the first few KK levels. We have done this by diagonalisation of the KK mass matrices (\ref{eq:KKgraviton})--(\ref{eq:KKscalar}), following the process described in general in section \ref{sec:PutUniv}, and for this solution in particular at the beginning of section~\ref{sec:ClassRSpec}. We have also computed the R-charges of these individual states as follow from the branchings (\ref{eq:SO3SBreaking2}), (\ref{eq:GroupTh}) for a few low $k$. These results suffice to find a pattern for all $k \geq 0$. All KK states of definite mass also have definite $\textrm{U}(1)_R \times \textrm{U}(1)_S$ charges, $(y_0, f)$. Here, $\textrm{U}(1)_R \subset \textrm{OSp}(4|2)$ is the R-symmetry of the $\cN=2$ SLAG solution, realised geometrically by $\partial_\psi$ in (\ref{eq:SLAGMetric}). This $\textrm{U}(1)_R$ is also identified with the SO(2) factor that appears in the SLAG entries of (\ref{eq:SO5subgroups}) and (\ref{eq:GroupTh}). The $\textrm{U}(1)_S$ factor is the Cartan subgroup of the generalised SO$(3)_S$ structure group of the SLAG solution defined in (\ref{eq:SO3SBreaking2}), (\ref{eq:GroupTh}). The definite $\textrm{U}(1)_S$ charges, already observed at $k=0$ in \cite{Pico:2025cmc}, come as a slight surprise, as this $\textrm{U}(1)_S$ is not a symmetry of the SLAG solution. In any case, these $\textrm{U}(1)_S$ charges are not conserved.

{
\renewcommand{\arraystretch}{1.4} % Adjust row spacing globally

\begin{table}%[H]
%\centering
\resizebox{\textwidth}{!}{
\begin{tabular}{l | l } 
\hline
\hline

\textbf{Multiplet} & $k=2$     \\
\hline

SGRAV 
&  $ [6,\pm4]_0 $      \\
\hline

LGRAV 
&  $\begin{array}{l}
[\tfrac12 (1 + \sqrt{97}),\pm2]_0 \oplus [\tfrac12 (1 + \sqrt{93}),\pm2]_{\pm2} \oplus \left( 2 \times [\tfrac12 (1 + \sqrt{89}),0]_0 \right) \\
\oplus [\tfrac12 (1 + \sqrt{85}),0]_{\pm2} \oplus [\tfrac12 (1 + \sqrt{73}),0]_{\pm4}
\end{array} $      \\
\hline

SGINO 
& $[\tfrac{13}{2},\pm5]_0$
        \\
\hline
LGINO 
&  $\begin{array}{l}
[\tfrac12 + \sqrt{35},\pm5]_{\pm2} 
\oplus \left( 2 \times [\tfrac12 + 2\sqrt{7},\pm3]_0 \right) 
\oplus \left( 2 \times [\tfrac12 + 3\sqrt{3},\pm3]_{\pm2} \right) 
\oplus [\tfrac12 + 2\sqrt{6},\pm3]_{\pm4} 

 \\
 
\oplus \left( 4 \times [\tfrac12 + 2\sqrt{6},\pm1]_0 \right)
\oplus \left( 4 \times [\tfrac12 + \sqrt{23},\pm1]_{\pm2} \right) 
\oplus \left( 2 \times  [\tfrac12 + 2\sqrt{5},\pm1]_{\pm4} \right) 
\oplus [\tfrac12 + \sqrt{15},\pm1]_{\pm6} 
\end{array} $
        \\
\hline

SVEC 
&  $[7,\pm6]_0$
\\
\hline

LVEC 
&  $\begin{array}{l}
\left( 2 \times [\tfrac12 (1 + \sqrt{129}),\pm4]_0 \right) 
\oplus [\tfrac12 (1 + 5\sqrt{5}),\pm4]_{\pm2} 
\oplus [\tfrac12 (1 + \sqrt{113}),\pm4]_{\pm4}  

\\

\oplus \left( 4 \times [\tfrac12(1 +  \sqrt{105}),\pm2]_0 \right) 
\oplus \left( 3 \times [\tfrac12 (1 + \sqrt{101}),\pm2]_{\pm2} \right) 
\oplus \left( 2 \times [\tfrac12 (1 + \sqrt{89}),\pm2]_{\pm4} \right) 
\\
\oplus [\tfrac12 (1 + \sqrt{69}),\pm2]_{\pm6} 

\oplus \left( 6 \times [\tfrac12 (1 + \sqrt{97}),0]_0 \right) 
\oplus \left( 4 \times [\tfrac12 (1 + \sqrt{93}),0]_{\pm2} \right) 
\\
\oplus \left( 3 \times [5,0]_{\pm4} \right)
\oplus [\tfrac12 (1 + \sqrt{61}),0]_{\pm6} 
\oplus [\tfrac12 (1 + \sqrt{33}),0]_{\pm8}
\end{array} $
\\
\hline
HYP 
& $ [8,\pm8]_0 $  
 \\
\hline
\hline

\textbf{Multiplet} & $k=1$     \\
\hline

SGRAV 
&  $ [4,\pm2]_0 $      \\
\hline

LGRAV 
&  $ [\tfrac12 (1 + \sqrt{41}),0]_0 \oplus  [\tfrac12 (1 + \sqrt{37}),0]_{\pm 2} $      \\
\hline

SGINO 
& $[\tfrac92,\pm 3]_0$  
        \\
\hline
LGINO 
& $[\tfrac12 + \sqrt{15},\pm3]_{\pm2} \oplus \left( 2 \times [\tfrac12 + 2\sqrt{3},\pm1]_0 \right) \oplus \left( 2 \times [\tfrac12 + \sqrt{11},\pm1]_{\pm2} \right) \oplus [\tfrac12 + 2\sqrt{2},\pm1]_{\pm4}$   
        \\
\hline

SVEC 
& $[5,\pm4]_0$ 
\\
\hline

LVEC 
& $ \begin{array}{l}
\left( 2 \times [\tfrac12 (1 + \sqrt{57}),\pm2]_0 \right) 
\oplus [\tfrac12 (1 + \sqrt{53}),\pm2]_{\pm2} 
\oplus [\tfrac12 (1 + \sqrt{41}),\pm2]_{\pm4}

\oplus  \left( 3 \times [4,0]_0 \right)  
\\
{\oplus}  \left( 2 \times [\tfrac12(1 +  3\sqrt{5}),0]_{\pm2} \right)
\oplus [\tfrac12 (1 + \sqrt{33}),0]_{\pm4} 
\oplus [\tfrac12 (1 + \sqrt{13}),0]_{\pm6}
\end{array}$ 
\\
\hline
HYP 
& $ [6,\pm6]_0 $  
 \\
\hline
\hline

\textbf{Multiplet} & $k=0$     \\
\hline

MGRAV 
&  $ [2,0]_0 $      \\
\hline

SGINO 
& $[\tfrac52,\pm 1]_0$  
        \\
\hline

LGINO 
& $[\tfrac12 + \sqrt{3},\pm 1]_{\pm 2}$   
        \\
\hline

MVEC$_*$ 
& $[1,0]_{\pm4}$   \\
\hline

SVEC 
& $[3,\pm 2]_0$ 
\\
\hline

LVEC 
& $ [\tfrac12 (1 + \sqrt{17}),0]_0$ 
\\
\hline
HYP 
& $ [4,\pm4]_0 \oplus [2,\pm2]_{\pm4} $  
 \\
\hline
\hline
\end{tabular}
}
\caption{\footnotesize{Universal putative spectrum of $\textrm{OSp}(4|2) \times \mathrm{U}(1)_S$ supermultiplets for the SLAG solution, for the first few KK levels $k=0,1,2$. The OSp$(4|2)$ supermultiplets are denoted with the acronyms of \cite{Klebanov:2008vq}. The entries collect the superconformal primary dimension $E_0$ and R-charge $y_0$ for each $\textrm{OSp}(4|2)$ supermultiplet, along with the non-conserved charge U$(1)_S$ $f$ (identical for all states in the same supermultiplet) as $[E_0,y_0]_f$. Supermultiplets marked with a $*$ subindex contain generalised mass eigenstates. An entry with $n\times$ in front is $n$ times repeated. An entry with multiple $\pm$ signs contains all possible sign combinations, see the main text. Level $k=0$ reproduces table 5 of \cite{Pico:2025cmc}.
}\normalsize}
\label{tab:N=8SLAGMultiplets}
\end{table}
}

Further, all states in the putative spectrum organise themselves KK level by KK level in representations of OSp$(4|2) \times \textrm{U}(1)_S$, namely, in OSp$(4|2)$ supermultiplets with the same U$(1)_S$ charge for all states therein. This was already noted at level $k=0$ in \cite{Pico:2025cmc}, and holds for all $k \geq 0$. The tables of appendix A of \cite{Klebanov:2008vq} are very helpful to allocate states into OSp$(4|2)$ multiplets. We borrow their notation for the latter. Supermultiplets are characterised by labels, GRAV, GINO, VEC or HYP that correlate with their top spin component, $s_{\textrm{top}} =2, \frac32 , 1, \frac12$, respectively. Additional labels L, S or M are attached if the supermultiplet is long, corresponding to the generic situation, or undergoes shortening or becomes massless. The dimensions and R-charges of each descendant state in the multiplet follow from those, $[E_0 , y_0]$, of the superconformal primary. The latter has spin $s_0 =1, \frac12, 0 , 0$ for the four types of multiplets listed above. All states in a given multiplet share the same non-conserved $\textrm{U}(1)_S$ charge $f$.

We find that the dimension $E_0$ of the superconformal primary with spin $s_0$ of an OSp$(4|2)$ multiplet, long, short or massless, present in the spectrum at KK level $k$ with R-charge $y_0$ and U$(1)_S$ charge $f$ is given by the formula
\begin{equation} \label{eq:SLAGN8Form}
E_0=\tfrac12 + \sqrt{\tfrac{17}{4}-s_0(s_0+1)+2k(k+3)+\tfrac12 y_0^2 -\tfrac14 f^2}\,.
\end{equation}
We have also been able to obtain closed formulae to describe the  multiplets that arise at fixed $k$, for all $k=1,2,3, \ldots$ Except for the massless MGRAV$[2,0]_0$ sitting at $k=0$, all these multiplets are generically long. However, unlike in section~\ref{sec:ClassRSpec}, where shortening patterns have already been factored in, some dimensions and R-charges in the listings below might saturate the superconformal bounds and become short. With this in mind, the complete list of (generically long) multiplets present in the putative KK spectrum is as follows. There are two generic LGRAV towers, starting at $k=1$ and  $k=2$, with superconformal primary charges $[E_0,y_0]_f$ given by 
\begin{equation} \label{eq:LGRAVFormula}
[2,0]_0\oplus \bigoplus_{k=1}^{\infty} \bigg( \bigoplus_{q=0}^{k} [E_0,4q-2k]_0 \oplus \bigoplus_{q=-k}^{k}[E_0,0]_{2q} \bigg) \oplus \bigoplus_{k=2}^{\infty} \bigoplus_{p=1}^{k-1}\bigoplus_{q=0}^p \bigoplus_{r=0}^{2(k-p)}[E_0,4q-2p]_{2r-2(k-p)}
\end{equation}
and $E_0$ given by the corresponding value of (\ref{eq:SLAGN8Form}) with $s_0=1$. In addition, there are eight generic LGINO towers, four of them starting at $k=1$ and the other four at $k=2$, with superconformal primary charges $[E_0,y_0]_f$ given by 
\begin{eqnarray} \label{eq:LGINOLVECFormula}
&  \bigoplus_{k=1}^{\infty} \Big( \bigoplus_{q=0}^{k} [E_0,4q-2k +m_1]_{m_2} \oplus \bigoplus_{q=-k}^{k}[E_0,m_1]_{2q+m_2} \Big)   \nonumber \\%[4pt]
& \qquad  
 \oplus \bigoplus_{k=2}^{\infty} \bigoplus_{p=1}^{k-1}\bigoplus_{q=0}^p \bigoplus_{r=0}^{2(k-p)}[E_0,4q-2p+m_1]_{2r-2(k-p)+m_2} \; . 
\end{eqnarray} \label{eq:IntGINO}
Here, $(m_1 , m_2)$ are integers fixed for each tower to one of the four possibilities
\begin{equation}
\textrm{LGINO: \quad  $ (m_1 , m_2)=(1 , 2)$, or $(1 , -2)$, or $(-1 , 2)$, or $(-1 ,- 2)$,}
\end{equation}
and each $E_0$ again given by the appropriate value of (\ref{eq:SLAGN8Form}) with, now, $s_0=\tfrac12$. Finally, there are ten generic LVEC towers, five of them starting at $k=1$ and the other five at $k=2$. The superconformal primary charges $[E_0,y_0]_f$ turn out to be given again by (\ref{eq:LGINOLVECFormula}), with now $m_2=0$ and $m_1$ fixed to one of the five possibilities
\begin{equation} \label{eq:IntVEC}
\textrm{LVEC: \quad  $ m_1 =0 $, or $2$, or $-2$, or $4 $, or $-4 $,}
\end{equation}
and $E_0$ given by the appropriate value of (\ref{eq:SLAGN8Form}) with $s_0=0$. 

We emphasise that the generically long multiplets (\ref{eq:LGRAVFormula})--(\ref{eq:IntVEC}) might still undergo shortening if the dimensions and charges hit the superconformal bound. In this case, (\ref{eq:SLAGN8Form}) still reproduces the relevant short dimensions, and the supermultiplets split according to the recombination rules
\begin{eqnarray} \label{eq:N=2Recomb}
&& \textrm{LGRAV}[|y_0|+2,y_0] =  \textrm{SGRAV}[|y_0|+2,y_0] \oplus \textrm{SGINO}[|y_0|+ \tfrac52, \text{sgn}(y_0) (|y_0| + 1)] \; , \nonumber \\[4pt]
&& \textrm{LGINO}[|y_0|+\tfrac32 ,y_0] =  \textrm{SGINO}[|y_0|+ \tfrac32,y_0] \oplus \textrm{SVEC}[|y_0|+ 2, \text{sgn}(y_0) (|y_0| + 1)] \; , \nonumber \\[4pt]
&& \textrm{LVEC}[|y_0|+1,y_0] = \textrm{SVEC}[|y_0|+1,y_0] \oplus\textrm{HYP}[|y_0|+2,\text{sgn}(y_0) (|y_0| + 2)]\; , 
\end{eqnarray}
We have taken these recombination rules into account in order to list, in table~\ref{tab:N=8SLAGMultiplets}, the actual mixture of (massless, at $k=0$), short and strictly long multiplets that arise in the putative spectrum (\ref{eq:LGRAVFormula})--(\ref{eq:IntVEC}) for the first few KK levels. In particular, the $k=0$ spectrum reproduces the result of \cite{Pico:2025cmc}. The asterisk in MVEC$_*$  there signifies that these multiplets contain (generalised) eigenvalues of the $k=0$ vector, (\ref{eq:KKvector}), and scalar, (\ref{eq:KKscalar}), mass matrices in Jordan chains. We do not observe this pathology in these or other KK mass matrices at the higher levels, $k=1,2,3$, that we have explicitly checked. This is so even if these mass matrices are not symmetric.

Keeping track of the recombinations (\ref{eq:N=2Recomb}), one can extract the short multiplets present in the putative spectrum (\ref{eq:LGRAVFormula})--(\ref{eq:IntVEC}) at all $k$. At KK level $k=0$,  the short multiplets are
\begin{eqnarray} \label{eq:SLAGN8ShortKKZero}
& \textrm{MGRAV}[2,0]_0 \oplus \textrm{SGINO}[\tfrac52,\pm1]_0 \oplus \textrm{MVEC}_*[1,0]_{\pm4} \oplus \textrm{SVEC}[3,\pm2]_{0} \nonumber \\[3pt]
& \oplus \textrm{HYP}[4,\pm4]_{0}\oplus \textrm{HYP}[2,\pm2]_{\pm4} \; ,
\end{eqnarray}
and, at all higher KK levels $k \geq 1$,
\begin{eqnarray} \label{eq:SLAGN8ShortKK}
& \textrm{SGRAV}[2k+2 ,\pm2k]_0 \oplus \textrm{SGINO}[2k+\tfrac52,\pm (2k+1)]_0 \nonumber \\[3pt]
& \oplus \textrm{SVEC}[2k+3,\pm (2k+2)]_{0} \oplus \textrm{HYP}[2k+4,\pm  (2k+4)]_{0} \; .
\end{eqnarray}
In (\ref{eq:SLAGN8ShortKKZero}), (\ref{eq:SLAGN8ShortKK}), as in table \ref{tab:N=8SLAGMultiplets}, an entry with multiple $\pm$ signs contains all possible sign combinations. For example $[4,\pm 4]_0 \equiv [4,4]_0 \oplus [4,-4]_0$, or $[2,\pm2]_{\pm4} \equiv [2,2]_{4} \oplus [2,2]_{-4} \oplus [2,-2]_{4} \oplus [2,-2]_{-4}$. The list (\ref{eq:SLAGN8ShortKK}) adds generically local SGINO and SVEC multiplets to the list (\ref{eq:ShortSLAGUniv}) of globally defined short multiplets.

As explained in section~\ref{sec:PutUniv}, the globally defined supermultiplet spectrum (\ref{eq:GRAVSLAGUniv})--(\ref{eq:SLAGForm}) discussed in section \ref{sec:ClassRSpec} arises from the generically local putative spectrum (\ref{eq:SLAGN8Form})--(\ref{eq:IntVEC}) by selecting the SO$(3)_S$-invariant states therein, with SO$(3)_S$ defined in (\ref{eq:SO3SBreaking2}), (\ref{eq:GroupTh}). Extracting the SO$(3)_S$-invariant spectrum is not as straightforward as it seems because, as we have already emphasised, the putative spectrum comes in representations of only OSp$(4|2) \times \textrm{U}(1)_S$ and not OSp$(4|2) \times \textrm{SO}(3)_S$. For this reason, the exercise needs to be done by expanding the putative spectrum in individual states, keeping track of $\textrm{U}(1)_S \subset \textrm{SO}(3)_S$ charges, and then recombining the $\textrm{SO}(3)_S$-singlet states back into OSp$(4|2)$ supermultiplets. Altogether, the global spectrum (\ref{eq:GRAVSLAGUniv})--(\ref{eq:SLAGForm}) contains modes in (\ref{eq:SLAGN8Form})--(\ref{eq:IntVEC}) with $f=0$, but not all of them: only $f=0$ modes make it to the global spectrum that do not sit along with $f\neq 0$ modes in non-singlet representations of $\textrm{SO}(3)_S$. From this point of view, it is remarkable that the $\textrm{SO}(3)_S$-invariant states in the putative spectrum can be repackaged in OSp$(4|2)$ representations. Of course, this is a consequence of the fact that the relevant supercharges commute with SO$(3)_S$.

%%%%%%%%%%%%%%%

\subsection{Putative A3C spectrum} \label{sec:N=8SpecA3C}

%%%%%%%%%%%%%%%

{
\renewcommand{\arraystretch}{1.2} % Adjust row spacing globally

\begin{table}%[H]
%\centering
\resizebox{\textwidth}{!}{
\begin{tabular}{l | l } 
\hline
\hline

\textbf{Multiplet} & $k=2$     \\
\hline

GRAV 
& $\left[ 1 + \tfrac{\sqrt{109}}{2} \right] \otimes [0]_0 

\oplus \left[ 1 + \tfrac{\sqrt{401}}{4} \right] \otimes [\tfrac12]_{\pm\tfrac12} 
 
\oplus \left[ 1 + \tfrac{\sqrt{89}}{2} \right] \otimes [1]_{0} 
\oplus \left[ 1 + \sqrt{21} \right] \otimes [1]_{\pm 1}$   \\
\hline \\
GINO 
& $
\begin{array}{l}
\left( 2 \times \left[ 1 + \sqrt{29} \right] \otimes [0]_{0} \right) 
\oplus \left( 2 \times \left[ 1 + \tfrac{\sqrt{111}}{2} \right] \otimes [0]_{\pm 1} \right) 

\oplus \left( 4 \times  \left[ 1 + \tfrac{\sqrt{429}}{4} \right] \otimes [\tfrac12]_{\pm \tfrac12} \right)
\oplus \left( 2 \times  \left[ 1 + \tfrac{\sqrt{389}}{4} \right] \otimes [\tfrac12]_{\pm \tfrac32} \right) 
\\
\oplus \left( 4 \times  \left[ 1 + 2\sqrt{6}  \right] \otimes [1]_{0} \right) 
\oplus \left( 2 \times  \left[ 1 + \tfrac{\sqrt{91}}{2} \right] \otimes [1]_{\pm1} \right) 
\oplus  \left[ 1 + \sqrt{19}  \right] \otimes [1]_{\pm 2}

\oplus \left( 2 \times \left[ 1 + \tfrac{\sqrt{309}}{4} \right] \otimes [\tfrac32]_{\pm \tfrac12} \right)
\\
\oplus \left[ 1 + \tfrac{\sqrt{269}}{4} \right] \otimes [\tfrac32]_{\pm \tfrac32} 
\end{array}$
  \\
\hline \\
VEC 
&  $
\begin{array}{l}
\left( 4 \times \left[ \tfrac{13}{2} \right]\otimes [0]_{0} \right) 
\oplus \left( 3 \times \left[ 1 + \sqrt{29} \right]\otimes [0]_{\pm 1} \right) 
\oplus \left[ 1 + \tfrac{\sqrt{101}}{2} \right]\otimes [0]_{\pm 2 } 

\oplus \left( 7 \times \left[ 1 + \tfrac{\sqrt{449}}{4} \right]\otimes [\tfrac12]_{\pm \tfrac12} \right)
\\
\oplus \left( 4 \times \left[ 1 + \tfrac{\sqrt{409}}{4} \right]\otimes [\tfrac12]_{\pm \tfrac32} \right) 
\oplus \left[ 1 + \tfrac{\sqrt{329}}{4} \right]\otimes [\tfrac12]_{\pm \tfrac52} 
 
\oplus \left( 7 \times \left[ 1 + \tfrac{\sqrt{101}}{2} \right]\otimes [1]_{0}  \right)  
\oplus \left( 5 \times \left[ 1 + 2\sqrt{6} \right]\otimes [1]_{\pm 1}  \right) 
\\
\oplus \left( 2 \times \left[ \tfrac{11}{2} \right]\otimes [1]_{\pm 2} \right)

\oplus  \left( 4 \times \left[ 1 + \tfrac{\sqrt{329}}{4} \right]\otimes [\tfrac32]_{\pm \tfrac12} \right)
\oplus  \left( 2 \times \left[ \tfrac{21}{4} \right]\otimes [\tfrac32]_{\pm \tfrac32} \right) 
\oplus \left[ 1 + \tfrac{\sqrt{209}}{4} \right]\otimes [\tfrac32]_{\pm \tfrac52} 

\\

\oplus \left[ 1 + \tfrac{\sqrt{61}}{2} \right]\otimes [2]_{0}
\oplus \left[ 1 + \sqrt{14} \right]\otimes [2]_{\pm 1} 
\end{array}
$
    \\
\hline \\
CHIRAL 
& $
\begin{array}{l}
\left( 6 \times \left[ 1 + \sqrt{31} \right]\otimes [0]_{0} \right)  
\oplus \left( 2 \times \left[ 1 + \tfrac{\sqrt{119}}{2} \right]\otimes [0]_{\pm 1} \right) 
\oplus \left( 2 \times \left[ 1 + \sqrt{26}  \right]\otimes [0]_{\pm 2} \right) 

\oplus \left(  6 \times \left[ 1 + \tfrac{\sqrt{461}}{4}  \right]\otimes [\tfrac12]_{\pm \tfrac12} \right)
\\
\oplus \left(  3 \times \left[ 1 + \tfrac{\sqrt{421}}{4}  \right]\otimes [\tfrac12]_{\pm \tfrac32} \right) 
\oplus  \left[ 1 + \tfrac{\sqrt{341}}{4} \right]\otimes [\tfrac12]_{\pm \tfrac52}  

\oplus \left( 6 \times \left[ 1 + \sqrt{26} \right]\otimes [1]_{0}  \right) 
\oplus \left( 6 \times \left[ 1 + \tfrac{3\sqrt{11}}{2} \right]\otimes [1]_{\pm 1}  \right)

\\
\oplus \left( 2 \times \left[ 1 + \sqrt{21} \right]\otimes [1]_{\pm 2}  \right)  
\oplus {\left[ 1 + \tfrac{\sqrt{59}}{2} \right]}\otimes [1]_{\pm 3}

\oplus \left( 3 \times \left[ 1 + \tfrac{\sqrt{341}}{4} \right]\otimes [\tfrac32]_{\pm\tfrac12}  \right)
\oplus \left( 2 \times \left[ 1 + \tfrac{\sqrt{301}}{4} \right]\otimes [\tfrac32]_{\pm\tfrac32}  \right) 

\\

\oplus \left( 2 \times \left[ 5 \right]\otimes [2]_{0} \right)
\oplus  \left[ 1 + \tfrac{\sqrt{59}}{2} \right]\otimes [2]_{\pm 1} 
\oplus {\left[ 1 + \sqrt{11} \right]}\otimes [2]_{\pm 2} 
\end{array}
$
\\
\hline
\hline

\textbf{Multiplet} & $k=1$     \\
\hline

GRAV 
& $ \left[ \tfrac92 \right] \otimes [0]_0  \oplus \left[ 1 + \tfrac{\sqrt{161}}{4} \right] \otimes [\tfrac12]_{\pm\tfrac12} $   \\
\hline \\
GINO 
& 

$\begin{array}{l}

\left[ 1 + \sqrt{14} \right]\otimes [0]_{0} 
\oplus \left[ 1 + \tfrac{\sqrt{51}}{2} \right]\otimes [0]_{\pm 1} 

\oplus \left( 2 \times \left[ 1 + \tfrac{3 \sqrt{21}}{4} \right]\otimes [\tfrac12]_{\pm\tfrac12} \right)
\oplus \left[ 1 + \tfrac{\sqrt{149}}{4} \right]\otimes [\tfrac12]_{\pm\tfrac32} 

\\

\oplus \left( 2 \times \left[ 4 \right]\otimes [1]_{0} \right) 
\oplus \left[ 1 + \tfrac{\sqrt{31}}{2} \right]\otimes [1]_{\pm 1} 
\end{array}
$
  \\
\hline \\
VEC &  
$\begin{array}{l}
\left( 3 \times \left[ 1 + \tfrac{\sqrt{61}}{2} \right]\otimes [0]_{0}\right)
\oplus  \left( 3 \times \left[ 1 + \sqrt{14} \right]\otimes [0]_{\pm 1}\right) 
\oplus \left[ 1 + \tfrac{\sqrt{41}}{2} \right]\otimes [0]_{\pm2} 

\oplus \left( 4 \times \left[ 1 + \tfrac{\sqrt{209}}{4} \right]\otimes [\tfrac12]_{\pm \tfrac12}\right) 
\\
\oplus \left( 2 \times \left[ \tfrac{17}{4} \right]\otimes [\tfrac12]_{\pm \tfrac32}\right)

\oplus \left( 3 \times \left[ 1 + \tfrac{\sqrt{41}}{2} \right]\otimes [1]_{0} \right)
\oplus \left( 2 \times \left[ 4 \right]\otimes [1]_{\pm 1} \right) 
\oplus \left[ 1 + \tfrac{\sqrt{21}}{2} \right]\otimes [1]_{\pm 2}

\oplus \left[ 1 + \tfrac{\sqrt{89}}{4}  \right]\otimes [\tfrac32]_{\pm \tfrac12} 
\end{array}$
    \\
\hline \\
CHIRAL & 
$
\begin{array}{l}
\left( 4 \times \left[ 5 \right]\otimes [0]_0 \right)
\oplus \left[ 1+\tfrac{\sqrt{59}}{2} \right]\otimes [0]_{\pm1} 
\oplus \left[ 1 + \sqrt{11} \right]\otimes [0]_{\pm 2}

\oplus \left( 4 \times \left[ 1+\tfrac{\sqrt{221}}{4} \right]\otimes [\tfrac12]_{\pm \tfrac12} \right) 
\\
\oplus \left( 2 \times \left[ 1+\tfrac{\sqrt{181}}{4} \right]\otimes [\tfrac12]_{\pm \tfrac32}\right)  
\oplus \left[ 1+\tfrac{\sqrt{101}}{4} \right] \otimes [\tfrac12]_{\pm \tfrac52}

\oplus \left( 2 \times \left[ 1+\sqrt{11} \right]\otimes [1]_{0} \right)
\oplus \left( 2 \times \left[ 1+\tfrac{\sqrt{39}}{2} \right]\otimes [1]_{\pm 1} \right) 

\\

\oplus \left[ 1+\tfrac{\sqrt{101}}{4} \right]\otimes [\tfrac32]_{\pm \tfrac12} 
\oplus \left[ 1+\tfrac{\sqrt{61}}{4} \right]\otimes [\tfrac32]_{\pm \tfrac32} 
\end{array}
$
\\
\hline
\hline

\textbf{Multiplet} & $k=0$     \\
\hline
MGRAV 
&  $\left[ \tfrac52 \right] \otimes [0]_0$  \\
\hline \\
GINO 
& $  [3] \otimes[ 0]_{0} 
\; \oplus \;
\left[  1 + \tfrac{\sqrt{11}}{2} \right] \otimes \left[0\right]_{\pm1}  
\; \oplus \;
\left[ 1 + \tfrac{\sqrt{29}}{4} \right] \otimes \left[\tfrac12\right]_{\pm\tfrac12}$
  \\
\hline \\
MVEC 
&  $\left[ \tfrac32 \right] \otimes [1]_0$ 
    \\
\hline \\
VEC 
&  $ \left[  1 + \tfrac{\sqrt{21}}{2} \right] \otimes[0]_{0}
\; \oplus \;
[3] \otimes[0]_{\pm1}
\; \oplus \;
\left[\tfrac{11}4 \right] \otimes \left[\tfrac12 \right]_{\pm \tfrac12}
\; \oplus \;
\left[\tfrac74\right] \otimes \left[\tfrac12\right]_{\pm \tfrac32}
$
    \\
\hline \\
CHIRAL 
& $ \left( 2 \times  \left[1+\sqrt{6} \right] \otimes [0]_0 \right)
\; \oplus \;
[2] \otimes[0]_{\pm 2}
\; \oplus \;
 \left[ 1 + \tfrac{\sqrt{61}}{4} \right] \otimes \left[\tfrac12 \right]_{\pm \tfrac12}
\; \oplus \;
 [2]\otimes [1]_{0}  
\; \oplus \;
 \left[ 1 + \tfrac{i}{2} \right] \otimes \left[1 \right]_{\pm 1}   $
\\
\hline
\hline
\end{tabular}
}
\caption{\footnotesize{Universal putative spectrum of $\textrm{OSp}(4|1) \times \textrm{SO}(3)_+ \times \textrm{U}(1)^\prime_S$ supermultiplets for the A3C solution, for the first few KK levels $k=0,1,2$. The OSp$(4|1)$ supermultiplets are denoted with the acronyms of \cite{Cesaro:2020soq}. The entries collect the superconformal primary dimension $E_0$ for each OSp$(4|1)$ supermultiplet, the spin $\ell$ of $\textrm{SO}(3)_+$ and the charge $m$ of $\textrm{U}(1)^\prime_S$ in the format $[ E_0 ] \otimes [\ell]_m$. An entry with $n\times$ in front is $n$ times repeated. Level $k=0$ reproduces table 6 of \cite{Pico:2025cmc}.
}\normalsize}
\label{tab:N=8A3CMultiplets}
\end{table}
}

We have similarly computed the putative KK spectrum for the $\cN=1$ A3C solution. The states arrange themselves KK level by KK level in representations of OSp$(4|1) \times \textrm{SO}(3)_+ \times \textrm{U}(1)_S^\prime$, even if only $\textrm{SO}(3)_+$, not $\textrm{U}(1)_S^\prime$, is a symmetry of the solution. This behaviour was already observed in \cite{Pico:2025cmc} at KK level $k=0$, and holds for all $k$. The $\textrm{SO}(3)_+$ factor here is defined in (\ref{eq:SO3SBreaking2}), (\ref{eq:GroupTh}), and $\textrm{U}(1)_S^\prime$ is the Cartan subgroup of $\textrm{SO}(3)_S^\prime$ defined in (\ref{eq:GroupTh}). Again, only $\textrm{SO}(3)_+$, not $\textrm{U}(1)_S^\prime$, has associated conserved (flavour) currents. These are contained in the $k=0$ MVEC$\left[ \tfrac32 \right] \otimes [1]_0$ multiplet indicated in table~\ref{tab:N=8A3CMultiplets}. 

Table 1 of \cite{Cesaro:2020soq} is very helpful to organise the states of definite mass and $\textrm{SO}(3)_+ \times \textrm{U}(1)_S^\prime$ charge in OSp$(4|1)$ multiplets, and we will use their terminology. In this notation, multiplets with top $s_{\textrm{top}} =2, \frac32 , 1, \frac12$ and superconformal primary spin $s_{0} =\frac32 , 1, \frac12 , 0$, are respectively denoted (M)GRAV, GINO, (M)VEC or CHIRAL. Recall that there is no R-symmetry in the $\cN=1$ case so, in particular, the multiplets cannot get shortened unless they are massless. This is indicated with an M in front of the acronym. The supermultiplets present in the spectrum are characterised by one of the four labels above, along with the superconformal primary dimension $E_0$, the $\textrm{SO}(3)_+$ spin $\ell$ and the non-conserved $\textrm{U}(1)_S^\prime$ charge $m$. We have employed the notation LABEL$[E_0] \otimes [\ell]_m$ to specify a supermultiplet with those quantum numbers.

The supermultiplets have been tabulated for the first few KK levels in table \ref{tab:N=8A3CMultiplets}. In particular, the graviton multiplets are described by the expression
\begin{equation} \label{eq:A3CGravs}
\textrm{MGRAV}[0]_0 \oplus \bigoplus_{k=1}^{\infty} \bigoplus_{p=0}^{k} \bigoplus_{q=-p}^{q=p} \textrm{GRAV}[E_0]\otimes [\tfrac{p}{2}]_{\frac{q}{2}} \; . 
\end{equation}
We have also been able to find a closed formula for the dimension $E_0$ of any of the four type of multiplets with superconformal primary spin $s_0$, arising at KK level $k$ with $\textrm{SO}(3)_+$ spin $\ell$ and $\textrm{U}(1)_S^\prime$ charge $m$. It reads
\begin{equation} \label{eq:A3CN8Form}
E_0= 1 + \sqrt{6-s_0(s_0+1)+\tfrac52 k(k+3) -\tfrac52 \ell (\ell+1) -\tfrac54 m^2}\,.
\end{equation}
In particular, $E_0$ in (\ref{eq:A3CGravs}) is given by (\ref{eq:A3CN8Form}) with $s_0 = \tfrac32$ and the indicated values of the $\textrm{SO}(3)_+ \times \textrm{U}(1)_S^\prime$ charges.

As already noted in \cite{Pico:2025cmc}, the A3C putative spectrum at KK level $k=0$ contains non-physical modes of complex mass. We do not observe these pathologies at higher KK levels, in spite of the KK mass matrices (\ref{eq:KKgraviton})--(\ref{eq:KKscalar}) still being non-symmetric. The universal, globally defined A3C supermultiplet spectrum presented in section \ref{sec:N=1Spec} is obtained by selecting the SO$(3)_S^\prime$-invariant modes within the putative spectrum described here, following a process similar to that described in appendix \ref{sec:N=8SpecSLAG} for the SLAG case. Again, it is remarkable that the global SO$(3)_S^\prime$-neutral states rearrange themselves in representations of OSp$(4|1) \times \textrm{SO}(3)_+$, given that the putative states do not come in representations of OSp$(4|1) \times \textrm{SO}(3)_+ \times \textrm{SO}(3)_S^\prime$.

%%%%%%%%%%%%%%%
%%%%%%%%%%%%%%%

\bibliography{references}

\providecommand{\href}[2]{#2}\begingroup\raggedright\begin{thebibliography}{10}

\bibitem{Dimofte:2011ju}
T.~Dimofte, D.~Gaiotto, and S.~Gukov, {\it {Gauge Theories Labelled by
  Three-Manifolds}},  {\em Commun. Math. Phys.} {\bf 325} (2014) 367--419,
  [\href{http://arxiv.org/abs/1108.4389}{{\tt arXiv:1108.4389}}].

\bibitem{Dimofte:2011py}
T.~Dimofte, D.~Gaiotto, and S.~Gukov, {\it {3-Manifolds and 3d Indices}},  {\em
  Adv. Theor. Math. Phys.} {\bf 17} (2013), no.~5 975--1076,
  [\href{http://arxiv.org/abs/1112.5179}{{\tt arXiv:1112.5179}}].

\bibitem{Dimofte:2014zga}
T.~Dimofte, {\it {Complex Chern{\textendash}Simons Theory at Level k via the
  3d{\textendash}3d Correspondence}},  {\em Commun. Math. Phys.} {\bf 339}
  (2015), no.~2 619--662, [\href{http://arxiv.org/abs/1409.0857}{{\tt
  arXiv:1409.0857}}].

\bibitem{Gang:2014ema}
D.~Gang, N.~Kim, and S.~Lee, {\it {Holography of 3d-3d correspondence at Large
  N}},  {\em JHEP} {\bf 04} (2015) 091,
  [\href{http://arxiv.org/abs/1409.6206}{{\tt arXiv:1409.6206}}].

\bibitem{Gang:2014qla}
D.~Gang, N.~Kim, and S.~Lee, {\it {Holography of wrapped M5-branes and
  Chern{\textendash}Simons theory}},  {\em Phys. Lett. B} {\bf 733} (2014)
  316--319, [\href{http://arxiv.org/abs/1401.3595}{{\tt arXiv:1401.3595}}].

\bibitem{Gauntlett:2006ux}
J.~P. Gauntlett, O.~A.~P. Mac~Conamhna, T.~Mateos, and D.~Waldram, {\it {AdS
  spacetimes from wrapped M5 branes}},  {\em JHEP} {\bf 11} (2006) 053,
  [\href{http://arxiv.org/abs/hep-th/0605146}{{\tt hep-th/0605146}}].

\bibitem{Maldacena:2000mw}
J.~M. Maldacena and C.~Nunez, {\it {Supergravity description of field theories
  on curved manifolds and a no go theorem}},  {\em Int. J. Mod. Phys. A} {\bf
  16} (2001) 822--855, [\href{http://arxiv.org/abs/hep-th/0007018}{{\tt
  hep-th/0007018}}].

\bibitem{Acharya:2000mu}
B.~S. Acharya, J.~P. Gauntlett, and N.~Kim, {\it {Five-branes wrapped on
  associative three cycles}},  {\em Phys. Rev. D} {\bf 63} (2001) 106003,
  [\href{http://arxiv.org/abs/hep-th/0011190}{{\tt hep-th/0011190}}].

\bibitem{Bobev:2019zmz}
N.~Bobev and P.~M. Crichigno, {\it {Universal spinning black holes and theories
  of class $ \mathcal{R} $}},  {\em JHEP} {\bf 12} (2019) 054,
  [\href{http://arxiv.org/abs/1909.05873}{{\tt arXiv:1909.05873}}].

\bibitem{Benini:2019dyp}
F.~Benini, D.~Gang, and L.~A. Pando~Zayas, {\it {Rotating Black Hole Entropy
  from M5 Branes}},  {\em JHEP} {\bf 03} (2020) 057,
  [\href{http://arxiv.org/abs/1909.11612}{{\tt arXiv:1909.11612}}].

\bibitem{Donos:2010ax}
A.~Donos, J.~P. Gauntlett, N.~Kim, and O.~Varela, {\it {Wrapped M5-branes,
  consistent truncations and AdS/CMT}},  {\em JHEP} {\bf 12} (2010) 003,
  [\href{http://arxiv.org/abs/1009.3805}{{\tt arXiv:1009.3805}}].

\bibitem{Maldacena:1997re}
J.~M. Maldacena, {\it {The Large N limit of superconformal field theories and
  supergravity}},  {\em Int. J. Theor. Phys.} {\bf 38} (1999) 1113--1133,
  [\href{http://arxiv.org/abs/hep-th/9711200}{{\tt hep-th/9711200}}]. [Adv.
  Theor. Math. Phys.2,231(1998)].

\bibitem{Gubser:1998bc}
S.~Gubser, I.~R. Klebanov, and A.~M. Polyakov, {\it {Gauge theory correlators
  from noncritical string theory}},  {\em Phys. Lett. B} {\bf 428} (1998)
  105--114, [\href{http://arxiv.org/abs/hep-th/9802109}{{\tt hep-th/9802109}}].

\bibitem{Witten:1998qj}
E.~Witten, {\it {Anti-de Sitter space and holography}},  {\em Adv. Theor. Math.
  Phys.} {\bf 2} (1998) 253--291,
  [\href{http://arxiv.org/abs/hep-th/9802150}{{\tt hep-th/9802150}}].

\bibitem{Kim:1985ez}
H.~Kim, L.~Romans, and P.~van Nieuwenhuizen, {\it {The Mass Spectrum of Chiral
  N=2 D=10 Supergravity on $S^5$}},  {\em Phys.Rev.} {\bf D32} (1985) 389.

\bibitem{Malek:2019eaz}
E.~Malek and H.~Samtleben, {\it {Kaluza-Klein Spectrometry for Supergravity}},
  {\em Phys. Rev. Lett.} {\bf 124} (2020), no.~10 101601,
  [\href{http://arxiv.org/abs/1911.12640}{{\tt arXiv:1911.12640}}].

\bibitem{Malek:2020yue}
E.~Malek and H.~Samtleben, {\it {Kaluza-Klein Spectrometry from Exceptional
  Field Theory}},  {\em Phys. Rev. D} {\bf 102} (2020), no.~10 106016,
  [\href{http://arxiv.org/abs/2009.03347}{{\tt arXiv:2009.03347}}].

\bibitem{Varela:2020wty}
O.~Varela, {\it {Super-Chern-Simons spectra from Exceptional Field Theory}},
  {\em JHEP} {\bf 04} (2021) 283, [\href{http://arxiv.org/abs/2010.09743}{{\tt
  arXiv:2010.09743}}].

\bibitem{Cesaro:2020soq}
M.~Ces{\`a}ro and O.~Varela, {\it {Kaluza-Klein fermion mass matrices from
  exceptional field theory and $ \mathcal{N} $ = 1 spectra}},  {\em JHEP} {\bf
  03} (2021) 138, [\href{http://arxiv.org/abs/2012.05249}{{\tt
  arXiv:2012.05249}}].

\bibitem{Hohm:2013pua}
O.~Hohm and H.~Samtleben, {\it {Exceptional Form of D=11 Supergravity}},  {\em
  Phys.Rev.Lett.} {\bf 111} (2013) 231601,
  [\href{http://arxiv.org/abs/1308.1673}{{\tt arXiv:1308.1673}}].

\bibitem{Hohm:2013vpa}
O.~Hohm and H.~Samtleben, {\it {Exceptional Field Theory I: $E_{6(6)}$
  covariant Form of M-Theory and Type IIB}},  {\em Phys. Rev.} {\bf D89}
  (2014), no.~6 066016, [\href{http://arxiv.org/abs/1312.0614}{{\tt
  arXiv:1312.0614}}].

\bibitem{Hohm:2013uia}
O.~Hohm and H.~Samtleben, {\it {Exceptional Field Theory II: E$_{7(7)}$}},
  {\em Phys.Rev.} {\bf D89} (2014), no.~6 066017,
  [\href{http://arxiv.org/abs/1312.4542}{{\tt arXiv:1312.4542}}].

\bibitem{Coimbra:2011nw}
A.~Coimbra, C.~Strickland-Constable, and D.~Waldram, {\it {Supergravity as
  Generalised Geometry I: Type II Theories}},  {\em JHEP} {\bf 11} (2011) 091,
  [\href{http://arxiv.org/abs/1107.1733}{{\tt arXiv:1107.1733}}].

\bibitem{Coimbra:2011ky}
A.~Coimbra, C.~Strickland-Constable, and D.~Waldram, {\it {$E_{d(d)} \times
  \mathbb{R}^+$ generalised geometry, connections and M theory}},  {\em JHEP}
  {\bf 02} (2014) 054, [\href{http://arxiv.org/abs/1112.3989}{{\tt
  arXiv:1112.3989}}].

\bibitem{Coimbra:2012af}
A.~Coimbra, C.~Strickland-Constable, and D.~Waldram, {\it {Supergravity as
  Generalised Geometry II: $E_{d(d)} \times \mathbb{R}^+$ and M theory}},  {\em
  JHEP} {\bf 03} (2014) 019, [\href{http://arxiv.org/abs/1212.1586}{{\tt
  arXiv:1212.1586}}].

\bibitem{Berman:2020tqn}
D.~S. Berman and C.~D.~A. Blair, {\it {The Geometry, Branes and Applications of
  Exceptional Field Theory}},  {\em Int. J. Mod. Phys. A} {\bf 35} (2020),
  no.~30 2030014, [\href{http://arxiv.org/abs/2006.09777}{{\tt
  arXiv:2006.09777}}].

\bibitem{Pico:2025cmc}
M.~Pico and O.~Varela, {\it {Consistent subsectors of maximal supergravity and
  wrapped M5-branes}},  \href{http://arxiv.org/abs/2511.15892}{{\tt
  arXiv:2511.15892}}.

\bibitem{Pico:2026rji}
M.~Pico and O.~Varela, {\it {Maximal trombone supergravity from wrapped
  M5-branes}},  \href{http://arxiv.org/abs/2601.07960}{{\tt arXiv:2601.07960}}.

\bibitem{deWit:2007mt}
B.~de~Wit, H.~Samtleben, and M.~Trigiante, {\it {The Maximal D=4
  supergravities}},  {\em JHEP} {\bf 0706} (2007) 049,
  [\href{http://arxiv.org/abs/0705.2101}{{\tt arXiv:0705.2101}}].

\bibitem{LeDiffon:2008sh}
A.~Le~Diffon and H.~Samtleben, {\it {Supergravities without an Action: Gauging
  the Trombone}},  {\em Nucl. Phys. B} {\bf 811} (2009) 1--35,
  [\href{http://arxiv.org/abs/0809.5180}{{\tt arXiv:0809.5180}}].

\bibitem{LeDiffon:2011wt}
A.~Le~Diffon, H.~Samtleben, and M.~Trigiante, {\it {N=8 Supergravity with Local
  Scaling Symmetry}},  {\em JHEP} {\bf 04} (2011) 079,
  [\href{http://arxiv.org/abs/1103.2785}{{\tt arXiv:1103.2785}}].

\bibitem{Bhattacharya:2024tjw}
R.~Bhattacharya, A.~Katyal, and O.~Varela, {\it {Class S Superconformal Indices
  from Maximal Supergravity}},  {\em Phys. Rev. Lett.} {\bf 134} (2025), no.~18
  181601, [\href{http://arxiv.org/abs/2411.16837}{{\tt arXiv:2411.16837}}].

\bibitem{Varela:2025xeb}
O.~Varela, {\it {Trombone gaugings of five-dimensional maximal supergravity}},
  {\em JHEP} {\bf 02} (2026) 163, [\href{http://arxiv.org/abs/2509.12391}{{\tt
  arXiv:2509.12391}}].

\bibitem{BKV2026}
R.~Bhattacharya, A.~Katyal, and O.~Varela, {\it {In progress}}, .

\bibitem{Cassani:2019vcl}
D.~Cassani, G.~Josse, M.~Petrini, and D.~Waldram, {\it {Systematics of
  consistent truncations from generalised geometry}},  {\em JHEP} {\bf 11}
  (2019) 017, [\href{http://arxiv.org/abs/1907.06730}{{\tt arXiv:1907.06730}}].

\bibitem{Cassani:2020cod}
D.~Cassani, G.~Josse, M.~Petrini, and D.~Waldram, {\it {$\mathcal{N} $ = 2
  consistent truncations from wrapped M5-branes}},  {\em JHEP} {\bf 02} (2021)
  232, [\href{http://arxiv.org/abs/2011.04775}{{\tt arXiv:2011.04775}}].

\bibitem{Blair:2024ofc}
C.~D.~A. Blair, M.~Pico, and O.~Varela, {\it {Infinite and finite consistent
  truncations on deformed generalised parallelisations}},  {\em JHEP} {\bf 09}
  (2024) 065, [\href{http://arxiv.org/abs/2407.01298}{{\tt arXiv:2407.01298}}].

\bibitem{Josse:2025uro}
G.~Josse, M.~Petrini, and M.~Pico, {\it {Consistent Truncations and Generalised
  Geometry: Scanning through Dimensions and Supersymmetry}},
  \href{http://arxiv.org/abs/2512.03027}{{\tt arXiv:2512.03027}}.

\bibitem{Berman:2012uy}
D.~S. Berman, E.~T. Musaev, D.~C. Thompson, and D.~C. Thompson, {\it {Duality
  Invariant M-theory: Gauged supergravities and Scherk-Schwarz reductions}},
  {\em JHEP} {\bf 10} (2012) 174, [\href{http://arxiv.org/abs/1208.0020}{{\tt
  arXiv:1208.0020}}].

\bibitem{Lee:2014mla}
K.~Lee, C.~Strickland-Constable, and D.~Waldram, {\it {Spheres, generalised
  parallelisability and consistent truncations}},  {\em Fortsch. Phys.} {\bf
  65} (2017), no.~10-11 1700048, [\href{http://arxiv.org/abs/1401.3360}{{\tt
  arXiv:1401.3360}}].

\bibitem{Hohm:2014qga}
O.~Hohm and H.~Samtleben, {\it {Consistent Kaluza-Klein Truncations via
  Exceptional Field Theory}},  {\em JHEP} {\bf 1501} (2015) 131,
  [\href{http://arxiv.org/abs/1410.8145}{{\tt arXiv:1410.8145}}].

\bibitem{Godazgar:2014nqa}
H.~Godazgar, M.~Godazgar, O.~Hohm, H.~Nicolai, and H.~Samtleben, {\it
  {Supersymmetric E$_{7(7)}$ Exceptional Field Theory}},  {\em JHEP} {\bf 09}
  (2014) 044, [\href{http://arxiv.org/abs/1406.3235}{{\tt arXiv:1406.3235}}].

\bibitem{Musaev:2014lna}
E.~Musaev and H.~Samtleben, {\it {Fermions and supersymmetry in E$_{6(6)}$
  exceptional field theory}},  {\em JHEP} {\bf 03} (2015) 027,
  [\href{http://arxiv.org/abs/1412.7286}{{\tt arXiv:1412.7286}}].

\bibitem{deWit:2004nw}
B.~de~Wit, H.~Samtleben, and M.~Trigiante, {\it {The Maximal D=5
  supergravities}},  {\em Nucl. Phys.} {\bf B716} (2005) 215--247,
  [\href{http://arxiv.org/abs/hep-th/0412173}{{\tt hep-th/0412173}}].

\bibitem{Berman:2019izh}
D.~S. Berman, C.~D.~A. Blair, and R.~Otsuki, {\it {Non-Riemannian geometry of
  M-theory}},  {\em JHEP} {\bf 07} (2019) 175,
  [\href{http://arxiv.org/abs/1902.01867}{{\tt arXiv:1902.01867}}].

\bibitem{Bachas:2011xa}
C.~Bachas and J.~Estes, {\it {Spin-2 spectrum of defect theories}},  {\em JHEP}
  {\bf 06} (2011) 005, [\href{http://arxiv.org/abs/1103.2800}{{\tt
  arXiv:1103.2800}}].

\bibitem{Gauntlett:2002rv}
J.~P. Gauntlett, N.~Kim, S.~Pakis, and D.~Waldram, {\it {M theory solutions
  with AdS factors}},  {\em Class. Quant. Grav.} {\bf 19} (2002) 3927--3946,
  [\href{http://arxiv.org/abs/hep-th/0202184}{{\tt hep-th/0202184}}].

\bibitem{Guarino:2015qaa}
A.~Guarino and O.~Varela, {\it {Dyonic ISO(7) supergravity and the duality
  hierarchy}},  {\em JHEP} {\bf 02} (2016) 079,
  [\href{http://arxiv.org/abs/1508.04432}{{\tt arXiv:1508.04432}}].

\bibitem{Klebanov:2008vq}
I.~Klebanov, T.~Klose, and A.~Murugan, {\it {AdS(4)/CFT(3) Squashed, Stretched
  and Warped}},  {\em JHEP} {\bf 03} (2009) 140,
  [\href{http://arxiv.org/abs/0809.3773}{{\tt arXiv:0809.3773}}].

\bibitem{Kinney:2005ej}
J.~Kinney, J.~M. Maldacena, S.~Minwalla, and S.~Raju, {\it {An Index for 4
  dimensional super conformal theories}},  {\em Commun. Math. Phys.} {\bf 275}
  (2007) 209--254, [\href{http://arxiv.org/abs/hep-th/0510251}{{\tt
  hep-th/0510251}}].

\bibitem{Kim:2021zlh}
H.~Kim and N.~Kim, {\it {On the superconformal index of Chern-Simons theories
  and their KK spectrometry}},  {\em JHEP} {\bf 10} (2021) 241,
  [\href{http://arxiv.org/abs/2108.07182}{{\tt arXiv:2108.07182}}].

\bibitem{Bobev:2020zov}
N.~Bobev, A.~M. Charles, D.~Gang, K.~Hristov, and V.~Reys, {\it
  {Higher-derivative supergravity, wrapped M5-branes, and theories of class $
  \mathrm{\mathcal{R}} $}},  {\em JHEP} {\bf 04} (2021) 058,
  [\href{http://arxiv.org/abs/2011.05971}{{\tt arXiv:2011.05971}}].

\bibitem{Malek:2020mlk}
E.~Malek, H.~Nicolai, and H.~Samtleben, {\it {Tachyonic Kaluza-Klein modes and
  the AdS swampland conjecture}},  {\em JHEP} {\bf 08} (2020) 159,
  [\href{http://arxiv.org/abs/2005.07713}{{\tt arXiv:2005.07713}}].

\bibitem{Guarino:2020flh}
A.~Guarino, E.~Malek, and H.~Samtleben, {\it {Stable Nonsupersymmetric
  Anti{\textendash}de Sitter Vacua of Massive IIA Supergravity}},  {\em Phys.
  Rev. Lett.} {\bf 126} (2021), no.~6 061601,
  [\href{http://arxiv.org/abs/2011.06600}{{\tt arXiv:2011.06600}}].

\bibitem{Bobev:2020lsk}
N.~Bobev, E.~Malek, B.~Robinson, H.~Samtleben, and J.~van Muiden, {\it
  {Kaluza-Klein Spectroscopy for the Leigh-Strassler SCFT}},  {\em JHEP} {\bf
  04} (2021) 208, [\href{http://arxiv.org/abs/2012.07089}{{\tt
  arXiv:2012.07089}}].

\bibitem{Giambrone:2021zvp}
A.~Giambrone, E.~Malek, H.~Samtleben, and M.~Trigiante, {\it {Global properties
  of the conformal manifold for S-fold backgrounds}},  {\em JHEP} {\bf 06}
  (2021), no.~111 111, [\href{http://arxiv.org/abs/2103.10797}{{\tt
  arXiv:2103.10797}}].

\bibitem{Cesaro:2021haf}
M.~Cesaro, G.~Larios, and O.~Varela, {\it {Supersymmetric spectroscopy on
  AdS$_{4} \times S^{7}$ and AdS$_{4} \times S^{6}$}},  {\em JHEP} {\bf 07}
  (2021) 094, [\href{http://arxiv.org/abs/2103.13408}{{\tt arXiv:2103.13408}}].
  [Erratum: JHEP 05, 084 (2024)].

\bibitem{Cesaro:2021tna}
M.~Ces{\`a}ro, G.~Larios, and O.~Varela, {\it {The spectrum of
  marginally-deformed $ \mathcal{N} $ = 2 CFTs with AdS$_{4}$ S-fold duals of
  type IIB}},  {\em JHEP} {\bf 12} (2021) 214,
  [\href{http://arxiv.org/abs/2109.11608}{{\tt arXiv:2109.11608}}].

\bibitem{Eloy:2021fhc}
C.~Eloy, G.~Larios, and H.~Samtleben, {\it {Triality and the consistent
  reductions on AdS$_{3} \times S^{3}$}},  {\em JHEP} {\bf 01} (2022) 055,
  [\href{http://arxiv.org/abs/2111.01167}{{\tt arXiv:2111.01167}}].

\bibitem{Cesaro:2022mbu}
M.~Ces{\`a}ro, G.~Larios, and O.~Varela, {\it {$ \mathcal{N} $ = 1 S-fold
  spectroscopy}},  {\em JHEP} {\bf 08} (2022) 242,
  [\href{http://arxiv.org/abs/2206.04064}{{\tt arXiv:2206.04064}}].

\bibitem{Bonifacio:2020xoc}
J.~Bonifacio and K.~Hinterbichler, {\it {Bootstrap Bounds on Closed Einstein
  Manifolds}},  {\em JHEP} {\bf 10} (2020) 069,
  [\href{http://arxiv.org/abs/2007.10337}{{\tt arXiv:2007.10337}}].

\bibitem{Bonifacio:2023ban}
J.~Bonifacio, D.~Mazac, and S.~Pal, {\it {Spectral Bounds on Hyperbolic
  3-Manifolds: Associativity and the Trace Formula}},  {\em Commun. Math.
  Phys.} {\bf 406} (2025) 51, [\href{http://arxiv.org/abs/2308.11174}{{\tt
  arXiv:2308.11174}}].

\bibitem{Gesteau:2023brw}
E.~Gesteau, S.~Pal, D.~Simmons-Duffin, and Y.~Xu, {\it {Bounds on spectral gaps
  of Hyperbolic spin surfaces}},  {\em J. Assoc. Math. Res.} {\bf 3} (2025),
  no.~1 72--139, [\href{http://arxiv.org/abs/2311.13330}{{\tt
  arXiv:2311.13330}}].

\bibitem{Pacheco:2008ps}
P.~Pires~Pacheco and D.~Waldram, {\it {M-theory, exceptional generalised
  geometry and superpotentials}},  {\em JHEP} {\bf 09} (2008) 123,
  [\href{http://arxiv.org/abs/0804.1362}{{\tt arXiv:0804.1362}}].

\end{thebibliography}\endgroup

%%%%%%%%%%%%%%%
%%%%%%%%%%%%%%%

\end{document}